\documentclass[superscriptaddress,prd,nofootinbib,preprintnumbers,longbibliography,11pt,eqsecnum]{revtex4-1}
\usepackage{amsmath} 
\usepackage{graphicx} 
\usepackage[caption = false]{subfig}
\usepackage{siunitx}
\usepackage{amsthm}
\usepackage{tensor}
\usepackage{amssymb} 
\usepackage{dsfont}
\usepackage{yfonts}
\usepackage{hyperref}
\usepackage{floatrow,dcolumn,rotating}
\usepackage{array,xcolor}
\usepackage{booktabs,multirow}
\usepackage[utf8]{inputenc}
\usepackage{mathtools}

\usepackage[normalem]{ulem}

\usepackage[all]{xy}
\usepackage{tikz}
\usetikzlibrary{arrows.meta}

\hypersetup{colorlinks=true,linkcolor=blue,citecolor=red,urlcolor=blue}

\usepackage{etoolbox}
\patchcmd{\section}
  {\centering}
  {\raggedright}
  {}
  {}
\patchcmd{\subsection}
  {\centering}
  {\raggedright}
  {}
  {}
\newcommand{\be}{\begin{equation}}
\newcommand{\ee}{\end{equation}}
\newcommand{\bea}{\begin{eqnarray}}
\newcommand{\eea}{\end{eqnarray}}

\newcommand{\nn}{\nonumber\\}

\def\vecb#1{\mathbf{#1}}

\def\vecb#1{\mathbf{#1}}

\def\CC{\mathcal{C}}

\def\CJ{\mathcal{J}}

\def\CL{\mathcal{L}}

\def\CN{\mathcal{N}}

\def\CT{\mathcal{T}}

\def\CW{\mathcal{W}}

\def\lgb{\lambda_{\scriptscriptstyle GB}}

\DeclareMathOperator{\vol}{vol}

\def\slashchar#1{\ensuremath{                               %
   \setbox0=\hbox{${}#1{}$}       
   \dimen0=\wd0                                 
   \setbox1=\hbox{/} \dimen1=\wd1               
   \ifdim\dimen0>\dimen1                        
      \rlap{\hbox to \dimen0{\hfil/\hfil}}      
      {}#1{}                                    
   \else                                        
      \rlap{\hbox to \dimen1{\hfil${}#1{}$\hfil}}   
   \fi}} 

\begin{document}

\title{Coupling Constant Corrections in a Holographic Model of \\Heavy Ion Collisions with Nonzero Baryon Number Density}

\author{\r{A}smund Folkestad}
\affiliation{Center for Theoretical Physics, Massachusetts Institute of Technology, \\Cambridge, MA 02139, USA}

\author{Sa\v{s}o Grozdanov}
\affiliation{Center for Theoretical Physics, Massachusetts Institute of Technology, \\Cambridge, MA 02139, USA}

\author{Krishna Rajagopal}
\affiliation{Center for Theoretical Physics, Massachusetts Institute of Technology, \\Cambridge, MA 02139, USA}

\author{Wilke van der Schee}
\affiliation{Institute for Theoretical Physics and Center for Extreme Matter and Emergent Phenomena,\\
Utrecht University, Leuvenlaan 4, 3584 CE Utrecht, The Netherlands}

\preprint{MIT-CTP/5136}
\begin{abstract}
\vspace{1cm}  
Sufficiently energetic collisions of heavy ions result in the formation of a droplet of a strongly coupled liquid state of QCD matter known as quark-gluon plasma. By using gauge-gravity duality (holography), a model of a rapidly hydrodynamizing and thermalizing process like this can be constructed by colliding sheets of energy density moving at the speed of light and tracking the subsequent evolution. In this work, we consider the dual gravitational description of such collisions in the most general bulk theory with a four-derivative gravitational action containing a dynamical metric and a gauge field in five dimensions. Introducing the bulk gauge field enables the analysis of collisions of sheets which carry nonzero ``baryon'' number density in addition to energy density.  Introducing the four-derivative terms enables consideration of such  collisions in a gauge theory with finite gauge coupling, working perturbatively in the inverse coupling. While the dynamics of energy and momentum in the presence of perturbative inverse-coupling corrections has been analyzed previously, here we are able to determine the effect of such finite coupling corrections on the dynamics of the density of a conserved global charge, which we take as a model for the dynamics of nonzero baryon number density. In accordance with expectations, as the coupling is reduced we observe that after the collisions less baryon density ends up stopped at mid-rapidity and more of it ends up moving near the lightcone.
\end{abstract}

\maketitle
\begingroup
\hypersetup{linkcolor=black}
\tableofcontents
\endgroup

\section{Introduction}\label{Sec:Intro}

Previous authors~\cite{Casalderrey-Solana:2016xfq} have completed a holographic analysis of collisions of sheets of energy density, incident at the speed of light, that carry a nonzero density of a conserved global charge in ${\cal N}=4$ $SU(N_c)$ supersymmetric Yang-Mills (SYM) 
theory in the limit of a large number of colors $N_c$ and infinite 't Hooft coupling $\lambda\equiv g^2 N_c$, with $g$ the gauge
coupling.
Although ${\mathcal N}=4$ SYM theory differs from QCD, there are many similarities between
the strongly coupled liquid phase that it features at any nonzero $T$ and the strongly coupled
liquid 
quark-gluon plasma phase of that is found in QCD over a range of temperatures.
 The range extends well above the crossover temperature at which
ordinary hadrons form and includes the range of temperatures explored via heavy ion collisions at RHIC and the LHC.  Because ${\mathcal N}=4$ SYM theory has 
a dual gravitational description that allows reliable insights into its properties and dynamics
at strong coupling it has often been used as a toy model with which to mimic the QCD dynamics 
via which QGP is formed and probed.  (For reviews, see Refs.~\cite{CasalderreySolana:2011us,Dewolfe:2013cua,Chesler:2016vft,Busza:2018rrf}.)
The nonzero density in the ${\mathcal N}=4$ SYM calculation, which is that associated with a global $U(1)_R$ symmetry of the gauge theory, can be used as a toy model for baryon number density in QCD. Hence, the collisions studied in Ref.~\cite{Casalderrey-Solana:2016xfq}
can be thought of as a toy model for
heavy ion collisions with nonzero baryon number density. 
The authors of~\cite{Casalderrey-Solana:2016xfq} found that the fluid energy, momentum, and baryon current all become well described by charged hydrodynamics at roughly the same time after the collision. 
They find very significant stopping of the baryon number, with the baryon density after the collision
piled up at mid-rapidity.
These results are quite different from what is seen in ultrarelativistic heavy ion collisions.
In the holographic calculations of Ref.~\cite{Casalderrey-Solana:2016xfq}, the baryon number distribution in the hydrodynamic fluid
produced after the collision has a rapidity distribution that is
narrowly {\bf sharply} peaked around zero, very different  than what is seen
in high energy heavy ion collisions in which the baryon number ends up near the lightcone, losing only about two units of rapidity rather than getting stopped at midrapidity. (For a review, see Ref.~\cite{Busza:2018rrf}.) Because the physics in the holographic model is strongly coupled at all length scales whereas in QCD
it is weakly coupled at short length scales, meaning at the earliest moments in a collision,
the baryon stopping is much greater in the holographic model. 
This study reveals the most striking qualitative {\it difference} between holographic collisions and high energy heavy ion collisions.  
Our goal in the present study is 
to test the hypothesis that this difference can be attributed to the weakness of the QCD coupling during the earliest moments of a collision.

Before continuing, we pause to review some basic facts known
from experimental measurements, to provide some context.
(This paragraph follows the discussion in Ref.~\cite{Busza:2018rrf} closely.) On average, in high energy nucleus-nucleus collisions, each participating nucleon
loses about two units of rapidity~\cite{Busza:1989px}, which is to
say 85\% of its incident energy goes into the creation of
the droplet of QGP that eventually ends up becoming
a very large number of particles, up to 30,000 in PbPb collisions
at the LHC, and their kinetic energy. These particles, and this energy, end up in a broad distribution centered at mid-rapidity.
However, the net proton (number of protons minus antiprotons) rapidity distribution in ultrarelativistic $AA$ collisions (which traces where the incident baryon number ends up) is not centered at
mid-rapidity. It
has a double-hump structure~\cite{Arsene:2009aa}, with each hump consisting
of hot baryon-number-rich matter moving at a speed of about two units of rapidity
below that of the incident beam, and having a net baryon density of
about 5-10 times that of normal nuclear matter~\cite{Busza:1983rj}.
At LHC, the beam rapidity is at $y=8.5$, denoting rapidity by $y$. So, in the center-of-mass frame, this hot baryon rich matter is found at rapidities $y \sim 6.5$ and $y \sim -6.5$. 
It is interesting to consider how the collision looks upon boosting to the frame in
which $y=6.5$ is at rest. In this frame, an incident disc that is Lorentz contracted
by a factor of about $\cosh(2)$ is hit by a disc that is Lorentz
contracted by about a factor of $\cosh(15)$ and brought approximately
to rest, compressed by roughly $2\cosh(2)\approx 7.5$.
A further consequence of these considerations is that the maximum value of the net baryon
density at mid-rapidity is produced when heavy ions collide with a lower center-of-mass energy per nucleon pair given by $\sqrt{s_{NN}}\approx 7$~GeV. Above this collision energy, the mid-rapidity net baryon density,
and so also the baryon chemical potential in the QGP produced at mid-rapidity,
decreases with energy. By top RHIC collision energies ($\sqrt{s_{NN}}=200$~GeV), and even more
so for LHC energies ($\sqrt{s_{NN}}=5.5$~TeV), both are essentially zero~\cite{Abelev:2013vea}.
This is all very different from ultrarelativistic holographic collisions in an infinitely strongly coupled gauge theory, in which after the collision, the baryon density ends up in a distribution that is centered on mid-rapidity and that is not dissimilar to the way in which the energy density is distributed~\cite{Casalderrey-Solana:2016xfq}.

To investigate the hypothesis that the distinction between how the baryon number density
ends up distributed in rapidity in heavy ion collisions vs.~in holographic collisions reflects the weakness of the QCD coupling early in the collision, we shall analyze the dynamics of the baryon number density in holographic collisions upon reducing the gauge coupling to a finite value, which is to say upon increasing the inverse coupling from zero to nonzero.
A starting point for such a study can be found in Ref.~\cite{Grozdanov:2016zjj}.
These authors have
performed a holographic study of coupling-constant-dependence in holographic collisions
in a hypothetical quantum field theory at large $N_c$ and 't Hooft coupling $\lambda$,
including inverse-$\lambda$ corrections to the large $\lambda$ limit, albeit in collisions with no baryon number.  The gauge theory here is typically assumed to be some deformation
of ${\cal N}=4$ SYM theory. Such collisions can then be thought of as a toy model
for heavy ion collisions with zero baryon number density that includes finite coupling corrections.
For the first time, these authors analyzed the implications in collisions of the effects of corrections to the assumption of infinitely
strong coupling, including inverse-coupling-constant corrections to leading order.
In the dual description, this amounts to colliding sheets of energy density 
in a gravitational theory with curvature-squared terms.  
They find that at intermediate coupling, less energy density is stopped at mid-rapidity and more ends up near the lightcone.  
They also find that when the coupling is decreased to the point that the
shear viscosity is increased by 80\%, the hydrodynamization time increases, but only by 25\%.

Our goal here is to extend the calculation of Ref.~\cite{Casalderrey-Solana:2016xfq} to include finite coupling corrections. Equivalently, our goal is to extend the calculation of Ref.~\cite{Grozdanov:2016zjj} to incorporate collisions of sheets of energy {\it and baryon number} density.
That is, we shall
include corrections that are leading order in the inverse-coupling in the analysis of 
holographic heavy ion collisions with baryon number density.

We find that decreasing the coupling constant
results in substantially less baryon stopping, with the baryon density stopped at mid-rapidity after 
the collision dropping,
and with more baryon number density ending up moving near the light cones after the collision.
In fact, it appears possible to dial the coupling constant down to a value such that the hydrodynamic plasma that forms at mid-rapidity 
has almost no baryon density, as is the case in ultrarelativistic heavy ion collisions.
Although driving the mid-rapidity baryon density down this far would require an inverse-coupling that is much larger than can be treated in our perturbative approach,
our result nevertheless provides substantive support to the hypothesis
that the excessive baryon stopping found in holographic collisions is an artifact of 
the fact that in these models the coupling at very early times cannot be made small whereas
the QCD coupling is weak in the earliest moments of an ultrarelativistic heavy ion collision.

A reader who is principally interested in our results, and in seeing how these results support
the hypothesis noted above, should skip ahead to Sections \ref{Sec:Results} and \ref{Sec:Discussion}.  In Section \ref{Sec:HolographicSetup} we describe the holographic model that we employ, and in Section \ref{Sec:Analysis} we detail how we analyze this model and how we conduct our numerical simulations.

\section{The holographic model}\label{Sec:HolographicSetup}

Quark-gluon plasma (QGP) is a hot, strongly coupled, liquid, deconfined,
state of matter with temperature $T$. Generically, in addition it has some nonzero baryon number density $\rho$. The baryon number chemical potential $\mu$ is  
thermodynamically conjugate to $\rho$. In this work, we 
begin from a
model for the formation and evolution of such a state in a 
non-Abelian gauge theory (${\cal N}=4$ SYM theory) with a large number of degrees of freedom $N_c$ by analyzing its holographically dual gravitational theory. In  the simplest holographic model, the combined dynamics of the energy-momentum and baryon number density of a four-dimensional state with an infinite 't Hooft coupling $\lambda$ is described by an asymptotically Anti-de Sitter (AdS) geometry in five-dimensional Einstein-Maxwell theory. Its bulk action is  
\begin{align}\label{EHM}
S = \frac{1}{2 \kappa_5^2} \int d^5 x  \sqrt{-g} \left(  R -2 \Lambda - \frac{L^2}{4} F_{\mu\nu} F^{\mu\nu} \right) ,
\end{align}
where $\Lambda \equiv  - 6 / L^2$ is the negative cosmological constant with $L$ the AdS radius, 
and where the five-dimensional Newton's constant $\kappa_5$ is inversely proportional to the number of colors in the four-dimensional gauge theory: $\kappa_5 \propto 1/N_c \ll 1$ --- a limit which allows for a study of the gravitational theory in the classical approximation to bulk (quantum) gravity. 
The four-dimensional boundary field theory energy-momentum tensor $T^{\mu\nu}$ is sourced by the five-dimensional metric perturbation and the four-dimensional 
conserved global current $J^\mu$ in the boundary theory 
is sourced by the gauge field in the bulk, according to the standard holographic dictionary. (See, e.g.,~Ref.~\cite{jorge-book}.) We will think of the dynamics of $J^\mu$ as a model for
the dynamics of baryon number current. 

The holographic ``experiment" of a collision of two sheets of energy --- and baryon number --- density incident at the speed of light 
proceeds as follows. First, an initial state with a pair of well-separated (along the direction of their motion) sheets of energy density and baryon number density 
is prepared on the asymptotic AdS boundary. 
We shall only consider sheets which are infinite in transverse extent and translationally invariant in
directions perpendicular to their direction of motion and
whose profiles (in the direction of their motion) are Gaussian.  (Both these simplifications can
be relaxed~\cite{Chesler:2015wra,Chesler:2015bba,Chesler:2015fpa}.)
The initial state is then evolved in a way such that the two sheets of energy and baryon number 
travel towards each other along the beam axis at the speed of light. The sheets collide, resulting in the formation of matter which rapidly hydrodynamizes and eventually thermalizes and equilibrates by forming a charged black hole in the bulk which is the holographic representation of strongly coupled QGP.  In the absence of any baryon number density, this setup has been extensively studied at infinite coupling; see for example Refs. \cite{Chesler:2010bi,Grumiller:2008va,Casalderrey-Solana:2013aba,vanderSchee:2013pia,Chesler:2013lia,Casalderrey-Solana:2013sxa,Chesler:2015wra,Chesler:2015fpa,vanderSchee:2015rta,Chesler:2015bba,Waeber:2019nqd} and the reviews in Refs.~\cite{jorge-book,Chesler:2015lsa,Heller:2016gbp,Busza:2018rrf}. 
Collisions of sheets of energy density with nonzero $\mu$ and consequently with nonzero
baryon density $\rho$ were studied via their dual description in terms of the five-dimensional Einstein-Maxwell theory \eqref{EHM} in Ref. \cite{Casalderrey-Solana:2016xfq}.

While the interactions between quarks and gluons in a realistic QCD plasma are strong, they are not infinite. This provides one motivation to move beyond infinite 't Hooft coupling $\lambda$.
The more important motivation for our consideration is that, regardless of how strong
the QCD coupling is in the liquid plasma that forms after a heavy ion collision, at the earliest
moments of an ultrarelativistic collision in QCD the coupling is weak. And, the question of where
the baryon number ends up at late times depends crucially on what happens during the very earliest moments, long before hydrodynamization.  For these reasons, we wish to investigate whether weakening the 't Hooft coupling $\lambda$ in the holographic model calculation changes where
the baryon number ends up at late times in a way that makes it look a little more realistic.

Inverse coupling constant corrections can be incorporated into the holographic calculation by perturbing the bulk theory with a series of higher-derivative terms. A well-understood such (top-down) example is the leading-order 't Hooft coupling correction to the dynamics of energy and momentum in $SU(N_c)$, $\CN = 4$ SYM theory. At zero baryon number density, the leading part of the corrected dimensionally reduced gravity action is 
\begin{align}\label{EHW}
S = \frac{1}{2 \kappa_5^2} \int d^5 x  \sqrt{-g} \left(  R - 2 \Lambda + \frac{1}{8} \alpha'^3 \zeta(3) \CW \right) ,
\end{align}
where $\CW$ is a contraction of Weyl tensors to the fourth power and $\zeta(z)$ is the Riemann zeta function. For details, see Refs.~\cite{Gubser:1998nz,Pawelczyk:1998pb,Grozdanov:2016vgg}. In type IIB string theory, the tension of the fundamental string sets the dual 't Hooft coupling of $\CN = 4$ SYM theory, i.e. $\alpha' \sim 1/ \sqrt{\lambda}$. Thus, a perturbative supergravity expansion in $\alpha'$, which starts at $\alpha'^3$ gives rise to field theory observables computed in a perturbative expansion to ${\cal O}(1 / \lambda^{3/2})$.

In a generic string theory, the corrections to the Einstein-Hilbert action start at ${\cal O}(\alpha')$ with curvature-squared terms $\sim R^2$ instead of $\CW \sim R^4$. The most general such action is 
\begin{align}\label{R2}
S = \frac{1}{2 \kappa_5^2} \int d^5 x  \sqrt{-g} \Bigl(  R - 2\Lambda + \alpha_1 L^2 R^2  + \alpha_2 L^2  R_{\mu\nu} R^{\mu\nu} + \alpha_3 L^2 R_{\mu\nu\rho\sigma} R^{\mu\nu\rho\sigma} \Bigr) .
\end{align}
The higher-derivative terms all need to be treated perturbatively. If we were doing a top-down construction starting from a known string theory with a known field theory dual, each of the $\alpha_i$ would represent some combinations of corrections that are $\sim \alpha'$ and corrections that are $\sim \ell_P^2 / L^2$, where $\ell_P$ is the Planck scale, with $\kappa_5^2 = \ell_P^3$. While $\alpha'$ corrections 
correspond to inverse coupling corrections in the field theory, $\ell_P^2 / L^2$ corrections correspond to $1/N_c$ corrections. (See Refs.~\cite{Grozdanov:2016fkt,Grozdanov:2016zjj}.) 
Here, in our bottom-up construction in which we are writing down the most general action to $R^2$ order without knowing the identity of its dual field theory, we will treat all the higher-derivative corrections as if they induce coupling constant corrections in the dual field theory. 
By performing a field redefinition of the metric $g_{\mu\nu}$ to ${\cal O}(\alpha_i)$ (see e.g. Refs.~\cite{Brigante:2007nu,Grozdanov:2014kva}), the action \eqref{R2} can be brought into the form
\begin{align}\label{GB}
S = \frac{1}{2 \kappa_5^2} \int d^5 x  \sqrt{-g} \left[  R - 2\Lambda  + \frac{\lgb L^2}{2} \left(R^2  - 4 R_{\mu\nu} R^{\mu\nu} +  R_{\mu\nu\rho\sigma} R^{\mu\nu\rho\sigma}  \right) \right] ,
\end{align}
where $\lgb = 2 \alpha_3$. This is the Einstein-Gauss-Bonnet action, which gives rise to second-order equations of motion. In this theory, the coupling $\lgb$ can in principle be treated non-perturbatively as long as it lies in the interval $\lgb \in (-\infty, 1/4]$~\cite{Brigante:2007nu,Grozdanov:2014kva}. In this work, we will treat $\lgb \sim \alpha'$ perturbatively and exploit  the two-derivative form of the resulting differential equations to simplify the complexity of numerical simulations. Collisions of sheets of energy density in this theory
were studied up to $\mathcal{O}(\lgb^2)$ in Ref.~\cite{Grozdanov:2016zjj}.

Now, in order to study coupling-dependent collisions at finite $\mu$, we add all possible four-derivative terms involving the gauge field and its coupling to the metric with the exception of Chern-Simons-type terms odd in $A_\mu$, which we explicitly set to zero as their presence would violate parity in both the gravitational theory and the field theory to which it is dual. 
We treat all other higher-derivative couplings as being of the same (perturbative) order as $\lgb$. 
To find the relevant terms in the bulk action, we start by writing the most general four-derivative action involving $g_{\mu\nu}$ 
and $A_\mu$~\cite{Myers:2009ij,Cremonini:2009sy,Cremonini:2009ih,Grozdanov:2016fkt}. 
Then, we perform the field redefinitions  of both $g_{\mu\nu}$ and $A_\mu$ to first order in higher-derivative couplings and up to two derivatives, and expand the action to  fourth order in derivatives. 
The final form of the four-derivative part of the action has only two remaining coupling constants: $\lgb$ from Eq. \eqref{GB} and a single new constant $\beta$ that characterizes the strength of the higher-derivative coupling between $g_{\mu\nu}$ and $A_\mu$.
A particularly convenient choice for the action that will be used in all of our analyses is~\cite{Grozdanov:2016fkt}\footnote{Note that if we wanted to include a density of axial charge or to study less symmetric collisions in which vorticity or a magnetic field could arise in the gauge theory plasma, and study consequent anomalous transport effects, we would need to extend the gravitational theory relative to the one that we shall employ in this paper, completing it by adding a Chern-Simons term to this 4+1-dimensional action~\cite{Gauntlett:2003fk,Erdmenger:2008rm,Casalderrey-Solana:2016xfq}. We shall not pursue any of these directions, meaning that this term would play no role in any of our analyses. For this reason we leave it out.}
\begin{align}\label{ActFin}
S &=  \frac{1}{2 \kappa_5^2} \int d^5 x \sqrt{-g}  \biggr[  R + \frac{12}{L^2} - \frac{\epsilon^2L^2}{4} F_{\mu\nu} F^{\mu\nu} +  \frac{\lgb L^2}{2} \left(R^2  - 4 R_{\mu\nu} R^{\mu\nu} +  R_{\mu\nu\rho\sigma} R^{\mu\nu\rho\sigma}  \right) \nn
& +  \beta \epsilon^2 L^4  \left( R F_{\mu\nu} F^{\mu\nu}   - 4 R^{\mu\nu} F_{\mu\rho} F_\nu^{~\rho} + R^{\mu\nu\rho\sigma} F_{\mu\nu} F_{\rho\sigma} \right)  \biggr] ,
\end{align}
which, in analogy with the Einstein-Gauss-Bonnet theory in Eq. \eqref{GB}, gives rise to purely second-order equations of motion for all components of $g_{\mu\nu}$ and $A_\mu$. 
Since both multiply four-derivative terms in the action, we shall assume throughout that $\beta$ is of the same order as $\lgb$. We furthermore assume that $\mu^2/T^2$ is perturbatively small and of the order of the $\lgb$ and $\beta$;  we shall discuss this further below.
We have introduced a control parameter $\epsilon$ as a device for later convenience; it could evidently be scaled away by redefining $A_\mu$.

Since $\mu/T$ is small in the QGP produced in ultrarelativistic heavy ion collisions, we
shall only consider the case where the incident sheets of energy before the collision have a baryon number density $\rho$ that is much smaller than
$\mathcal{E}^{3/4}$, with $\mathcal{E}$ their energy density. Consequently,
$\mu/T$ is small in the hydrodynamic fluid produced after the collision.
This assumption allows additional simplifications, as we explain below.
Small $\rho$ in the boundary theory means small $A_\mu$ in the dual Einstein-Maxwell theory,
and in particular means that the backreaction of the Maxwell field $A_\mu$ on the five-dimensional 
geometry is small.  We can implement this assumption by treating $\epsilon$ as small.
In the boundary theory it determines the size of the baryon number density and the chemical potential. In particular we have that $J^{\mu} \propto \epsilon$ and $\mu\propto \epsilon$.  
To see this, let us for a moment consider the action with the normalization of the Maxwell field used in Eq.~\eqref{EHM}. 
Then $\rho$ and $\mu$ are set by the size of the bulk gauge field $|A_\mu|$ which has a mass dimension of $1/L$.
If we assume that $\rho$ and $\mu$ are small and parametrically ${\cal O}(\epsilon)$, we have that $|A_\mu| \sim \epsilon/L$. 
It is then convenient to explicitly scale $A_\mu \rightarrow \epsilon A_{\mu}$ and take $A_{\mu}$ of order $\sim 1/L$, which introduces powers of $\epsilon$ into the action as in Eq.~\eqref{ActFin}.  
Note that the assumption of the smallness of $A_\mu$ in Eq.~\eqref{EHM} and consequently the smallness of $\epsilon$ in Eq.~\eqref{ActFin} is an assumption about the
smallness of $\rho$ in the incident sheets of energy and baryon number density.
Thus, $\epsilon$ in Eq.~\eqref{ActFin} is a control parameter governing the initial conditions
in our calculation, not a coupling constant in the theory itself.  The only coupling constants arising in the higher-derivative terms in Eq.~\eqref{ActFin} are $\lgb$ and $\beta$.

Introducing $\epsilon$ as we do gives us a convenient way to take the smallness of $\rho$
into account in a combined perturbative expansion:
we shall assume that $\epsilon^2$ is of order $\lgb$ so that we can arrange our calculations in a single perturbation series in $\lgb$. 
We will then determine $T^{\mu\nu}$ and $J^{\mu}/\epsilon$ to leading order, ${\cal O}(\lgb)$. 
Our motivation here is simply that we wish to take $\lgb$, $\beta$ and $\epsilon$ (and consequently $\rho$) to all be small and work to lowest nontrivial order in a combined expansion: the particular choice of scaling $\epsilon^2 \sim |\lgb|$ is arbitrary; others could be investigated.
However, if we had instead
assumed $\epsilon \sim |\lgb|$ 
this would not have captured the physics that we aim to elucidate as in that case
the Maxwell field would have no backreaction at ${\cal O}(\lgb)$, and we would effectively be working at zero chemical potential rather than at small chemical potential. 
Since in our scheme $\epsilon^2$$, \lgb$ and $\beta$ scale in the same way we will frequently express $\epsilon^2$ and $\beta$ in terms of $\lgb$ as
\begin{align}\label{DefEps12}
\epsilon^2 = c_\epsilon \lgb, \qquad \beta = c_\beta \lgb ,
\end{align}
where our choice of scaling means that both $c_\epsilon$ and $c_\beta$ are ${\cal O}(1)$ when we count powers of $\epsilon$. 

With the scaling that we have described, the action Eq.~\eqref{ActFin} yields 
the most general equations of motion to linear order in $\lgb$.
The Maxwell ($F^2$) and Gauss-Bonnet ($R^2$) terms are of the same order (since 
$\epsilon^2\sim\lgb$)
while the term proportional to $\beta\epsilon^2$ ($RF^2$) is subleading. However, the 
factor $\epsilon^2$ cancels in the Maxwell equations, so the $\beta$ term provides the leading correction to the Maxwell equations. Further details regarding the construction of the action Eq.~\eqref{ActFin} and the  field redefinitions involved are discussed in Appendix~\ref{Sec:ActFD}. 
We close our discussion here by noting that 
if we had not assumed that $\epsilon$ (and consequently $\rho$ and $\mu/T$) is small
then we would have had to include additional  $F^4$ terms in the action \eqref{ActFin}.

In this work, we want to analyze the effects of coupling constant corrections on the dynamics of energy-momentum and baryon density, which are encoded in one-point functions of the energy-momentum tensor $T^{\mu\nu}$ and a current $J^\mu$, respectively. Both conserved tensors need to be calculated to first order in our combined expansion in small $\lgb$, $\beta$ and $\epsilon^2$. This means that
$T^{\mu\nu}$ and $J^\mu$ will 
take the form
\begin{align}
T^{\mu\nu} &= \frac{T^4}{\kappa_5^2} \left( \CT^{\mu\nu} + \frac{\mu^2}{T^2}  \delta \CT^{\mu\nu}_{(1)} + \lgb \delta \CT^{\mu\nu}_{(2)} \right) , \label{TexpE}\\
J^\mu &= \frac{\epsilon T^3}{\kappa_5^2} \left( \CJ^\mu + \frac{\mu^2}{T^2} \delta \CJ^\mu_{(1)} + \lgb \delta \CJ^\mu_{(2)} + \beta \delta \CJ^\mu_{(3)} \right) , \label{JexpE} 
\end{align}
where we note that $\mu^2/T^2 \propto \epsilon^2$ and where
all of $\CT^{\mu\nu}$, $\CJ^\mu$, and all of the $\delta \CT^{\mu\nu}$  and $\delta \CJ^\mu$ 
are zeroth order in small quantities.
We shall obtain solutions for the time evolution of these quantities by developing and solving the equations of motion for the dual gravitational Einstein-Maxwell-Gauss-Bonnet theory in Sections \ref{Sec:Analysis} and \ref{Sec:Results}.

We shall also wish to compare the dynamics of $T^{\mu\nu}$ and $J^\mu$ that
we obtain holographically to what we obtain from hydrodynamics.
To this end, we shall need the
hydrodynamic expansion of the two conserved tensors to first order in derivatives, which is given by
\begin{align}
    T^{\mu\nu} &= \mathcal{E} u^\mu u^\nu + p \Delta^{\mu\nu}  - \eta \Delta^{\mu\alpha} \Delta^{\nu\beta} \left( \partial_\alpha u_\beta + \partial_\beta u_\alpha - \frac{2}{3} \eta_{\alpha\beta }\partial_\lambda u^\lambda \right) , \label{SEhydro}\\
    J^\mu &= \rho u^\mu - \sigma T \Delta^{\mu\nu} \partial_\nu \left(\frac{\mu}{T} \right) , \label{Jhydro} 
\end{align}
where $u^\mu$ is the velocity field normalized to $u^2 = -1$, $\mu$ and $T$ are treated as hydrodynamic near-equilibrium quantities, and $\Delta^{\mu\nu} = u^\mu u^\nu + \eta^{\mu\nu} $ is the projector in directions transverse to $u^\mu$. The thermodynamic quantities appearing
in these expressions
are the energy density $\mathcal{E}$, pressure $p$ and baryon number density $\rho$. Assuming local equilibrium, they satisfy the thermodynamic relation
\begin{align}\label{Thermo} 
\mathcal{E}+ p = s T + \mu \rho ,
\end{align}
with $s$ the entropy density, and all of them vary in space and time on wavelengths that are long compared to $1/T$.
Because ${\cal N}=4$ SYM theory and its deformations considered in this work are conformal, the equation of state is given by $\mathcal{E}= 3 p$. The two transport coefficients that enter into the hydrodynamic constitutive 
relations \eqref{TexpE} and \eqref{JexpE}
are the shear viscosity $\eta$ and the baryon number 
conductivity $\sigma$. (Recall that $J^\mu$ is not the electric current, but the baryon current.) However, if $J^\mu$ were to be weakly gauged, with the introduction of an additional  $U(1)$ gauge field, then $\sigma$ would indeed become the electrical conductivity of the strongly coupled field theory. 
For further details on relativistic hydrodynamics, see Ref.~\cite{Kovtun:2012rj}. 

We shall compute $\mathcal{E}$ and $p$ to linear order in  $\lgb$, meaning to ${\cal O}(\epsilon^2)$, and in so doing will precisely reproduce the analysis of finite-coupling corrections in the absence of any baryon number density from Ref.~\cite{Grozdanov:2016zjj}. 
Note that there are no ${\cal O}(\epsilon^3)$ terms ($\sim \mu^3$ or $\sim \lgb \mu$) in $\mathcal{E}$ and $p$, since these thermodynamic quantities are even in $\mu$.
We shall compute $\rho$, which at infinite coupling is proportional to $\epsilon$ for small $\epsilon\sim \mu$, 
to order $\epsilon \lgb \sim \epsilon^3$. In terms of the gravity couplings $\lgb$ and $\beta$, only $\lgb$ will influence the energy density $\mathcal{E}$ and pressure $p$, while both $\lgb$ and $\beta$ will affect the baryon number density $\rho$. 
Working to ${\cal O}(\epsilon \lgb)$ in the analysis of $\rho$ is necessary in order for us to find the leading-order finite-coupling corrections to the baryon number density, which is the central focus of this paper. 
For later convenience,
we define the ${\cal O}(1)$ quantities $\bar{\rho} \equiv \rho/\epsilon$, $\bar{\mu}\equiv \mu/\epsilon$ and $\bar{J}^{\mu} \equiv J^{\mu}/\epsilon$. Note that all these quantities are zeroth order in $\epsilon$, meaning in particular that they are finite 
in the $\epsilon\rightarrow0$ limit.

Before we turn to a detailed analysis of the action \eqref{ActFin} in the next Section, we 
close this section by motivating a qualitative sense for what the relevant range
of choices might be for the four-derivative coupling parameters $\lgb$ and $\beta$ in our
gravitational action.
This certainly cannot be done in a quantitative way because although we do not know exactly what
deformation of ${\cal N}=4$ SYM theory our bottom-up gravitational construction is dual 
to,
we know
that it is not dual to QCD, meaning that we cannot apply constraints coming from phenomenology or lattice QCD calculations quantitatively.
As noted below Eq.~\eqref{R2}, if we had derived the action \eqref{ActFin} from a top-down stringy construction we would know precisely which gauge theory it is dual to and in such a construction both $\lgb$ and $\beta$ would have been fixed in terms of $\alpha'$ and $\ell_P^2/L^2$. Instead, in our bottom-up approach reasonable values for these quantities can be estimated by  asking that $\lgb$- and $\beta$-dependent observables take on reasonable values. 
We need at least some rough sense of what values of these parameters we should
use when we later plot our results, and to this end we shall look at what we know about the
shear viscosity to entropy density ratio $\eta/s$  and the baryon number susceptibility $\chi$ in QCD and ask how this compares with their dependence on $\lgb$ and $\beta$ in the model
theory that we are employing.
Our goal is to set these couplings to 
values such that the magnitude of their effects on $\eta/s$ and $\chi$ in our model theory corresponds to the magnitude of the effect of the finiteness of the QCD gauge coupling on these observables,
in the hope that doing so will allow us to use our model theory to 
get a qualitative impression of the effect of the finiteness
of the gauge coupling on the dynamics of $\rho$ in collisions.

We start from a relatively well-established fact that when $\lgb R^2$ terms in the gravitational dual are treated as enabling finite coupling corrections in the gauge theory, it is natural to take $\lgb$ to have a negative sign. (See e.g. Refs.~\cite{Grozdanov:2016vgg,Grozdanov:2016zjj,Andrade:2016rln,DiNunno:2017obv,Atashi:2016fai}.) 
One way to motivate this sign is to look at the result for $\eta/s$
in (zero baryon number) Einstein-Gauss-Bonnet \eqref{GB}, where it is known nonperturbatively in $\lgb$ and is given by \cite{Kats:2007mq,Brigante:2007nu}
\begin{align}
\frac{\eta}{s} = \frac{1}{4\pi } \left( 1 - 4 \lgb \right) ,
\end{align}
meaning that if we choose $\lgb$ negative $\eta/s$ increases with increasing $|\lgb|$, meaning with
increasing inverse-gauge-coupling, or in other words, with decreasing gauge coupling.
This is the natural sign, given that in a weakly coupled gauge theory the leading
order dependence of $\eta/s$ on the gauge coupling $g$ at small $g$ is given
by $\eta/s \sim {\rm const}/[ g^4 \ln(1/g) ]$~\cite{Arnold:2000dr}, meaning that at weak coupling $\eta/s$ increases with decreasing gauge coupling and diverges as $g\to 0$.
We will typically choose $\lgb = -0.2$, which corresponds to an 80\% increase in the value of 
$\eta/s$ relative to its value when $\lgb=0$ and the gauge theory has infinite coupling.
There is nothing sacred about this particular value, in particular given that $\eta/s$ in QCD will in
reality have some temperature dependence, but this choice puts $\eta/s$ within the range estimated by comparing hydrodynamic calculations of the anisotropic expansion of the droplets of QGP produced in off-center heavy ion collisions with experimental data~\cite{Romatschke:2007mq,Schenke:2010rr,Bernhard:2016tnd}; 
for reviews, see Refs.~\cite{Heinz:2013th,Romatschke:2017ejr,Busza:2018rrf}.

At small but nonzero baryon number density, perturbative results in both the four-derivative couplings found in the Maxwell-Einstein-Gauss-Bonnet action \eqref{ActFin} give rise to corrections to the ratio of $\eta/s$ that take the form 
\begin{align}\label{etaOverS}
\frac{\eta}{s} = \frac{1}{4\pi} \left[ 1 - 4 \lgb  + {\cal O}\left(\lgb^2 \right) + {\cal O}\left(\lgb \mu^2/T^2 \right)+ {\cal O}\left(\beta \mu^2/T^2 \right)+ \ldots \right] ;
\end{align}
the ellipsis denotes terms of $\sim {\cal O}(\lgb^3)$ and beyond. 
Sufficiently little is known about the $\mu$-dependence of $\eta/s$ in QCD that we will 
not attempt to use a calculation of the $\mu$-dependent contributions
to (\ref{etaOverS}) to estimate what range of $\beta$ we might investigate.
Another reason why attempting this would be perilous is that the presently unknown contributions to $\eta/s$ that are
of order $\lgb^2$
are parametrically just as significant, given the scaling that we are using, as the known $\epsilon^2\beta$ and $\epsilon^2\lgb$ terms, both of which can be computed from the field redefinitions discussed in Appendix \ref{Sec:ActFD} and the calculations of Refs. \cite{Myers:2009ij,Cremonini:2009sy}.

We shall estimate a reasonable range for $\beta$ by looking at its effects on
the baryon number susceptibility $\chi$. Note that in our model theory, $\chi\propto N_c^2$ because
all degrees of freedom are in color-adjoint representations whereas in QCD, $\chi\propto N_c N_f$ because baryon number is carried by quarks, which come in $N_c$ colors and $N_f$ flavors.
We can scale out this difference, however, by computing the ratio of $\chi$ in the theory with 
finite coupling to $\chi$ in the non-interacting limit in both QCD and our model theory, since the
count of the number of degrees of freedom that carry baryon number cancels in this ratio.
Beginning with our model theory, we note that 
we will show
in Sec.~\ref{Sec:Analysis} (see Eq.~\eqref{eq:rhomuexp}) that
to linear order in $\mu$ and to lowest order in $\lgb$ and $\beta$ the baryon number density
is given (for a choice of normalization, see Section \ref{Sec:Analysis}) by
\begin{equation}
    \rho = \frac{1}{2} \pi^2 T^2 \mu \left(1+3\lgb + 16\beta\right),
    \label{eq:rho}
\end{equation}
giving a susceptibility
\begin{equation}
    \chi \equiv \frac{ \partial \rho }{ \partial \mu }\Big|_{\mu=0} = \frac12 \pi^2 T^2 \left( 1 + 3\lgb + 16\beta \right).
\end{equation}
From Refs.~\cite{Son:2006em,Teaney:2006nc,jorge-book} we know that in $\mathcal{N}=4$ SYM theory 
we have $\chi_{\infty}/\chi_{0} = \frac{ 1 }{ 2 }$ where $\chi_{\infty}$ is the susceptibility in the infinite coupling limit (corresponding to $\lgb=\beta=0$ in the above), while $\chi_0$ is the susceptibility in the non-interacting limit. This gives $\chi_0 = \pi^2 T^2$ in our normalization scheme. 
Now we can calculate the ratio $\chi/\chi_0$, namely the ratio between $\chi$ at the intermediate coupling that corresponds to given values of $\lgb$ and  $\beta$ to $\chi$ in the noninteracting limit, obtaining
\begin{equation}
\frac{ \chi }{ \chi_0 } = \frac{ 1 }{ 2 }\left( 1 + 3\lgb + 16\beta \right)  .
\label{eq:chi_over_chi0}
\end{equation}
This can be compared to lattice calculations of the same ratio in 
QCD~\cite{Borsanyi2012}, which indicate that 
$\frac{ \chi }{ \chi_0 }\big|_{\rm QCD}\in [0.8, 0.9]$ for temperatures from \SIrange{250}{400}{\mega\electronvolt}, corresponding to the temperatures of the strongly coupled QGP liquid produced in heavy ion collisions at RHIC and the LHC.
Setting $\lgb=-0.2$ and varying $\frac{ \chi }{ \chi_0 }\big|_{\rm QCD}$ in the given range we find 
$\beta \in [0.08, 0.09]$.
If we further vary $\lgb$ in the broad range $[-\frac{ 1 }{ 2 }, 0]$, corresponding to values of $\frac{ \eta }{ s }$ between $\frac{ 1 }{ 4\pi }$ and $\frac{ 3 }{ 4\pi }$, we find that $\beta$ should lie in the broad range 
\begin{equation}\label{eq:beta-range}
\beta \in \left[0.04, 0.15 \right] .
\end{equation}
That said, we caution that working only to linear order in $\beta$ may not be a good approximation, in particular at the upper end of this range.  Working only to this order we cannot reliably estimate the range of reliability of this linear approximation, but absent further information it is reasonable to guess from results like (\ref{eq:chi_over_chi0}) that it starts to break down once $16\beta$ gets as large as $1$.  
Another way of making a reasonable guess is to note from \eqref{eq:chi_over_chi0} that
values of $\lgb = -0.2$ and $\beta = 0.1$ imply that $\rho$ increases by a factor 2, suggesting that the linear approximation is breaking down. 
And, if we express $\mu$ 
as a function of $\rho$ and $T$ we see that to the order we are working it is proportional
to $1 - 3\lgb - 16\beta$, which for these particular values would be zero and hence clearly outside the linear regime. 
In the results that we shall present in 
Section \ref{Sec:Results}, we will see that values of $\beta$ in the 
range (\ref{eq:beta-range}) can have a large effect on the baryon number distribution, 
in some cases to a degree that supports our guess that working to linear order in $\beta$
may not be sufficient.

Summarizing these considerations, it seems reasonable to investigate the consequences
of choosing $\lgb$ small and negative and $\beta$ small and positive in the action \eqref{ActFin} on the dynamics of $\rho$ in
collisions.  We shall present results in Section \ref{Sec:Results} with this in mind, but first, in the next Section, we must describe how we analyze the dynamics governed by the action \eqref{ActFin}.

\section{Analysis of the model and details of the numerical setup}\label{Sec:Analysis}

In this Section we proceed with the analysis of the action (\ref{ActFin}).
We shall present the equations of motion, describe the initial conditions that we employ and 
the numerical methods that we use to solve for the time evolution, provide the dictionary that connects bulk quantities obtained via solving these holographic evolution equations to field theory observables, and close by remarking upon the hydrodynamics and thermodynamics of the model.

The action (\ref{ActFin}) is the most general 4+1-dimensional four-derivative theory containing an asymptotically AdS metric and a gauge field which is consistent with working to first order in $\lgb$ and assuming the scaling $\lgb\sim \beta \sim \epsilon^2 \sim \mu^2 / T^2$. The equations of motion that follow from variations of $g_{\mu\nu}$ and $A_\mu$ were derived in Appendix E of Ref.~\cite{Grozdanov:2016fkt}. The resulting Einstein equations are
\begin{equation}\label{EQEin}
R_{\mu\nu} - \frac{1}{2} g_{\mu\nu} R - \frac{6}{L^2} g_{\mu\nu} = T^{\rm A}_{\mu\nu} + T^{\rm GB}_{\mu\nu}  + T^{\rm \beta}_{\mu\nu},
\end{equation}
where the perturbatively small terms on the right-hand-side of the equation are
\begin{align}
T^{\rm A}_{\mu\nu} &= - \frac{\epsilon^2 L^2}{8}\left( g_{\mu\nu} F^2 - 4 F_{\mu\lambda} F_{\nu}^{~\lambda} \right), \label{TA} \\
T^{\rm GB}_{\mu\nu} &= \frac{\lgb}{4} g_{\mu\nu} \left( R^2 - 4 R_{\mu\nu} R^{\mu\nu} + R_{\mu\nu\rho\sigma} R^{\mu\nu\rho\sigma}  \right) \nn  
&- \lgb  \left( R R_{\mu\nu} - 2 R_{\mu\alpha} R_{\nu}^{~\alpha} - 2 R_{\mu\alpha\nu\beta} R^{\alpha\beta} + R_{\mu\alpha\beta\gamma} R_\nu^{\alpha\beta\gamma}  \right), \label{TGB} \\
T^{\rm \beta}_{\mu\nu} &= \frac{\epsilon^2 L^4 \beta}{2} \left[ g_{\mu\nu} R F^2 - 4 R F_{\mu\alpha} F_\nu^{~\alpha} - 2 R_{\mu\nu} F^2 + 2 \nabla_\mu \nabla_\nu F^2 - 2 g_{\mu\nu} \Box F^2 \right] \nn
&- 2 \epsilon^2 L^4 \beta  \left[ g_{\mu\nu} R^{\alpha\beta} F_{\alpha\lambda} F_\beta^{~\lambda} -4 R_{\mu\alpha} F_{\nu\beta} F^{\alpha\beta} - 2 R_{\alpha\beta} F_\mu^{~\alpha} F_\nu^{~\beta} - \Box \left(F_{\mu\alpha} F_\nu^{~\alpha} \right)  \right. \nn
&\left.- g_{\mu\nu} \nabla_\alpha \nabla_\beta \left( F^\alpha_{~\lambda} F^{\beta\lambda}\right) + \nabla_\alpha \nabla_\mu \left(F_{\nu\beta} F^{\alpha\beta} \right) + \nabla_\alpha \nabla_\nu \left( F_{\mu\beta} F^{\alpha\beta} \right) \right]  \nn
&+ \frac{\epsilon^2 L^4 \beta}{2} \left[ g_{\mu\nu} R^{\alpha\beta\gamma\delta} F_{\alpha\beta} F_{\gamma\delta} -6 R_{\mu\alpha\beta\gamma} F_\nu^{~\alpha} F^{\beta\gamma} - 4 \nabla^\beta \nabla^\alpha \left(F_{\mu\alpha} F_{\nu\beta} \right) \right]  .\label{TBeta}
\end{align}
The Maxwell equations, as modified by the four-derivative terms, are
\begin{align}
\nabla_\nu F^{\mu\nu} =  4 \beta L^2 \nabla_\nu \left(R F^{\mu\nu} + R^{\mu\nu\rho\sigma} F_{\rho\sigma} - 2 R^{\mu\rho} F_{\rho}^{~\nu} + 2 R^{\nu\rho}F_\rho^{~\mu}   \right)  . \label{EQMax}
\end{align}
Both the equations of motion \eqref{EQEin} and \eqref{EQMax} contain at most second derivatives with respect to each of the coordinates. 
The fact that, with our scaling, $T^\beta_{\mu\nu} \sim \mathcal{O}(\epsilon^2 \beta) \sim {\cal O}(\lgb^2)$ implies that to leading order in  $\lgb$  the $\beta$-term in the action only affects the Maxwell equations \eqref{EQMax}.
We shall employ the convention
\begin{align}\label{ChoiceOfUnits}
    L = 1+\frac{1}{2}\lgb + {\cal O}(\lgb^2)  . 
\end{align}
This can be thought of as choosing our units, although looked at that way it looks unconventional. We make this choice for later convenience  because with this choice the non-normalizable modes of the metric near the boundary are unaffected by our $\lgb$ corrections.

Equations \eqref{EQEin} and \eqref{EQMax} have a solution of the following form:
\begin{align}
    d s^2 &= \frac{ 1 }{ u^2 }\left( - d x_+ d x_- + u^4 \left[h_0(x_+) + \lgb h_1(x_+) - \frac{1 }{ 3 } e^2 u^2 a_{+(0)}(x_+)^2 \right] d x_+^2 + d\vecb{x}_\perp^2 + d u^2  \right), \label{shockwavemetric} \\
    A &= u^2 \left[a_{+ (0)} (x_+) + \lgb a_{+(1)}(x_+)\right] dx_+ , \label{shockwavegaugefield}
\end{align}
where $x_{\pm} = t\pm z$ are lightcone coordinates and where the functions $h_i (x_+)$ and $a_{+(i)}(x_+)$ can be chosen arbitrarily.  This metric describes a gravitational wave moving at the speed of light in the $z$-direction in an asymptotically AdS spacetime. 
On the boundary it represents a sheet of baryon and energy density moving at the speed of light. 
A wave moving in the opposite direction is obtained by exchanging $x_+ \leftrightarrow x_-$. The wave solution is planar (and translation-invariant)  in the two-dimensional $\vecb{x}_{\perp}$ plane. The functions $h_i (x_+)$ and $a_{+(i)}(x_+)$ determine the profile of the energy density and
baryon density, respectively, along the ``beam'' direction. In our computation, we use a pair of well-separated left- and right-moving wave solutions, namely \eqref{shockwavemetric}--\eqref{shockwavegaugefield} and their counterparts with $x_+ \leftrightarrow x_-$,
as initial conditions.   Specifying our initial conditions therefore requires choosing the eight functions
$h_i(x_{\pm})$ and $a_{\pm(i)}(x_\pm)$.  We choose
the four profile functions $h_0(x_\pm)$ and $a_{+(0)}(x_\pm)$ all to be Gaussians with the same width $w$: 
\begin{equation}
    h_0(x_{\pm}) = \frac{ m^3 }{ \sqrt{2\pi w^2} }\exp\left(-\frac{ x_\pm^2 }{ 2w^2  } \right), \qquad a_{\pm(0)}(x_{\pm}) = \frac{ h(x_{\pm}) }{ m },
\end{equation}
where with this choice $m^3$ is the energy density per unit transverse area of each
incident sheet of energy.
We then choose the four first order functions  $h_1(x_\pm)$ and $a_{\pm(1)}(x_\pm)$ needed to complete the specification of our initial conditions such that the total ${\cal O}(\lgb)$ correction to 
the expectation value of the stress-energy tensor and the conserved $U(1)$ current vanishes in our initial conditions,
so that our initial conditions are independent of $\beta$ and $\lgb$. 
This will allow us to compare simulations with different $\beta$ and $\lgb$ properly. 
The parameter $w$ sets the width of the sheets of energy and baryon number density. We shall
show results for two different choices of $w$ which we shall refer to as thick and thin sheets of energy density.   The two values of $w$ at which we shall quote results
 are $m w = 0.1$ and $m w=1.5$, for thin and thick sheets respectively.

In our numerical solution of the equations of motion, we use the ingoing Eddington-Finkelstein coordinates \cite{Chesler:2010bi} in which the metric takes the form 
\begin{equation}
    d s^2 = 2 d t d r - C(t,r,z) d t^2 +2F (t,r,z) d z d t + S(t,r,z)^2\left( e^{-2B(t,r,z)}d z^2 + e^{B(t,r,z)}d\vecb{x}_\perp^2\right), \label{eq:eddFink}
\end{equation}
where the functions $C$, $F$, $S$, $B$ depend on the boundary time $t$, the radial holographic coordinate $r$, and the coordinate $z$ that lies along the direction of motion  of the incident sheets of energy density, the ``beam direction''. The two coordinates collectively denoted as $\vecb{x}_\perp$ are perpendicular to $z$ in spatial $\mathbb{R}^3$. The metric (\ref{eq:eddFink}) is based on assuming translational symmetry in this plane, appropriate for our analysis of the collision of sheets that are infinite in extent and translationally invariant in these transverse directions. 
The conformal boundary is at $r=\infty$. For the gauge field, we choose the radial gauge, meaning that its form reads
\begin{align}
A = A_t (r,t,z) \, dt + A_z (r,t,z) \, dz.
\end{align}
In the numerical evolution, we work directly with the field strength $F_{\mu\nu}$. 
The coordinates used in Eq.~\eqref{eq:eddFink} allow us to rewrite the full set of Einstein and Maxwell equations as a nested set of ordinary differential equations (ODEs) using the characteristic formulation of general relativity~\cite{Chesler:2010bi, Bondi1960}. Details and examples of the numerical procedure in the (unperturbed) characteristic formulation can be found in a number of past publications (see e.g. Ref. \cite{Chesler:2013lia, vanderSchee:2014qwa,Attems:2017ezz,Waeber:2019nqd}). These include in particular the (tricky) coordinate transformation from \eqref{shockwavemetric} to \eqref{eq:eddFink}, which luckily is straightforward to generalize to our $\lgb$-corrected set-up (see also \cite{Grozdanov:2016zjj}).

For our specific case of Gauss-Bonnet gravity with a Maxwell field, several aspects are noteworthy. To accommodate a perturbative expansion in $\lgb$ (with our assumed scaling) we expand the metric functions, the gauge field, and later also various thermodynamic quantities as power series in $\lgb$, namely $f=\sum_n f_{(n)}\lgb^n$ (cf. Eq. \eqref{DefEps12}). For the metric this implies
\begin{align}
    C(r, t, z) &= C_{(0)}(r,t,z)+\lgb  C_{(1)}(r,t,z)+{\cal O}\left(\lgb^2\right),
\end{align}
and similarly for all other functions. At every time step in the numerical evolution (solving Eqs. \eqref{EQEin} and \eqref{EQMax}), we then split the 
computation into two steps. First, we evaluate the unperturbed functions $\{C_{(0)},F_{(0)},\ldots\}$ and their derivatives according to the nested scheme involving a set of homogeneous differential equations. Second, we solve a nested set of nonhomogeneous differential equations for the first-order perturbations $\{C_{(1)},F_{(1)},\ldots\}$ and their derivatives. The same second-order differential operators act on the perturbations $\{C_{(1)},F_{(1)},\ldots\}$ as in the first step on $\{C_{(0)},F_{(0)},\ldots\}$. The nonhomogeneous source terms needed to solve for $\{C_{(1)},F_{(1)},\ldots\}$ are obtained using the zeroth-order solution computed in the first step. The only subtlety in the procedure involves the computation of the following time derivatives: $\partial_t^2 B_{(0)}$, $\partial_t F_{(0)rt}$ and $\partial_t A_{0}$. These specific time derivatives do not arise in the differential operators of the homogeneous differential equations, meaning that they are not needed in the nested scheme for evolving the metric and gauge field, or in the computation of the constraints. They do, however, appear in the source terms in the nonhomogeneous equations for $\{C_{(1)},F_{(1)},\ldots\}$.  This means that they affect the perturbed first-order solutions, which means that we need to keep track of 
$\partial_t B_{(0)}$, $F_{(0)rt}$ and $A_0$
at every step of the evolution and evaluate their time derivatives. Finally, we note that we need not compute the correction to the location of the apparent horizon: since we work perturbatively, the code will be stable if the horizon is located at $r=1 + \mathcal{O}(\lgb)$. (Note, however, that these corrections were computed in Ref.~\cite{Grozdanov:2016zjj} in order to see the effects of $\lgb$ corrections on the entropy density.) Of course, the equations generated by the source terms are much longer than the unperturbed equations, which means that the code runs roughly a factor ten times slower. The calculation of the time evolution of the collision of thin sheets of energy density then takes about two weeks to perform on an ordinary desktop computer.

To obtain the boundary field theory quantities that we are interested in, namely the field theory energy-momentum tensor and the baryon number current, from the five-dimensional bulk metric and gauge field functions, we shall require the near-boundary solutions of \eqref{EQEin} and \eqref{EQMax}, which take the form
\begin{align}
    A_\mu(r, t, z) &= \frac{a_{(0)\mu}(t,z)+\lgb   a_{ (1) \mu}(t,z)}{r^2}+{\cal O}\left(1/r^3\right),\label{Amuexpansion}\\
    C(r, t, z) &= r^2+\frac{c_{(0)}(t,z)+\lgb   c_{(1)}(t,z)}{r^2}+{\cal O}\left(1/r^3\right),\label{Cexpansion}\\
    B(r, t, z) &= \frac{b_{(0)}(t,z)+\lgb   b_{(1)}(t,z)}{r^4}+{\cal O}\left(1/r^5\right),\label{Bexpansion}\\
    F(r, t, z) &= \frac{f_{(0)}(t,z)+\lgb   f_{(1)}(t,z)}{r^2}+{\cal O}\left(1/r^3\right),\label{Fexpansion}\\
    S(r, t, z) &= r+{\cal O}\left(1/r^5\right).
\end{align}
The functions of $t$ and $z$ that appear here are the normalizable modes of the gauge field and metric near the AdS boundary; they depend upon the bulk dynamics and therefore must be found from the evolution of the full metric, as we have described above. 
In order to determine the field theory
quantities of interest what we will need to extract from our evolution of the full metric
are the functions of $t$ and $z$ appearing on the RHS of \eqref{Amuexpansion}, \eqref{Cexpansion},
\eqref{Bexpansion} and \eqref{Fexpansion}. To make this extraction easier,
when we solve for the time evolution we recast the equations for $A_\mu$, $C$, $B$ and $F$ as equations for $r^2 A_\mu$, $r^2 (C - r^2)$, $r^4 B$ and $r^2 F$, and solve them
numerically in that form~\cite{vanderSchee:2014qwa}.

The main quantities in the field theory whose dynamics we wish to determine and 
study are the one-point functions of the boundary energy-momentum tensor $T^{\mu\nu}$ and the conserved $U(1)$ (baryon number) current $J^{\mu}$. We will work with the rescaled quantities
\begin{align}\label{rescaledCurrents}
    \mathcal{T}^{\mu\nu} = \frac{ \kappa_5^2 }{2  L_0^3 } T^{\mu\nu}, \qquad \mathcal{J}^{\mu} =  \frac{ \kappa_5^2 }{2  L_0^3 } J^{\mu} .
\end{align}
The normalization factor of $\kappa_5^2 / 2 L_0^3 $ is given by $ 2 \pi^2 / N_c^2 $ in $\CN=4$ SYM theory at infinite coupling, i.e. when $\lgb = \beta = 0$. $L_0$ is the $\lgb=0$ value of the AdS radius, which for convenience we have set equal to 1, see (\ref{ChoiceOfUnits}) \cite{Grozdanov:2016zjj}. 
A particularly straightforward way of ensuring that these normalization conventions are satisfied is to set $\kappa_5^2/8\pi = 1/4\pi$ in all formulae. We will do so below.

 The relationships that determine the 
 energy-momentum tensor for the boundary gauge theory 
 from the functions appearing in  Eqs.~(\ref{Cexpansion})--(\ref{Fexpansion}) that are determined by
 the time evolution in the bulk gravitational theory
  can be obtained to ${\cal O}(\lgb)$ in the same way as they were previously obtained in  Einstein-Gauss-Bonnet theory without a bulk gauge field, and take the form~\cite{Brihaye:2008kh,Grozdanov:2016fkt} 
\begin{equation}\label{Tdict}
    \mathcal{T}^{\mu\nu} = \mathcal{T}_{(0)}^{\mu\nu} + \lgb \mathcal{T}_{(1)}^{\mu\nu}, \\
\end{equation}
with 
\begin{align}
   \mathcal{T}_{(0)}^{\mu\nu} &= 
    \left(
    \begin{array}{cccc}
        -\frac{3 c_{(0)}}{4} & f_{(0)} & 0 & 0 \\
        f_{(0)} & -2 b_{(0)}\!-\!\frac{c_{(0)}}{4} & 0 & 0 \\
        0 & 0 & b_{(0)}\!-\!\frac{c_{(0)}}{4} & 0 \\
        0 & 0 & 0 & b_{(0)}\!-\!\frac{c_{(0)}}{4} \\
    \end{array}
    \right) , \label{SE0}\\
   \mathcal{T}_{(1)}^{\mu\nu} &= 
   \left(
    \begin{array}{cccc}
        \frac{3 c_{(0)}}{2}\!-\!\frac{3 c_{(1)}}{4} & f_{(1)}\!-\!2 f_{(0)} & 0 & 0 \\
        f_{(1)}\!-\!2 f_{(0)} & 4 b_{(0)} \! -\!2 b_{(1)}\!+\!\frac{c_{(0)}}{2}\!-\!\frac{c_{(1)}}{4} & 0 & 0 \\
        0 & 0 & -2 b_{(0)}\!+\!b_{(1)}\!+\!\frac{c_{(0)}}{2}\!-\!\frac{c_{(1)}}{4} & 0 \\
        0 & 0 & 0 & -2 b_{(0)}\!+\!b_{(1)}\!+\!\frac{c_{(0)}}{2}\!-\!\frac{c_{(1)}}{4} \\
    \end{array}
    \right).\label{SE1}
\end{align}

We must in addition find the relationship that determines the conserved $U(1)$ (baryon number) current
in the boundary field theory from the functions appearing in Eq.~(\ref{Amuexpansion}).
The conserved boundary current can be identified as the boundary value of the canonical momentum with respect to $r$ of the bulk Maxwell field. We see this as follows. The gauge field equations of motion in the bulk imply that 
\begin{equation}
    \nabla_\mu \Pi^{\mu\nu}=0, \qquad {\rm where}\quad \Pi^{\mu\nu}\equiv \frac{ \partial \mathcal{L}}{\partial(\nabla_\mu A_\nu)}
\end{equation}
is the canonical momentum conjugate 
to the bulk gauge field $A_\nu$, satisfying $\Pi^{\mu\nu}=-\Pi^{\nu\mu}$, and with $\mathcal{L}$ being the Lagrangian corresponding to the action \eqref{ActFin}. 
The covariant divergence of an antisymmetric tensor satisfies $\nabla_{\mu}\Pi^{\nu\mu}=\frac{ 1 }{ \sqrt{-g} }\partial_\mu (\sqrt{-g} \Pi^{\nu\mu}) $.
Hence, 
\begin{equation}\label{GreekAndLatin}
    -\nabla_\mu \Pi^{\mu r}= \nabla_\mu \Pi^{r \mu} 
    = \frac{ 1 }{ \sqrt{-g} }\partial_a\left(\sqrt{-g}\Pi^{ra}\right) = 0 ,
\end{equation}
where, to avoid confusion, in this paragraph alone we have found it necessary
to use Latin indices $a, b,\ldots$ for the indices that 
run over the four-dimensional boundary coordinates only. 
From (\ref{GreekAndLatin}) we see that $\sqrt{-g}\,\Pi^{ra}$ is a conserved current within each constant-$r$ slice. 
We can therefore define the conserved boundary current as
\begin{equation}
    J^a = \lim_{r\rightarrow \infty}\sqrt{-g} \Pi^{ra}.
\end{equation}
Inserting the near-boundary expansion of the gauge field (\ref{Amuexpansion}) and the metric functions (\ref{Cexpansion})--({\ref{Fexpansion}) into this expression, we find
\begin{equation}\label{Jdict}
    \mathcal{J}^a = \epsilon\left[\mathcal{J}^a_{(0)} + \lgb\mathcal{J}^a_{(1)} + {\cal O}(\lgb^2)\right], 
\end{equation}
with 
\begin{equation}
    \mathcal{J}_{(0)}^a = \frac{ 1 }{ 2 } a_{(0)}^a, \qquad  \mathcal{J}_{(1)}^a = \frac{ 1 }{ 2 }(1+24 c_{\beta}) a_{ (0)}^a + \frac{ 1 }{ 2 } a_{(1)}^a , \label{J01}
\end{equation}
which is the relationship that we need. 
We find no divergences in the $r\rightarrow\infty$ limit of $\sqrt{-g}\,\Pi^{ra}$, so no counterterms are needed. Note also that we have computed $\Pi^{\mu\nu}$ by taking the derivative with respect to $\epsilon A_\mu$, so that $\mathcal{J}\sim O(\epsilon)$, which is equivalent to first calculating $\Pi^{\mu\nu}$ and then recaling $A_\mu \rightarrow \epsilon A_\mu$, as described in the previous Section. We compute the baryon number density from $\rho = \mathcal{J}^0$. 
Finally, in (\ref{Jdict}) and (\ref{J01}) we can safely replace the Latin index $a$
by a Greek index $\mu$, returning to the notation that we have used elsewhere.

Having described the equations of motion, the initial conditions and the scheme for computing the time evolution, we can now perform collisions between two incident sheets of energy and baryon number and extract the resulting $\mathcal{T}^{\mu\nu}$ and $\mathcal{J}^{\mu}$. 
We shall present results in the next Section.

We anticipate that the matter produced in such collisions will rapidly hydrodynamize, becoming
strongly coupled liquid plasma whose subsequent expansion and cooling dynamics is well-described by relativistic viscous hydrodynamics, including a conserved (baryon number) current. In order to check that the plasma does indeed hydrodynamize, after presenting our results 
we shall need to define the standard hydrodynamic variables (including the baryon number density) 
in terms of $\mathcal{T}^{\mu\nu}$ and $\mathcal{J}^{\mu}$ and test the validity of the hydrodynamic approximation \eqref{SEhydro}--\eqref{Jhydro} 
by comparing solutions to these hydrodynamic equations to the full results that we shall obtain
by solving the holographic equations of motion that determine the
functions defined in \eqref{Amuexpansion}--\eqref{Fexpansion} and appearing in the energy-momentum tensor \eqref{Tdict} and the current \eqref{Jdict}.

In order to make a comparison to hydrodynamics, we need to obtain the local fluid velocity $u^{\mu}$ and then the local energy density $\mathcal{E}_{\rm loc}$  and the 
local baryon number density $\rho_{\rm loc} $ defined in the rest frame of the fluid. We proceed
as follows.
At a boundary point $x$ of the bulk spacetime, we define $u_{\mu}$ as the timelike eigenvector of $- \mathcal{T}\indices{^{\mu}_{\nu}}(x)$, hence, 
\begin{equation}\label{uDef}
    \mathcal{T}\indices{^{\mu}_{\nu}} u^{\nu} = -\mathcal{E}_{\rm loc} u^{\mu}.
\end{equation}
The local baryon number density is then 
\begin{equation}\label{rhoDef}
\rho_{\rm loc} = -J_{\mu}u^{\mu}.
\end{equation} 
To evaluate \eqref{SEhydro} and \eqref{Jhydro} we also need the equilibrium expressions for $T$ and $\mu$ as functions of $\mathcal{E}$ and $\rho$. (We shall use these expressions to obtain the
local $T$ and $\mu$ from  $\mathcal{E}_{\rm loc}$ and
$\rho_{\rm loc} $.)   We obtain the relationships that we need 
from the black brane solution of the bulk theory gravitational equations of motion, as this is the dual of the equilibrium state in the boundary field theory. To linear order in $\lgb$, the black brane solution is given by 
\begin{align}
    A &= -\left[g_0 + \frac{Q_0}{2 r^2} + \lgb\left(g_1 + \frac{Q_1}{2 r^2}+c_{\beta} \frac{4 k_0 Q_0 }{r^6}\right)\right] dt, \quad B=0, \quad F=0,  \\
    C &= r^2-\frac{k_0}{r^2} + \lgb \left( -\frac{k_1}{r^2}+c_{\epsilon}\frac{Q_0^2 }{12 r^4}+\frac{k_0^2}{r^6} \right), \quad S^2 = r^2,
\end{align}
where  $g_0$, $g_1$, $k_0$, $k_1$, $Q_0$ and $Q_1$ are free parameters. The Hawking temperature $T$, the chemical potential $\mu$, and the entropy density $s$ of the black brane solution are given by 
\begin{align}
    T &= \frac{ C'(r_+) }{ 4\pi } = \frac{k_0+r_0^4}{2 \pi  r_0^3} 
    + \lgb \left(\frac{6 k_1 r_0^4-18 k_0 r_1 r_0^3-18 k_0^2-Q_0^2 r_0^2 a +6 r_1 r_0^7}{12 \pi  r_0^7} \right), \label{Tdef} \\ 
    \mu/\epsilon &= A_t(\infty) - A_t(r_+) = \frac{Q_0}{2 r_0^2} 
    + \lgb \left( \frac{8 k_0 Q_0 b +Q_1 r_0^4-2 Q_0 r_1 r_0^3}{2 r_0^6} \right),  \label{mudef} \\
    s &= \frac{ 1 }{ \vol(\mathbb{R}^3) }\left(\frac{ \Sigma }{ \kappa_5^2/2\pi }\right) = \frac{1}{\kappa_5^2/2\pi}\left( r_0^3 + 3 \lgb r_0^2 r_1 \right),  \label{sdef} 
\end{align}
where the horizon radius $r_+ \equiv r_{0} + \lgb r_{1} + {\cal O}(\lgb^2)$ is defined as the largest solution to $C(r) = 0$ and where $\Sigma$ is the black brane horizon area. Note that the entropy density of a black brane in Gauss-Bonnet gravity satisfies the same Bekenstein-Hawking formula as in pure Einstein theory~\cite{Rong-Gen:2002}, and, as 
we discuss in Appendix~\ref{Sec:ActFD} we find
that the same is true when we add the Maxwell and Maxwell-curvature coupling terms.
Recall, however, that what we need is expressions that relate $T$ and $\mu$ to 
$\mathcal{E}$ and $\rho$.
We can use \eqref{SE0}, \eqref{SE1} and \eqref{J01} to express the energy density $\mathcal{E}$, pressure $p$ and baryon number density $\rho$ as functions of $k_i$, $Q_i$ and $r_i$. 
Then, we can solve equations \eqref{Tdef}--\eqref{sdef} together with $C(r_+)=0$ for  $k_i$, $Q_i$ and $r_i$ in terms of $T$ and $\bar{\mu} \equiv \mu/\epsilon$. 
Plugging this into our expressions for $\mathcal{E}$, $p$ and $\rho$ we find
\begin{align}
    \mathcal{E} &= \frac{3}{4} \pi ^4 T^4 \left[1 + \lgb \left(3 +  c_{\epsilon} \frac{\bar{\mu}^2}{\pi ^2 T^2}\right)\right], \label{eq:edensexp} \\ 
    \bar{\rho} &= \frac{1}{2} \pi ^2 \bar{\mu}T^2 \left[ 1 + \lgb \left(3+\frac{ 1 }{ 3 } c_{\epsilon}\frac{\bar{\mu}^2 }{\pi ^2 T^2}+16 c_{\beta}\right)\right], \label{eq:rhomuexp}
\end{align}
where we remind that $\bar{\rho}\equiv\rho/\epsilon$. 
Finally, we can solve these expressions 
for $\bar\mu$ and $T$ 
which yields
\begin{align}
    T & = \frac{ \sqrt{2} \mathcal{E}^{\frac{ 1 }{ 4 }} }{ \pi 3^{\frac{ 1 }{ 4 }} } \left[1 - \lgb\left(\frac{ 3 }{ 4 } + c_{\epsilon}\frac{ 3\sqrt{3} }{ 8 }\left(\frac{ \bar{\rho} }{ \mathcal{E}^{\frac{ 3 }{ 4 }} } \right)^2 \right) \right] ,\label{Tresult} \\
    \bar{\mu} &= \frac{ \sqrt{3} \bar{\rho} }{ \sqrt{\mathcal{E}} }\left[1 - \lgb \left(\frac{ 3 }{ 2 } + 16c_{\beta} - c_{\epsilon} \frac{ \sqrt{3} }{ 4 } \left(\frac{ \bar{\rho} }{ \mathcal{E}^{\frac{ 3 }{ 4 }} } \right)^2 \right) \right] .\label{muresult}
\end{align}
The expressions (\ref{Tresult}) and (\ref{muresult}) are
the expressions that we shall use in order to obtain the local $T$ and $\mu$ 
from  $\mathcal{E}_{\rm loc}$ and
$\rho_{\rm loc}$, so that we can then use (\ref{uDef}) and (\ref{rhoDef}) in the evaluation of the 
expressions \eqref{SEhydro} and \eqref{Jhydro} for the two conserved tensors in the hydrodynamic
description of the fluid.

In order to complete the evaluation of \eqref{SEhydro} and \eqref{Jhydro} we also need the transport coefficients $\eta/s$ and $\sigma/T$.  We discussed the first of these already in the previous Section; to the order at which we are working, it is given by
\begin{align}\label{etaOverS2}
\frac{\eta}{s} = \frac{1}{4\pi} \left( 1 - 4 \lgb \right). 
\end{align}
The conductivity can be obtained by using the relation $\sigma = D \chi$, where $\chi$ is the baryon susceptibility and $D$ the baryon diffusion constant.
$D$  was calculated nonperturbatively in both couplings
at $\mu=0$ in our theory in Ref.~\cite{Grozdanov:2016fkt} and perturbatively
at $\mu\neq0$ in a theory related to ours by field redefinitions in Ref.~\cite{Myers:2009ij}. 
Appendix~\ref{Sec:ActFD} contains a discussion on how to convert the result of Ref.~\cite{Myers:2009ij} to our theory. 
The result, with the conventions given in Eqs.~\eqref{ChoiceOfUnits} and \eqref{rescaledCurrents}, is
\begin{equation}
    \sigma = \frac{ \pi T }{ 4 }\left[ 1+\lgb \left(2 + 32c_{\beta} - c_{\epsilon} \frac{ 5 }{ 6 \pi^2 } \left( \frac{ \bar{\mu} }{ T } \right)^2 \right)\right]. \label{eq:conductivitydef}
\end{equation}

We shall take $\mathcal{E}_{\rm loc}$ and
$\rho_{\rm loc}$ from our holographic simulation, use (\ref{Tresult}) and (\ref{muresult}) to obtain $T$ and $\mu$, and then use (\ref{uDef}) and (\ref{rhoDef}) to evaluate 
\eqref{SEhydro} and \eqref{Jhydro}}.  From these hydrodynamic solutions we can obtain hydrodynamic ``predictions'' for the time evolution of the longitudinal and transverse pressure as well as for the baryon
number density and current, all of which we can compare to our full results for the evolution of
these quantities, obtained from our holographic calculation.  This comparison will allow us to assess the degree to which the matter produced in these collisions does in fact hydrodynamize.

As a final consistency check 
we have verified that thermodynamic relation
\begin{equation}
    \mathcal{E} = s T + \mu \rho - p \label{eq:thermocheck}
\end{equation}
is indeed satisfied. Note that the $\beta$-dependent correction to $\mathcal{J}^\mu$ in Eq.~\eqref{J01} contributes to \eqref{eq:thermocheck} only at ${\cal O}(\lgb^2)$, since $\rho$ and $\mu$ are   ${\cal O}(\epsilon)$, so that the leading $\beta$-independent contribution of $\mu\rho$ is already ${\cal O}(\lgb)$. We have computed a black brane solution and an expression for $\mathcal{T}^{\mu\nu}$ also to ${\cal O}(\lgb^2)$ and verified that \eqref{eq:thermocheck} holds to second order as well.

\section{Results}\label{Sec:Results}

In this section we present the results of our numerical calculations of the collisions of thick and thin sheets of energy and baryon number density, as described in Section \ref{Sec:Analysis}.

\subsection{Colliding thick and thin sheets}

We begin the discussion by considering the case with $\epsilon=0$. 
This means that we compute the stress-energy and baryon number density while neglecting 
the back-reaction of the baryon number density on the energy-momentum tensor. 
In equilibrium terms, this limit corresponds to the situation in which the ${\cal O}\left(\frac{ \mu^2 }{ T^2 }\right)$ corrections to $\bar{\rho}$ and $\mathcal{E}$ are neglected.
On the bulk side of the duality, setting $\epsilon=0$ means that we neglect the backreaction of the gauge field dynamics on the Einstein equations \eqref{EQEin}. The evolution of energy and momentum in the $\epsilon=0$ system is therefore completely equivalent to what was studied in Ref.~\cite{Grozdanov:2016zjj} in Einstein-Gauss-Bonnet theory. Note
that $\epsilon$ cancels from the Maxwell equation \eqref{EQMax}, which implies that the dynamics of 
the baryon number density whose time evolution we shall follow
remains (highly) nontrivial even at $\epsilon=0$.

\begin{figure}
    \centering
    \subfloat[independent of $\beta$ at $O(\lgb)$]{{\includegraphics[width=0.47\textwidth]{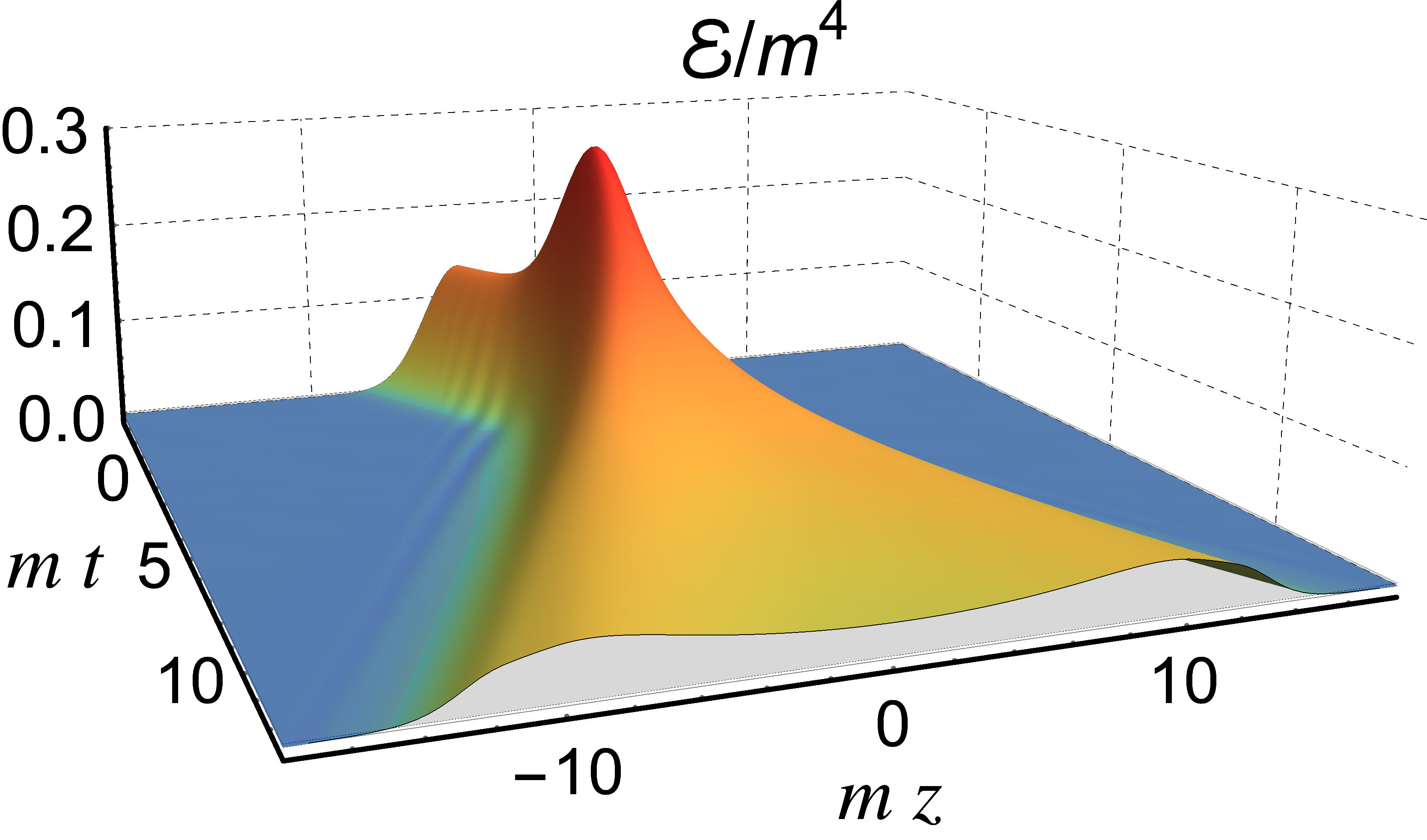}} \label{fig:edens3D}}%
    \\
    \subfloat[$\beta=0$]{
    {\includegraphics[width=0.47\textwidth]{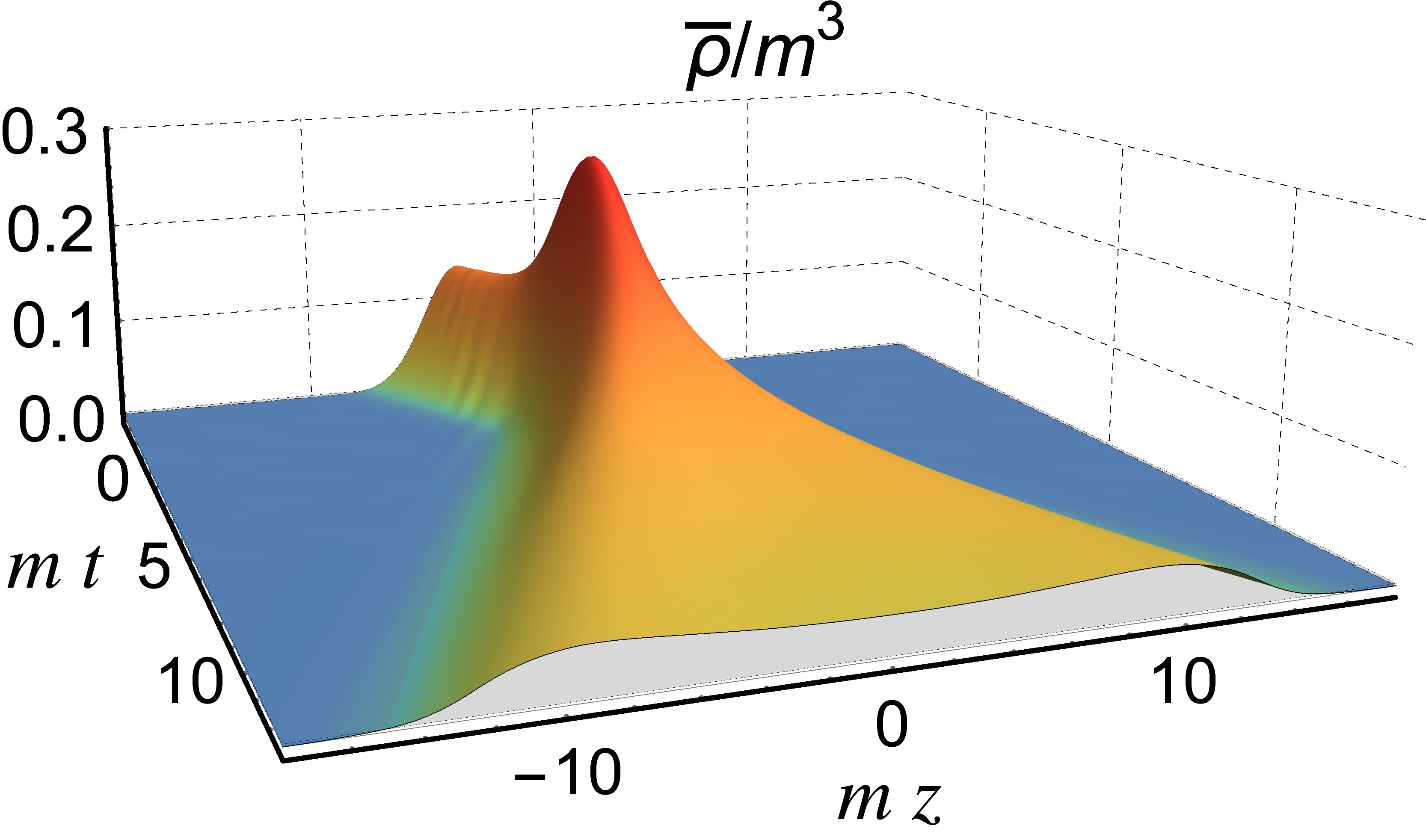}} \label{fig:charge3Dbeta0}}%
    \hspace{0.02\textwidth}
    \subfloat[$\beta=0.1$]{{\includegraphics[width=0.47\textwidth]{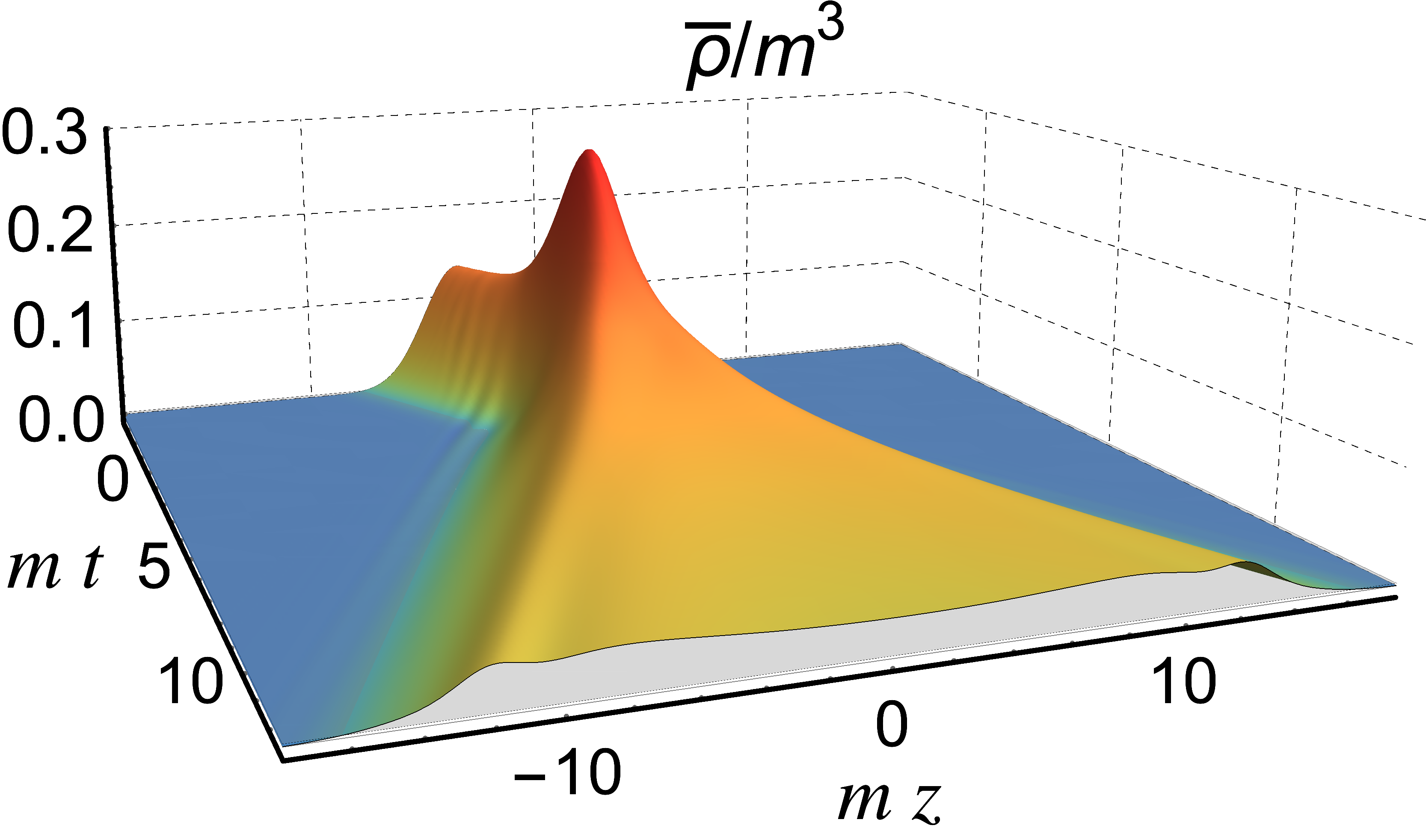}}\label{fig:charge3Dbeta0dot1}}%
    \caption{Spacetime dynamics of the energy density (top panel) and baryon number density (bottom panels) in collisions of thick sheets (width $w=1.5/m$) at $\lgb=0$ in the $\epsilon = 0$ limit (no back-reaction of the baryon number on the energy density). The two lower panels show the time evolution
    of the baryon number density at two different values of $\beta$. In each of the three panels, the horizontal axes are the time $t$ and the spatial coordinate $z$ along which the incident sheets are moving, in both cases measured in units of $m^{-1}$.  The energy and baryon number density (measured in units of $m^4$ and $m^3$ respectively) peak
    around $z=t=0$, the spacetime point at which the collision is centered.  At late times,
    we see the plasma, carrying both energy density and baryon density, extending between the two light-cones.  Although it is not visible by virtue of the perspective we have chosen, at early times before the collision there is nothing between the incident sheets. }%
    \label{fig:wide3D}%
\end{figure}

\begin{figure}
    \centering
    \subfloat[independent of $\beta$ at $O(\lgb)$]{{\includegraphics[width=0.47\textwidth]{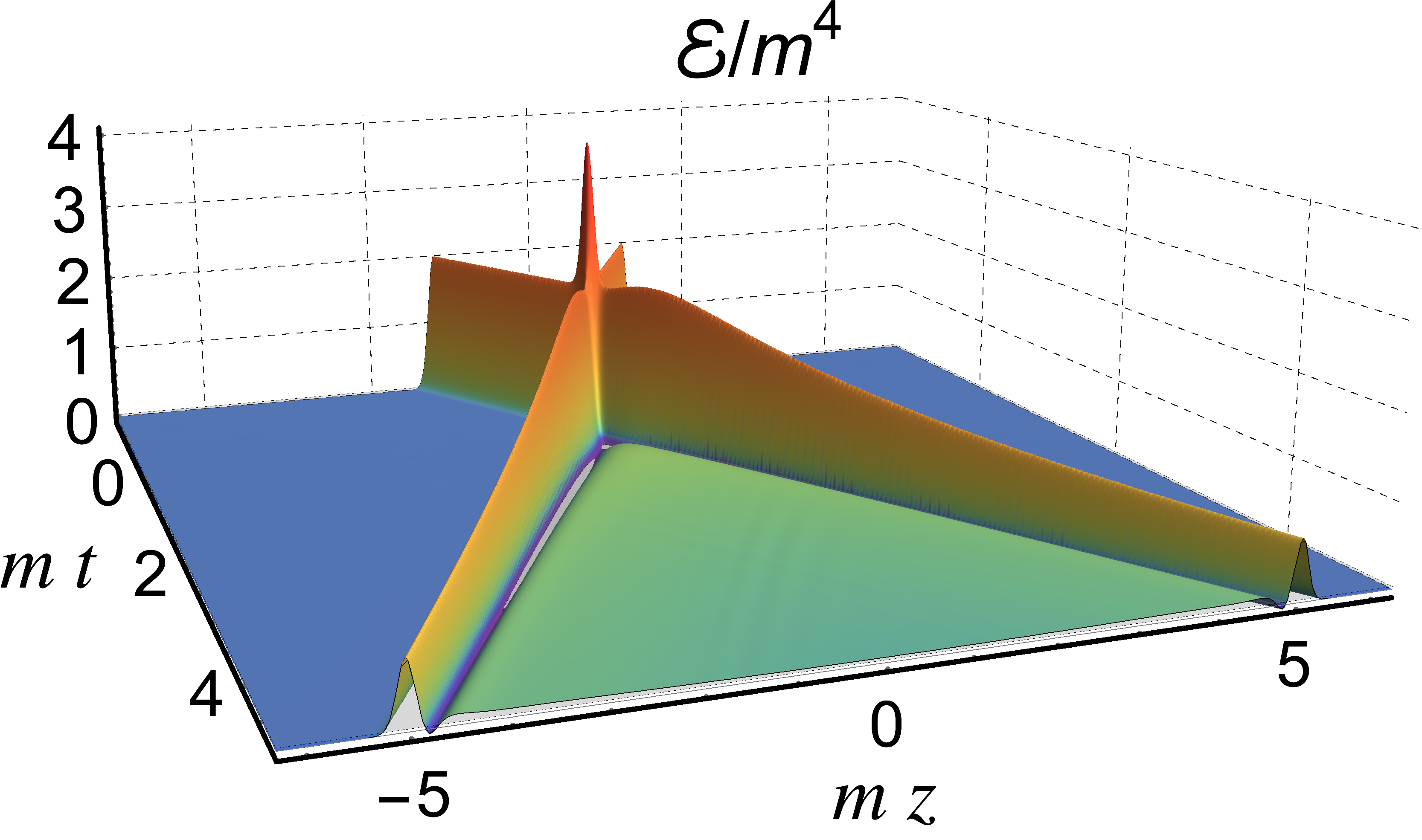}} \label{fig:edens3D_narrow}}
    \\
    \subfloat[$\beta=0$]{
    {\includegraphics[width=0.47\textwidth]{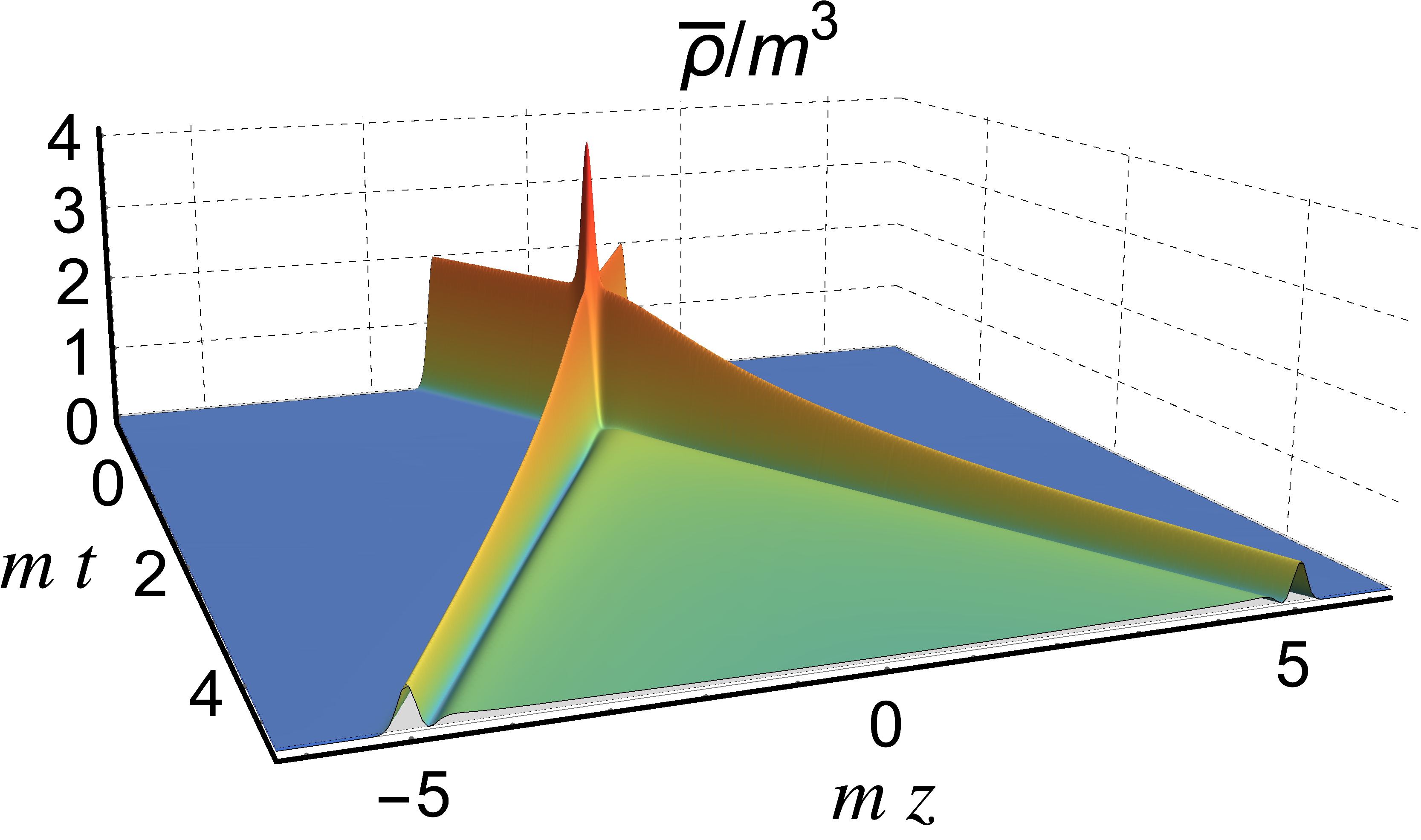}} \label{fig:charge3Dbeta0_narrow}}
\hspace{0.02\textwidth}
    \subfloat[$\beta=0.025$]{{\includegraphics[width=0.47\textwidth]{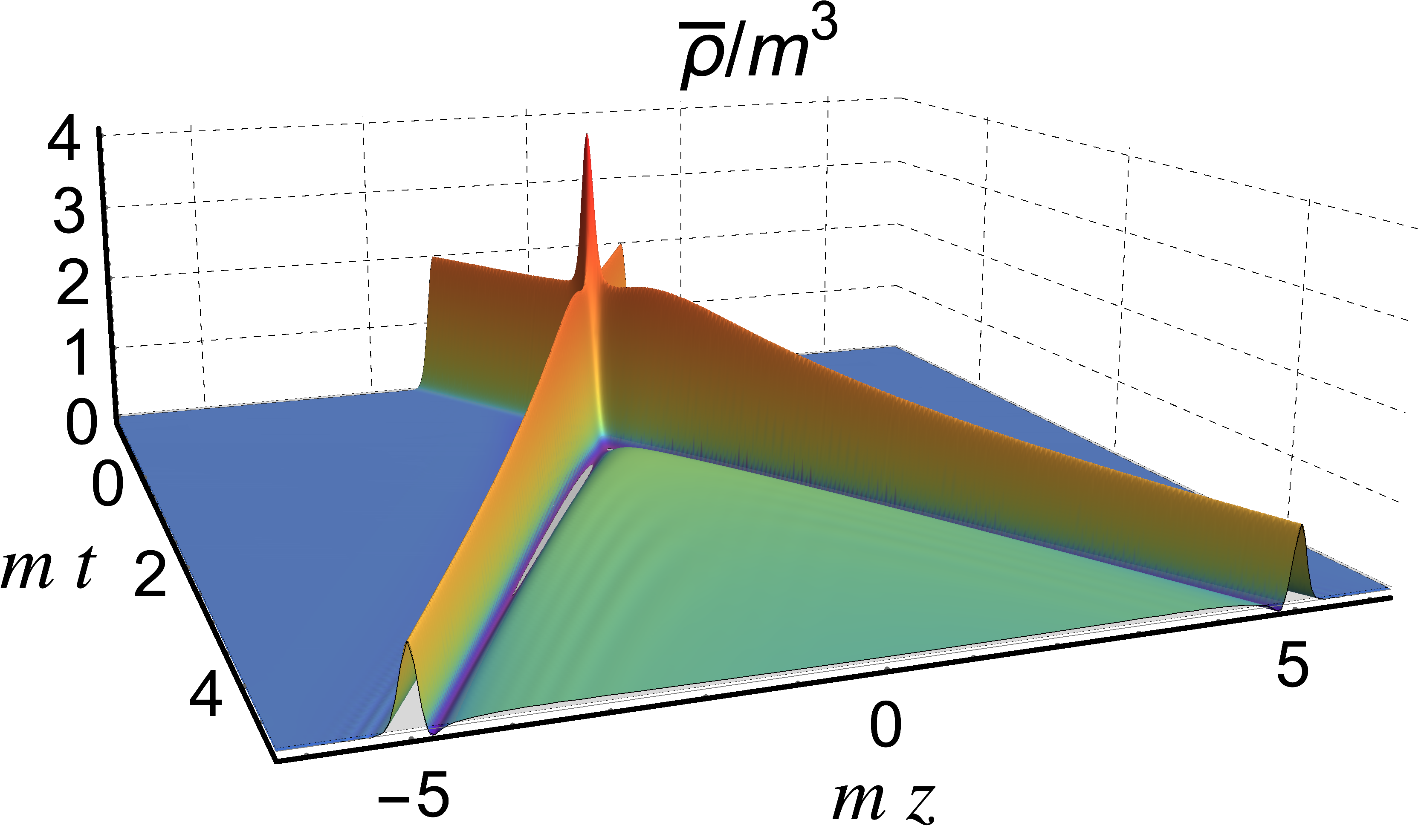}}\label{fig:charge3Dbeta0dot025_narrow}}%
    \caption{Spacetime dynamics of the energy density (top panel) and baryon number density (bottom panels) in collisions of thin sheets (width $w=0.1/m$) at $\lgb=-0.2$ in the $\epsilon=0$ limit.  The lower panels are at two different values of $\beta$ }
    \label{fig:narrow3D}
\end{figure}

To begin with a view of our results that is both illustrative and instructive, 
we show in Figs.~\ref{fig:wide3D} and \ref{fig:narrow3D} the evolution of energy and baryon number densities with $\epsilon=0$ and $\lgb = -0.2$ for thick and thin sheets, respectively. To the order that we are working, the
energy density has no dependence on $\beta$. We plot the baryon number density for two values of $\beta$ in each case, which we choose to be $\beta\in\{0, 0.1\}$ for thick sheets 
and $\beta\in\{0, 0.025\}$ for thin sheets. 
We choose the (smaller) values of $\beta$ that we use for thin sheets in order to be safely within the range of values of $\beta$ where all the qualitative considerations that we discussed
at the end of Section~\ref{Sec:HolographicSetup} indicate that working to linear order in $\beta$ is likely a safe approximation.  And indeed, if we increase $\beta$ further to $\beta=0.05$ we find that {\it after} the collision of thin sheets the baryon number on the lightcone increases for a time, which seems unphysical and thus suggests that working to linear order in $\beta$ is inadequate.  That said, in the case of collisions of thick sheets we find that $\beta$'s in this range
have only modest effects and so for this case we have been more bold and extended our
calculations up to $\beta=0.1$. It will take future higher-order calculations to reliably estimate up to what $\beta$ our linear approximation is under control.

Consistent with our general expectations about the phenomenology of heavy ion collisions, 
after dialing the coupling down from infinity towards intermediate values
we find that the inverse coupling constant corrections that we have computed do
indeed cause the collisions to become
more transparent to both energy density and baryon number density in the sense
that a larger amount of both energy and baryon number ends up moving close
to the light-cone after the collision at reduced coupling. 
With $\epsilon=0$ our results for the energy density are the same
as in Ref.~\cite{Grozdanov:2016zjj}, so we refer the reader 
to that work for a full analysis of the results shown in the upper panels of
Figs.~\ref{fig:wide3D} and \ref{fig:narrow3D}. We shall focus on 
the baryon number dynamics.

\subsection{Effect of $\beta$ and $\lgb$ on baryon number dynamics}

\begin{figure}[ht]
\centering
\includegraphics[width=0.69\textwidth]{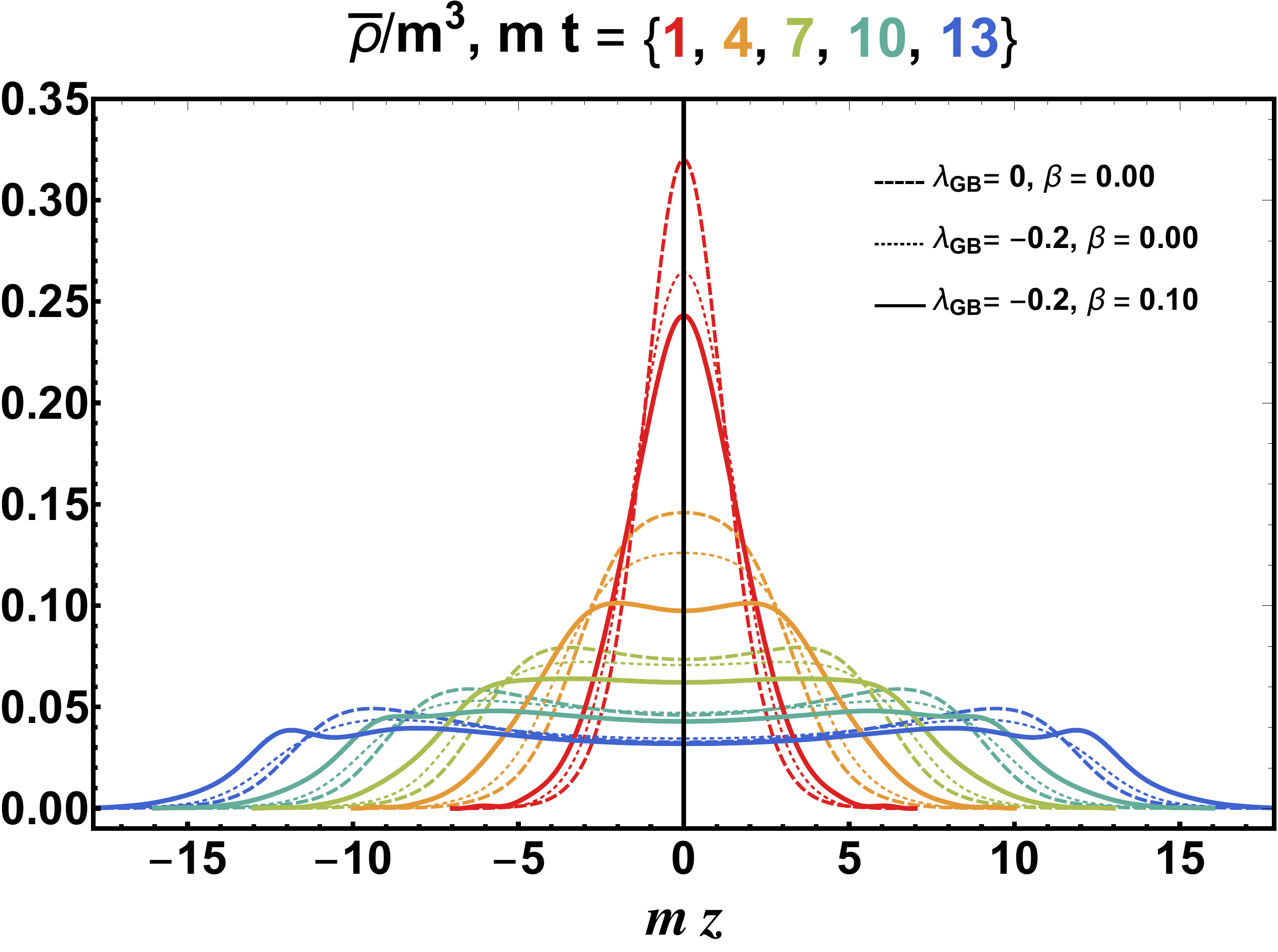}
\caption{Time evolution of the baryon number density 
after the collision of thick sheets with $\epsilon=0$, as in Fig.~\ref{fig:wide3D}. Colors denote  
snapshots in time. Solid/thin-dashed/thick-dashed curves denote
different values of $\lgb$ and $\beta$.}
\label{fig:wide_time_ev}
\end{figure}

\begin{figure}
\centering
\includegraphics[width=0.69\textwidth]{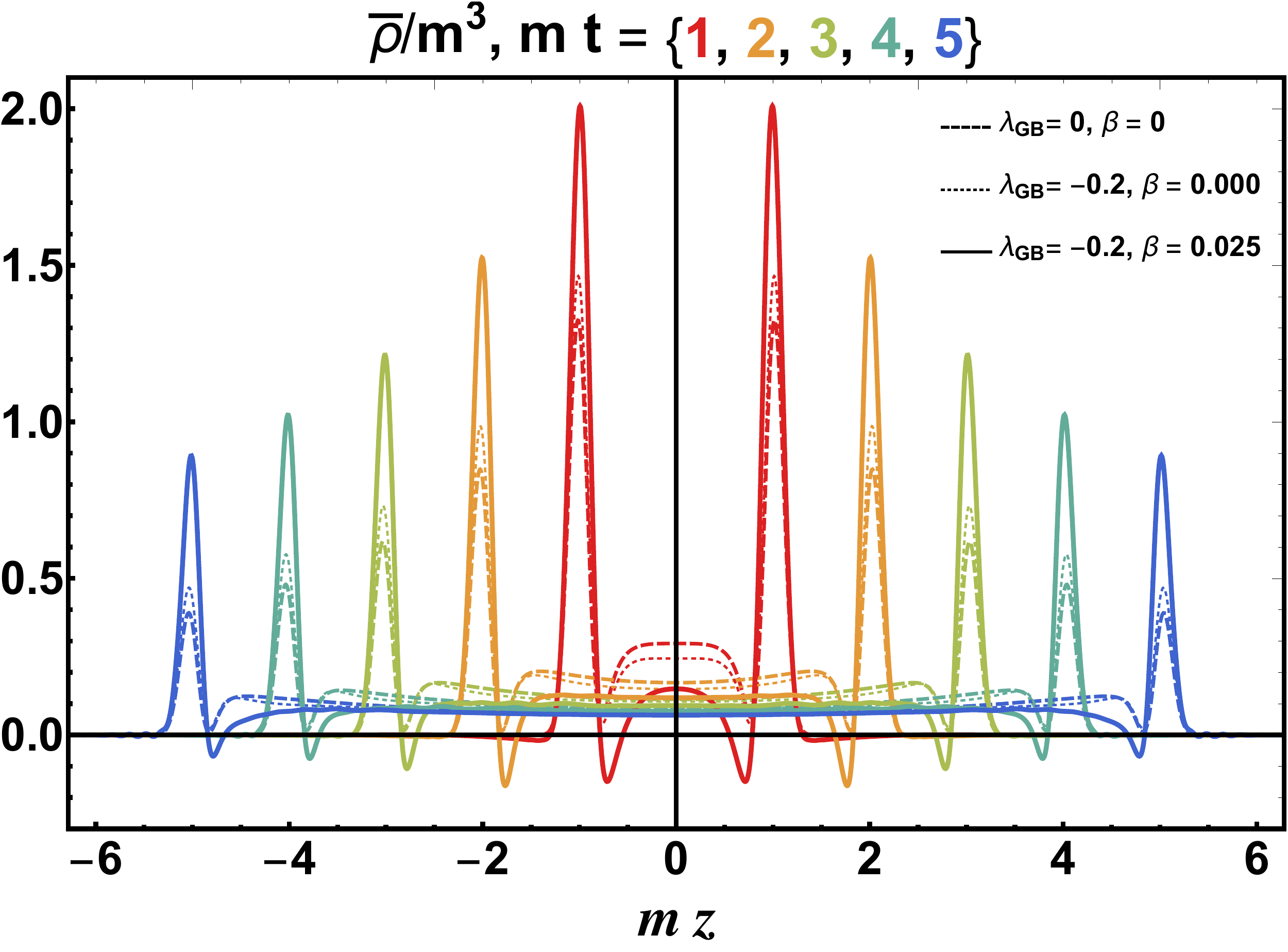}
\caption{Time evolution of the baryon number density after the 
collision of thin sheets with $\epsilon=0$, as in Fig.~\ref{fig:narrow3D}.  Colors denote\ snapshots in time; different line-styles denote different values of the couplings.}
\label{fig:narrow_time_ev}
\end{figure}

In this work, we are primarily concerned with the analysis of the effects of $\beta$ and $\lgb$ on the dynamics of baryon number density. In Figs.~\ref{fig:wide_time_ev} and \ref{fig:narrow_time_ev} we show snapshots that capture the evolution of the baryon number density profiles after the collisions (for $t \geq 0$) of thick and thin sheets, respectively. For comparison, at each time step 
we show a curve (solid) describing the baryon number dynamics in the
full theory, with $\lgb=-0.2$ and $\beta$ nonzero (0.1 for the thick sheets; 0.025
for the thin sheets) as well as a thin-dashed curve 
for the theory with $\lgb=-0.2$ and $\beta=0$ (whose gravitational
dual is Einstein-Maxwell-Gauss-Bonnet theory) and
a thick-dashed curve for the theory with $\lgb=\beta=0$ (whose gravitational dual is 
Einstein-Maxwell theory).  The snapshots at different times are denoted by different colors.

\begin{figure}[ht]
\centering
\includegraphics[width=0.7\textwidth]{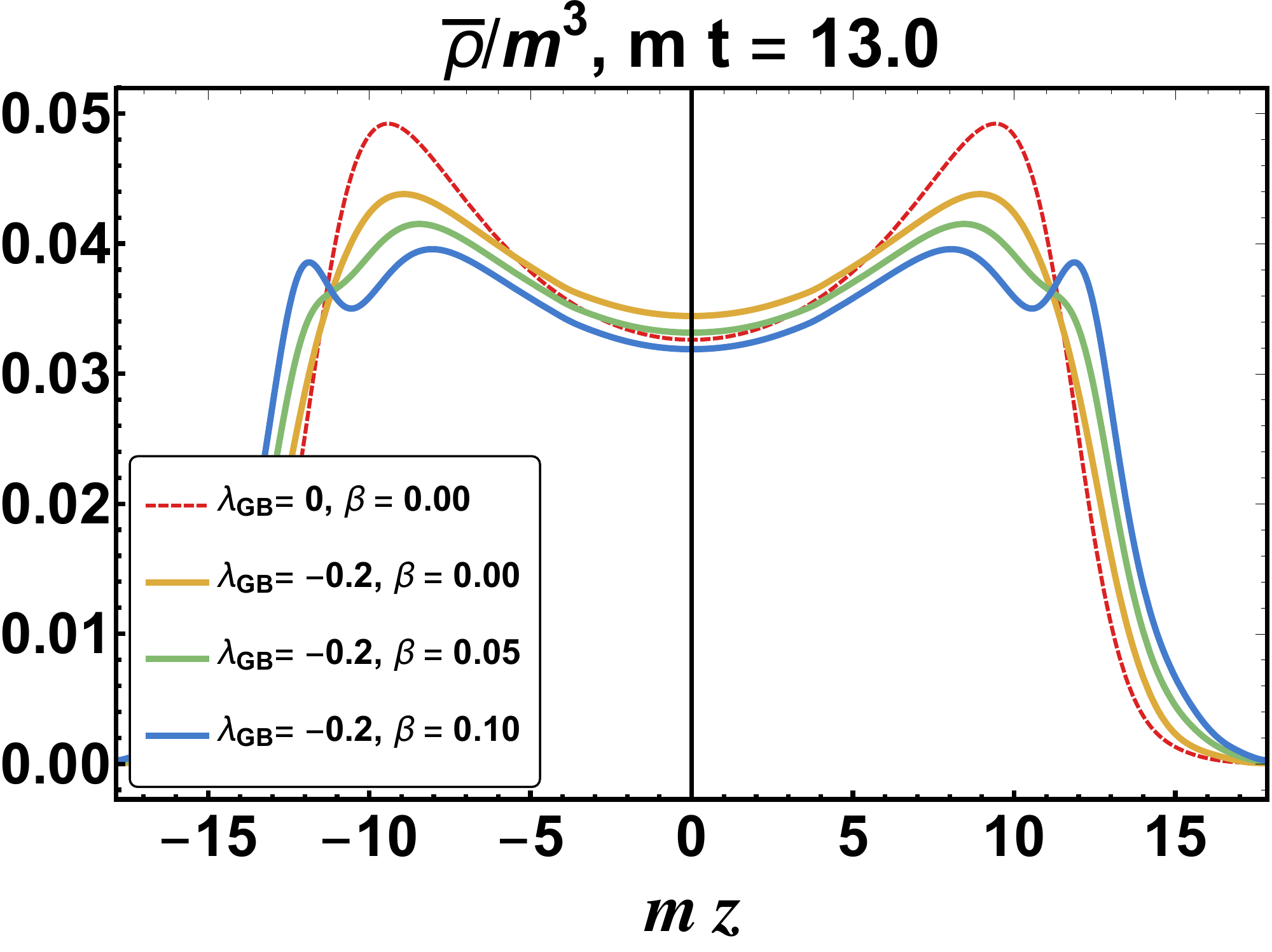}
\caption{$\beta$-dependence of the baryon number density after a collision between
thick sheets at the late time $t=13/m$  with $\epsilon=0$, plotted as a function of $mz$. }
\label{fig:wide_charge_betascan}
\end{figure}

\begin{figure}
\centering
\includegraphics[width=0.7\textwidth]{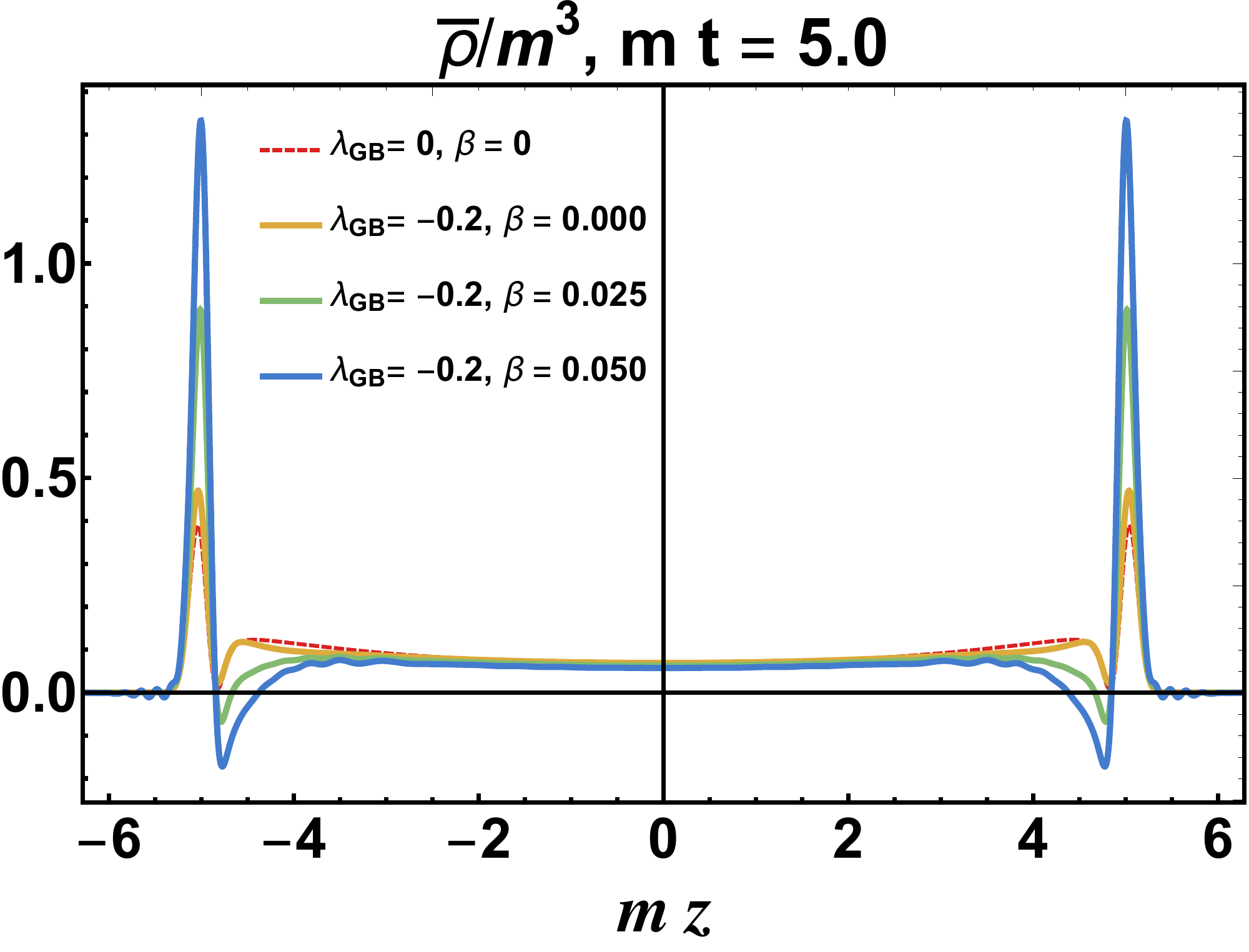}
\caption{$\beta$-dependence of the baryon number density after a collision between thin sheets at the late time $t=5/m$  with $\epsilon=0$, plotted as a function of $mz$.}
\label{fig:narrow_charge_betascan}
\end{figure}

In the case of the thick sheets, we see that turning on only $\lgb=-0.2$ with $ \beta=0$ slightly broadens the baryon number density profile at all times, indicating a somewhat increased transparency, similar to what is seen for the energy density~\cite{Grozdanov:2016zjj}. Even at late times, as depicted in Fig.~\ref{fig:wide_charge_betascan}, the effect is small and the qualitative picture remains unchanged. We see that for both zero and nonzero $\lgb$ when $\beta=0$, the thick sheets come to a full stop after which point the plasma expands. 

When we turn on a nonzero $\beta$ (specifically $\beta=0.1$), we see 
in Figs.~\ref{fig:wide_time_ev} and \ref{fig:wide_charge_betascan} 
that a small bump of baryon number appears on the lightcone, indicating that not all of the baryon number is fully stopped in the collision.  In Fig.~\ref{fig:wide_charge_betascan} we can
see this bump develop as we compare the baryon number density
at late times ($t=13/m$, the latest time plotted
in Fig.~\ref{fig:wide_time_ev}) in collisions of thick sheets with increasing values of $\beta$
at a fixed value of $\lgb$.  

For the thin sheets at infinite coupling, we see in Fig.~\ref{fig:narrow_time_ev} 
that a significant amount of baryon number remains on the lightcone after the collision, 
and baryon number is gradually deposited in the plasma. The change in the baryon number 
density profile upon setting $\lgb=-0.2$ and  $\beta=0$ is modest, 
while for the case of $\lgb=-0.2$ and  $\beta=0.025$ ({\em not} $\beta = 0.1$ as for thick sheets), a significant increase of baryon number is seen on the lightcone. 
This is clear from Fig.~\ref{fig:narrow_time_ev}, where at the latest plotted times, the peak baryon number density 
near the lightcone is more than doubled compared to the infinitely coupled limit. 
In Fig.~\ref{fig:narrow_charge_betascan}, we highlight the $\beta$-dependence of
this increased transparency by focusing on the $\beta$-dependence of the
baryon number density distribution in collisions of thin sheets at $t=5/m$, the 
latest time plotted in Fig.~\ref{fig:narrow_time_ev}.  Increasing $\beta$ to
$\beta=0.05$ further increases the effect. 
We find that if we were to set $\beta \approx 0.3$, which is far too large
for the linear approximation that we are using to be reliable, 
the baryon number density around $z=0$ is reduced to the point that it
vanishes at the late time shown  in Fig.~\ref{fig:narrow_charge_betascan}. 
(In the case of the collision of thick sheets, as in Fig.~\ref{fig:wide_time_ev}, our calculation
would indicate that the baryon number density around $z=0$ is reduced to near
vanishing at late times after the collision for $\beta \approx 1$, which is even farther beyond the regime where the linear approximation that we are using
can be trusted.)

To make this discussion more precise, we define a quantity $R_{\rm LC}(t)$ to be the fraction of the total baryon number at a given time that is found within $2w$ (with $w$ being the width of each incident sheet) of one lightcone or the other:
\begin{equation}\label{eq:LCdef}
    R_{\rm LC}(t) = \frac{ \int_{I(t)} d z \rho(t, z)}{  \int_{-\infty}^{\infty} d z \rho(t, z)  },
\end{equation}
where $I(t)\equiv [-t-2w, -t+2w]\cup[t-2w, t+2w]$. 
We find that at $t=13/m$ after the collision of thick sheets in Figs.~\ref{fig:wide_time_ev} and \ref{fig:wide_charge_betascan}, 
$R_{\rm LC}^{\rm(thick)} \in \{0.21, 0.22, 0.28\}$ for $(\lgb, \beta) \in \{(0, 0), (-0.2, 0), (-0.2, 0.1)\}$. 
And, we find that at $t=5/m$ after the collisions of thin sheets in Figs.~\ref{fig:narrow_time_ev} and \ref{fig:narrow_charge_betascan},
$R_{\rm LC}^{\rm (thin)} \in \{0.16, 0.20, 0.36\}$ for $(\lgb, \beta) \in \{(0, 0), (-0.2, 0), (-0.2, 0.025)\}$. 
We note that increasing $\beta$ from 0 to 0.1 at fixed $\lgb=-0.2$ increases $R_{\rm LC}^{\rm thick}$ from
0.22 to 0.28 while increasing $\beta$ from 0 to 0.025 at fixed $\lgb=-0.2$ increases $R_{\rm LC}^{\rm thin}$
from 0.20 to 0.36.
In both cases, increasing $\beta$ increases the transparency of the collisions
to baryon number density, with a higher fraction of the total baryon number ending up
near the lightcones. And, we see that the relative importance of $\beta$ increases
with decreasing sheet width: the increase in transparency becomes more sensitive to $\beta$
as the sheet width decreases.

\begin{figure}
\centering
\includegraphics[width=0.48\textwidth]{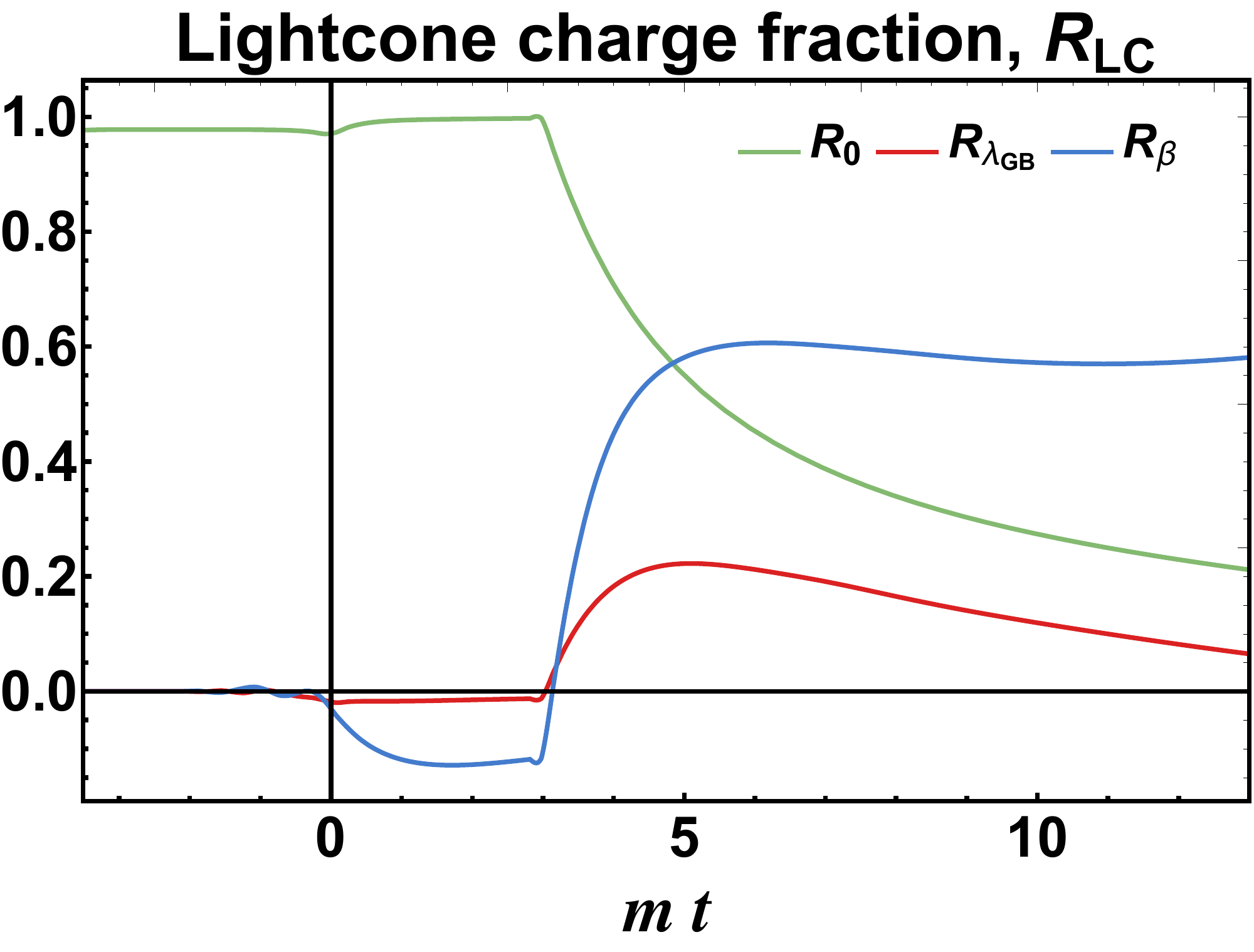}
\includegraphics[width=0.47\textwidth]{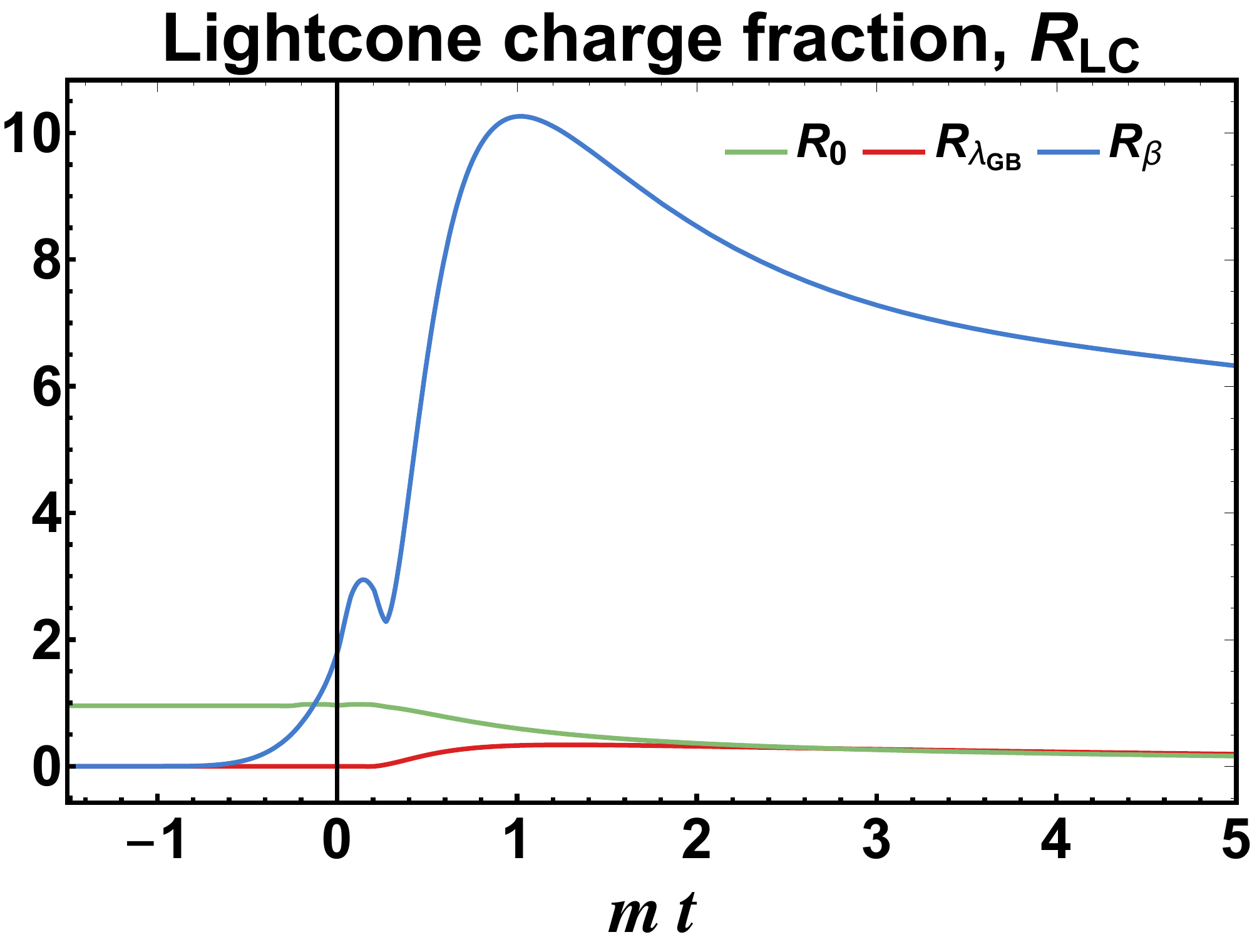}
\caption{Time dependence of the fitted coefficients $R_0$, $R_\beta$ and $R_{\lgb}$ 
defined in (\ref{eq:dcoefdef}) that describe the evolution of the three 
contributions to the fraction of the total baryon number that
ends up near one or other lightcone, $R_{\rm LC}$ defined in (\ref{eq:LCdef}),
after collisions between thick (left) and thin (right) sheets.}
\label{fig:fitcoefs}
\end{figure}

In fact, the first-order perturbative equations that describe the time evolution and that we have solved to obtain these results are linear in both $\lgb$ and $\beta$. This means
that the baryon number density $\rho$, at fixed $(z, t)$, is a linear function of $\beta$ and $\lgb$ (and of $\epsilon^2$, which has so far been set to zero). It follows that $R_{\rm LC}(t)$ is also linear in the same variables, and we can define the coefficients $R_i$ as 
\begin{equation}\label{eq:dcoefdef}
    R_{\rm LC}(t) = R_0(t) + R_{\beta}(t) \beta - R_{\lgb}(t) \lgb.
\end{equation} 
In Fig.~\ref{fig:fitcoefs} we show the time evolution for the coefficients $R_0$, $R_\beta$ and $R_{\lgb}$. Indeed, we observe that the $\beta$ term is responsible for the dominant contribution to the late time baryon number on the lightcone and that the $\beta$ term has a much greater effect for thin sheets. In particular, at the latest times plotted in Fig.~\ref{fig:fitcoefs} (which are the
times at which Figs.~\ref{fig:wide_charge_betascan} and \ref{fig:narrow_charge_betascan} are plotted), we find that 
\begin{align}
    R_{\rm LC}^{\rm(thick)}(mt=13) = 0.21 + 0.58\beta - 0.07 \lgb, \label{Rthick}\\
    R_{\rm LC}^{\rm(thin)}(mt=5)  = 0.16 + 6.32\beta - 0.19 \lgb. \label{Rthin}
\end{align}
Comparing Eqs.~\eqref{Rthick} and \eqref{Rthin} demonstrates that 
the effects of $\lgb$ and $\beta$ depend on the width of the sheets. 
Their influence on the dynamics of baryon number transport and hence
on its distribution in the final state is substantial in the 
collision of thin sheets, and 
becomes less for thicker sheets.
For example, if we set $\lgb=-0.2$ and compare $R_{\rm LC}$ with 
$\beta=0.05$ to $R_{\rm LC}$ with $\beta=0$, for thick sheets $R_{\rm LC}$ only increases
from 22\% to 25\% while for thin sheets it increases from 20\%  to 51\%.

From the point of view of the gravitational 
action \eqref{ActFin}, it is natural to ascribe this behavior to the 
fact that the two bulk couplings $\lgb$ and $\beta$ which represent inverse-coupling
corrections in the gauge theory multiply higher-derivative terms in the gravitational action. As such, 
when derivatives of the fields are larger (in the gauge theory and consequently in the bulk description)
the consequences of the terms in the action multiplied by these couplings 
should become larger, larger
relative to results obtained at infinite coupling in the gauge theory
when the gravitational action is the two-derivative Einstein-Maxwell action. 
This is consistent with 
what we have found by direct calculation of the time evolution.
What is particularly interesting is that the effect of $\beta$ on the dynamics 
is much more strongly dependent on the width of the sheets than that of $\lgb$. 
The effect of $\beta$ is illustrated in Figs.~\ref{fig:wide_charge_betascan} and \ref{fig:narrow_charge_betascan}.
Since $\mathcal{E}$ and $\rho$ are linear in  $\lgb$, $\beta$ and $\epsilon^2$, the trends 
seen upon varying $\beta$ in these Figures 
are also obtained for any other values of $\lgb$ and $\epsilon^2$, when working to linear order.

It is interesting to speculate on the qualitative implications of these results for QCD.
The collision of thick sheets can be seen as a model for the collision of less ultrarelativistic, and therefore
less Lorentz-contracted, heavy ions, with the thin sheets then modelling more ultrarelativistic
heavy ion collisions, with higher beam energy.  We now recall from our discussion
in the Introduction that as the beam energy
of a heavy ion collision is increased the baryon transparency increases: 
the baryon number density deposited in the QGP
produced at mid-rapidity decreases and more of the baryon number ends up at higher 
rapidities, closer and closer to the lightcone.  This means that as the beam energy
of heavy ion collisions is increased, the baryon number density  in these QCD collisions
becomes more and more different from what is seen in holographic collisions at infinite
coupling.
In our calculation, increased $\beta$ corresponds to studying
holographic collisions with a reduced gauge coupling.
It is therefore tempting to speculate that the ``reason'' that a given value of $\beta$ 
makes a much bigger difference in our calculations (causing 
a much bigger increase in baryon transparency)
for collisions of thin sheets than for collisions of thick sheets is that
the infinite-coupling results for thin sheets are ``more wrong'' than those for thick sheets, and
hence receive larger finite-coupling corrections.

We close our discussion of these results by noting that the values of $\beta$ that
we have employed are around what we described as the reasonable range, \eqref{eq:beta-range}, based upon consideration of the effect of $\beta$ on the baryon susceptibility in equilibrium.
But,  at the same time, these values of $\beta$ 
are in a regime where, as we expected based upon the discussion 
at the end of 
Section~\ref{Sec:HolographicSetup}, we find $\beta$-induced effects that are so large that
the linear approximation is near or perhaps somewhat beyond its regime of validity.
Given how different the baryon number distribution is in ultrarelativistic heavy ion collisions
in QCD relative to that obtained in holographic calculations at infinite coupling,
it is quite pleasing that reasonable values of $\beta$ have a substantial effect, in particular
in the case of collisions of thin sheets.  It is even more pleasing that these effects go in 
the right direction, increasing the baryon transparency and leading to more baryon number
ending up near the lightcones after the collision.   

A necessary consequence of
these pleasing conclusions, however, is that further corrections that are higher
order in $\beta$ (as well as all corrections that come in up to the same 
higher order in the inverse-gauge-coupling including those that are higher order in $\lgb$) are likely important.  Cranking up $\beta$ increases the 
baryon number on the lightcones and if we work only to linear order this increase is linear in $\beta$. This increase must be tamed at large enough $\beta$ by corrections that are higher order in $\beta$. We leave the calculation of higher-order-in-inverse-coupling corrections to future work.

\subsection{Spacetime rapidity distribution}

\begin{figure}
\centering
\includegraphics[width=0.48\textwidth]{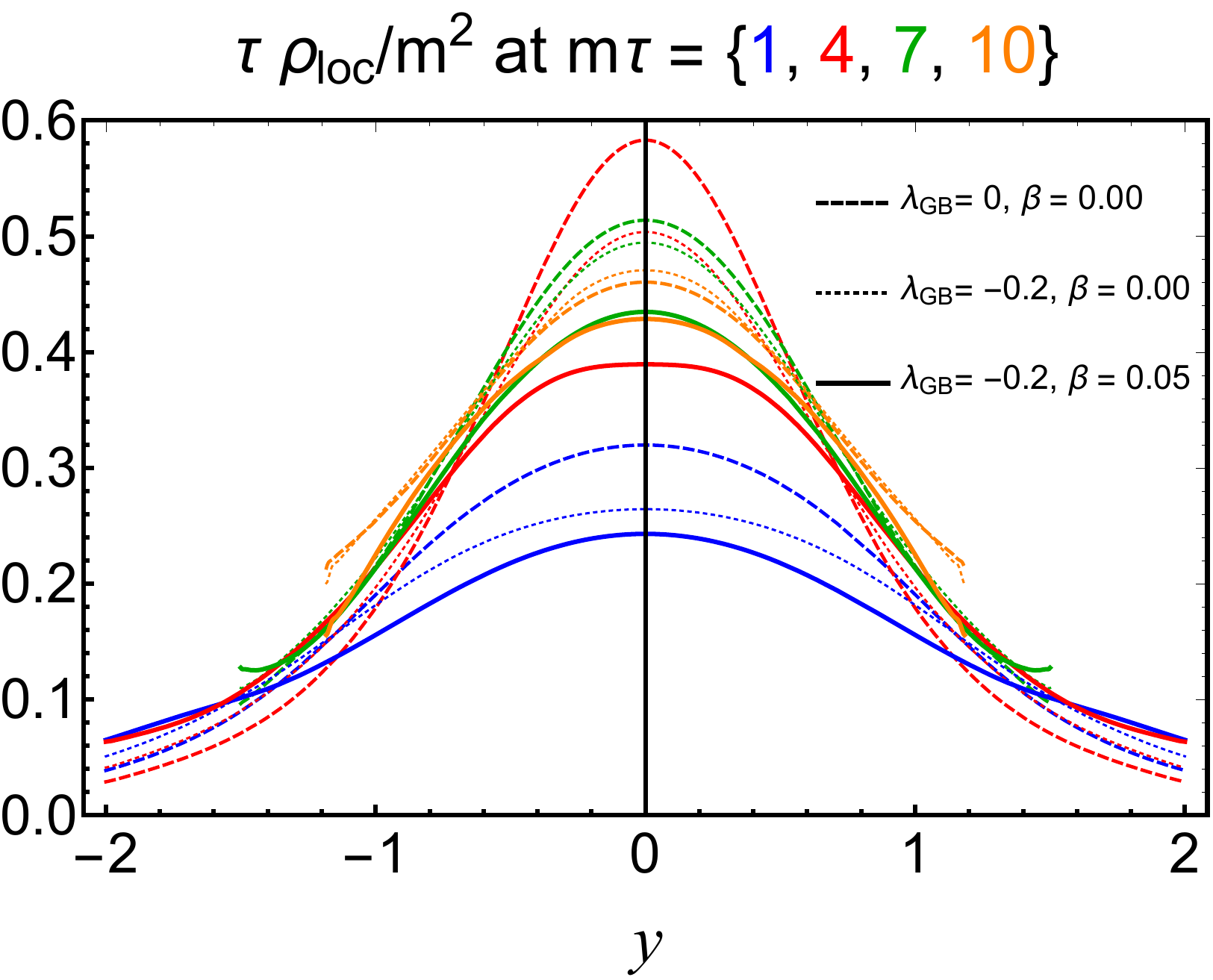}
\includegraphics[width=0.48\textwidth]{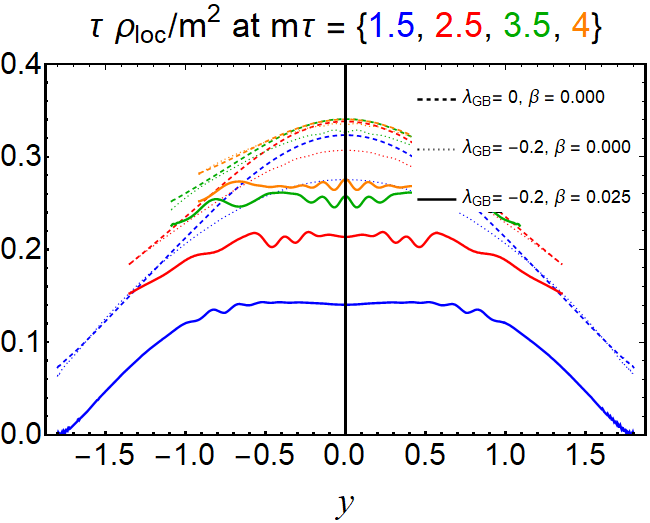}
\caption{The baryon number density in the local fluid rest frame, multiplied by the proper time $\tau$, 
as a function of the spacetime rapidity $y$
plotted at several different proper times after the collision of 
thick sheets (left) and thin sheets (right). 
The presence of the $\lgb$ corrections makes the rapidity distribution of the
baryon number 
wider and lower, with more baryon number at higher rapidity. The effect of $\beta$ is similar, but much stronger. 
Note also that at mid-rapidity the product $\tau \rho_{\rm loc}$ is approximately constant 
at late times, as would be expected if the dynamics approaches boost-invariance at late times.
The wiggles in the results from our $\beta=0.025$ calculations plotted in the right panel are artifacts
of our numerical discretization; as we go to finer discretizations, the wiggles decrease while in other respects the results do not change.
\label{fig:rapidity}}
\end{figure}

Next, to further illustrate our results
we transform from Minkowski coordinates $(t,z)$ to proper time $\tau=\sqrt{t^2-z^2}$
and spacetime rapidity $y={\rm arctanh}(z/t)$, compute the baryon number density
in the local fluid rest frame $\rho_{\rm loc}$, and
plot the spacetime rapidity distribution of $\rho_{\rm loc}$, 
shown in Fig.~\ref{fig:rapidity}. 
For collisions of both thick and thin sheets, we see that reducing the 
gauge coupling reduces the baryon number density in the local fluid rest frame at mid-rapidity 
and broadens the rapidity distribution. This is similar to the 
effect of $\lgb$ on the rapidity profile of the energy density analyzed 
in Ref.~\cite{Grozdanov:2016zjj}. For the baryon number, however, there are in addition 
effects introduced by the $\beta$-dependent terms. And, as we saw before, the
consequences of reasonable nonzero values of 
$\beta$ can be large. 
In particular, at $\lgb=-0.2$ and $\beta=0.025$, at a proper time $\tau = 2.5/m$ after the collision
of thin sheets the baryon number 
density at mid-rapidity is reduced by 56\%, whereas with the same $\lgb$ but $\beta=0$ this reduction is only 16\%. 
The magnitude of the effects that we see with $\lgb=-0.2$ and $\beta=0.025$ further suggests
that with these values of the couplings in the bulk gravitational action we are 
describing collisions in a gauge theory with an intermediate value of the gauge coupling.
It is also interesting to note that the product $\tau \rho_{\rm loc}$ at later times is approximately constant, as would be expected if the dynamics becomes boost invariant at late
times.

We note of course that baryon number is conserved, meaning that since the total baryon number present in the range of $y$ that we plot in Fig.~\ref{fig:rapidity} decreases as $\beta$ is increased,
the baryon number density at (much) higher rapidities or on the lightcone increases with increasing $\beta$.  These results thus  further confirm the increasing transparency of the initial moments of these collisions to baryon number with increasing $\beta$.   As we noted previously, the baryon density found near $y=0$ (namely near $z=0$) at late times in our calculations can be reduced to the point that it is close to vanishing in the collision of thin (thick) sheets if we increase $\beta$ to $\approx 0.3$ ($\approx 1$),
but these values of $\beta$ are well beyond those for which our linear approximation can be trusted.

\subsection{Hydrodynamization}

\begin{figure}
\centering
\includegraphics[width=0.48\textwidth]{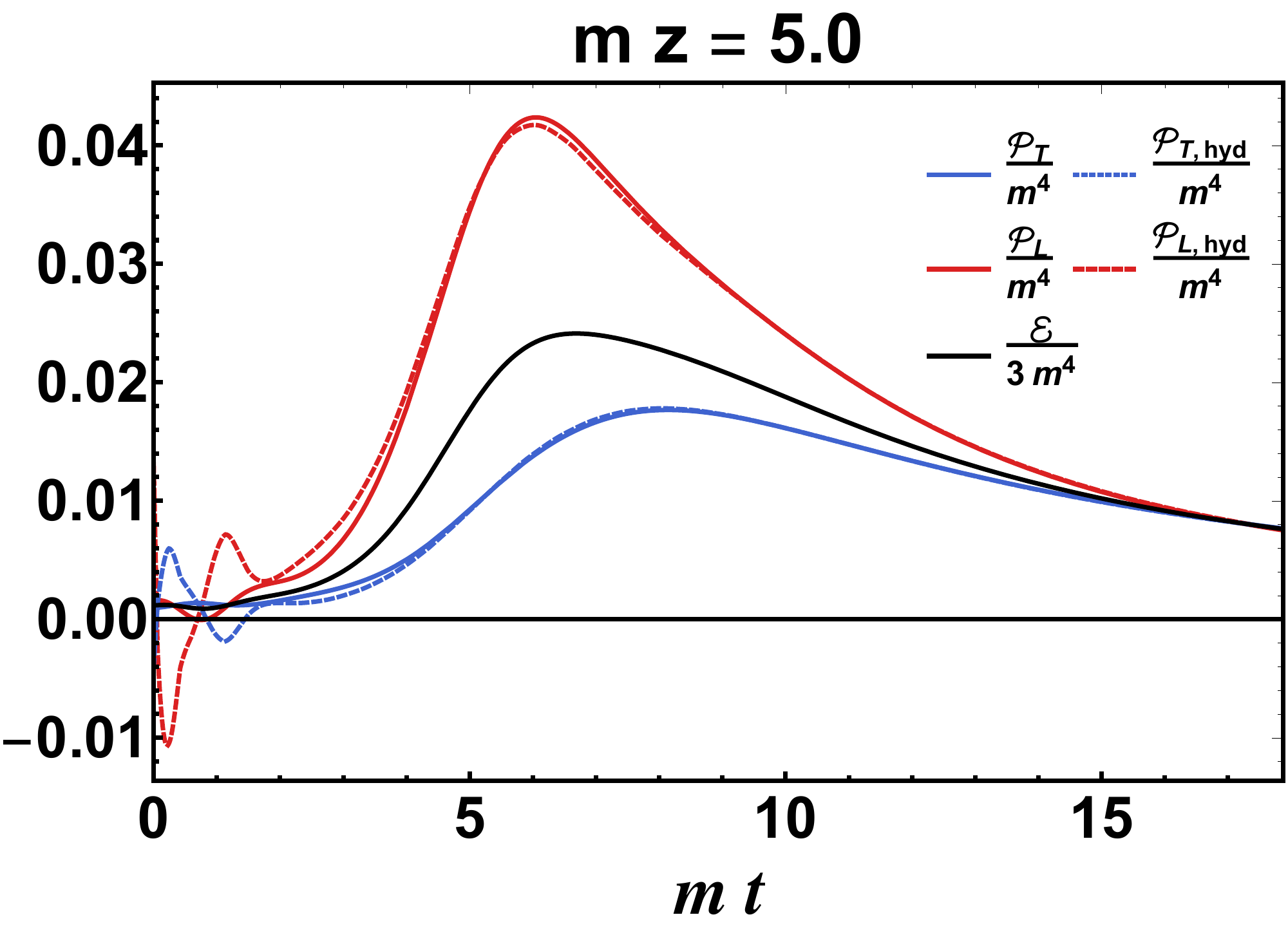}
\includegraphics[width=0.48\textwidth]{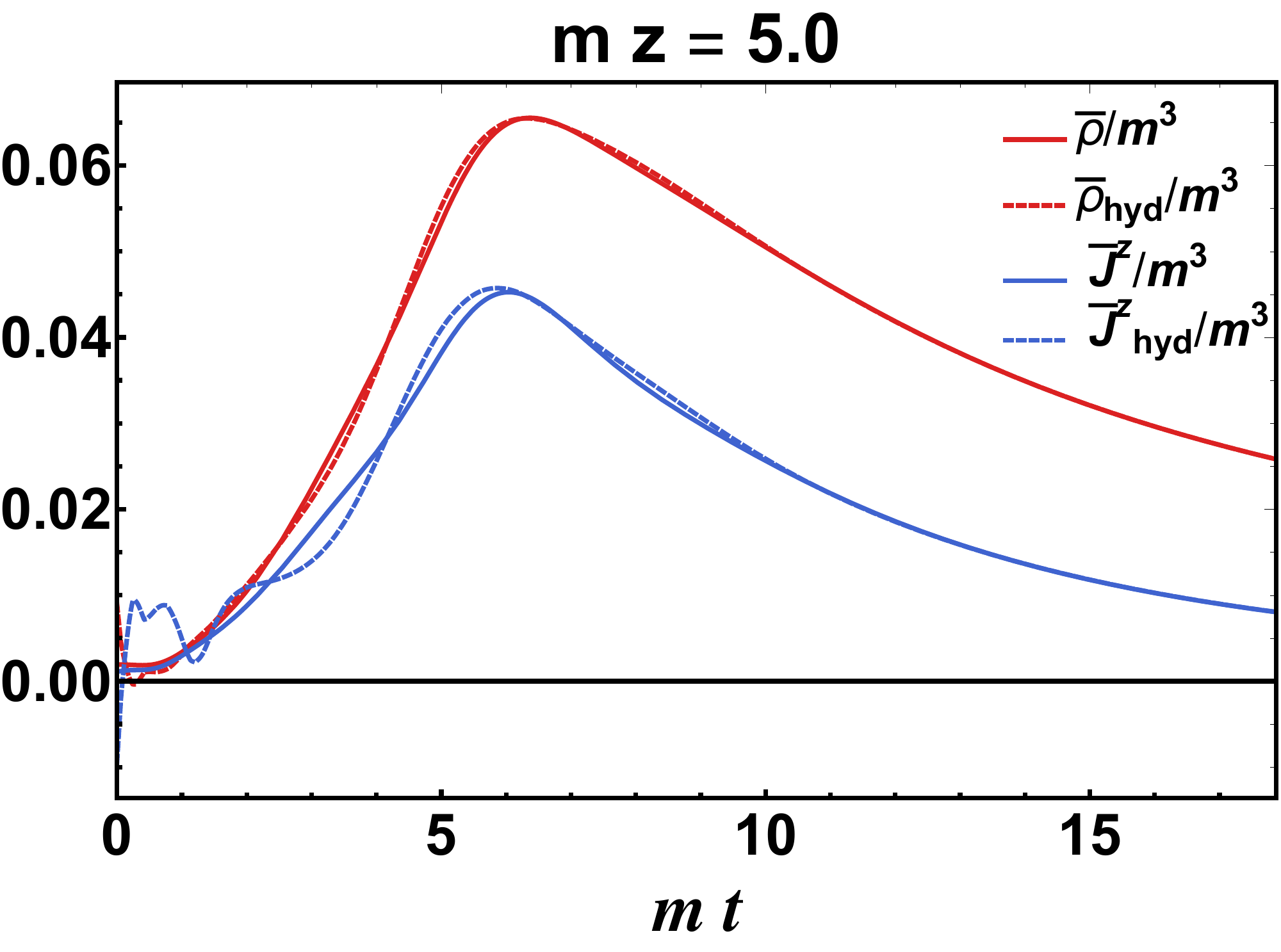}
\caption{Comparison of $\mathcal{P}_L$, $\mathcal{P}_T$, the lab frame baryon number 
density $\rho\equiv J^t$ and current $J^z$ for thick sheets extracted directly from simulation versus the values obtained via
the hydrodynamic constitutive relations. The calculation is done  with $\epsilon=0$, $\lgb=-0.2$ and $\beta=0.05$. The calculation is done at $z=5/m$, meaning that the lightcone
is at $t=5/m$. 
We also plot $\mathcal{E}$ and $\rho$, which
we obtain from the full simulation and use as inputs to the hydrodynamic constitutive
relations; the outputs of these relations are then the hydrodynamic ``predictions''
for $\mathcal{P}_L$, $\mathcal{P}_T$, and $J^z$.  We also compare the hydrodynamic
prediction for $\rho$ obtained from $\rho_{\rm loc}$ via the hydrodynamic 
constitutive relation to that which we take from the full simulation.  The
hydrodynamic predictions for $\rho$ and $J^z$ are related
through $u^\mu$ in that $u_\mu J^{\mu} = u_\mu J^{\mu}_{\rm hyd}=-\rho_{\rm loc}$.
}
\label{fig:wide_hydro}
\end{figure}

\begin{figure}
\centering
\includegraphics[width=0.48\textwidth]{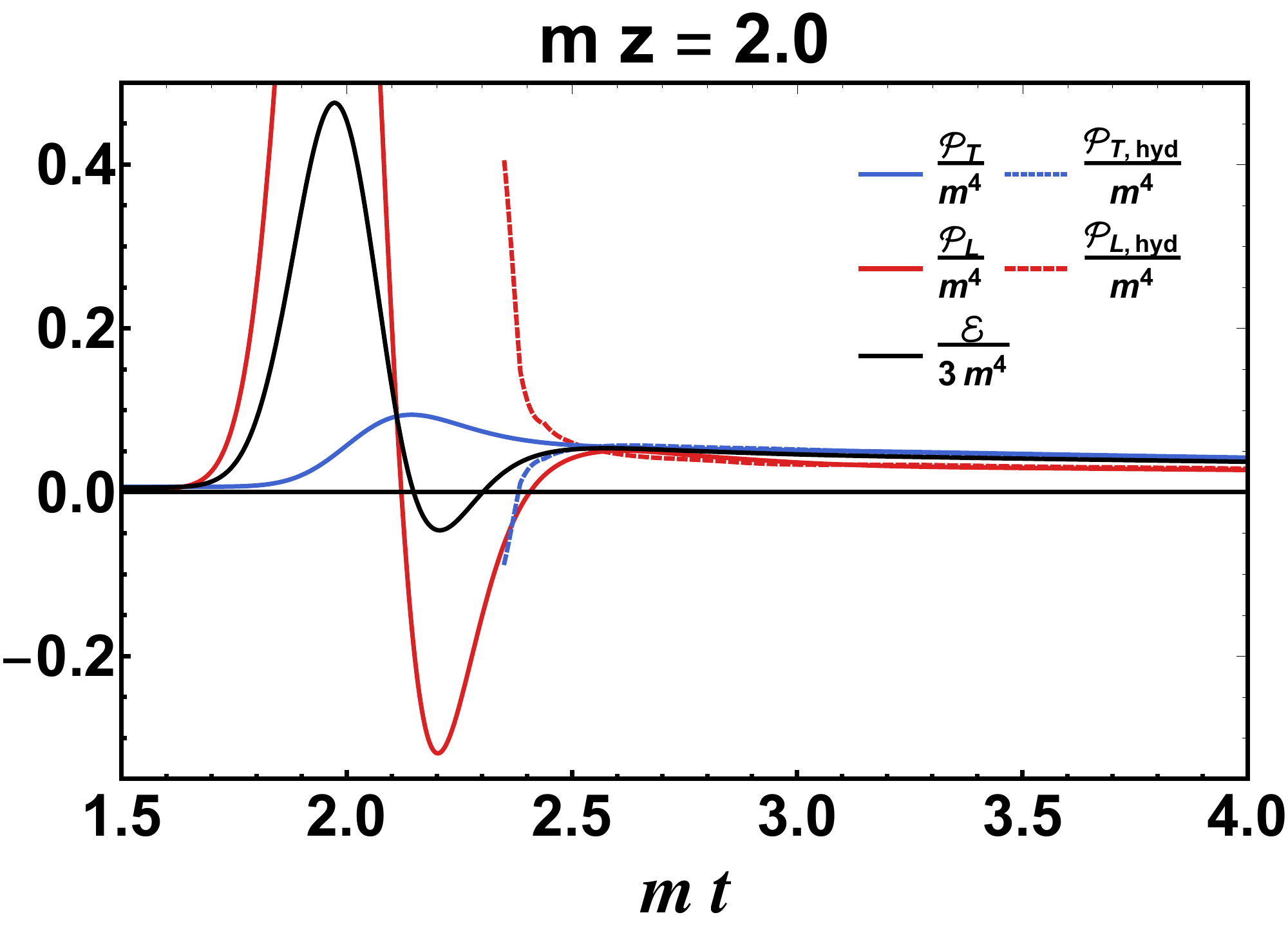}
\includegraphics[width=0.48\textwidth]{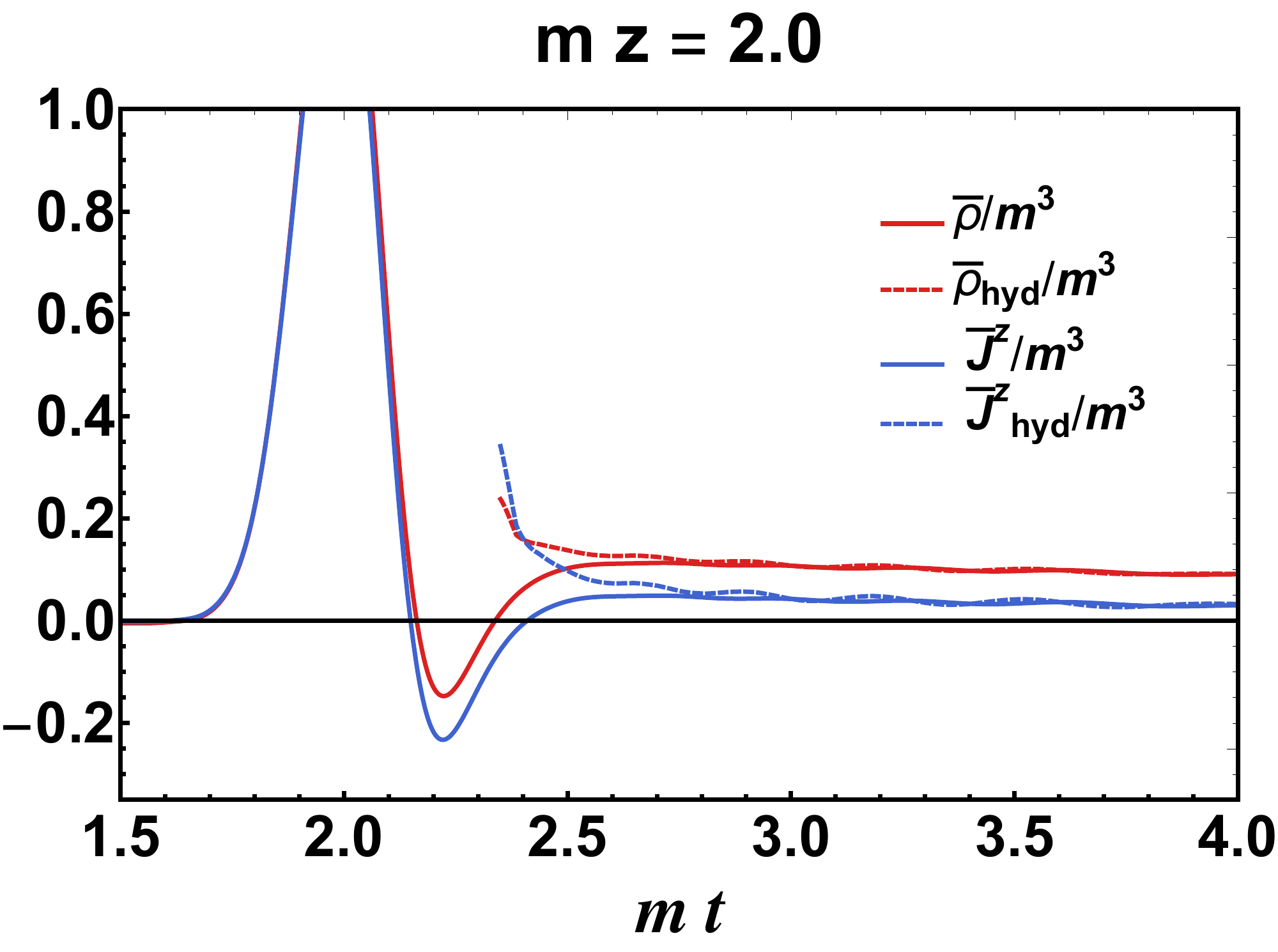}
\caption{Comparison of $\mathcal{P}_L$, $\mathcal{P}_T$, $\rho$ and $J^z$ for thin sheets extracted directly from simulation versus values  ``predicted'' 
from the hydrodynamic constitutive relations, as described above. The calculation is done with $\epsilon=0$, $\lgb=-0.2$ and $\beta=0.025$. The calculation is done at $z=2/m$, meaning that the lightcone is at $t=2/m$.
For completeness we also plot $\mathcal{E}$.
}
\label{fig:narrow_hydro}
\end{figure}

As promised in Sections \ref{Sec:HolographicSetup} and \ref{Sec:Analysis}, we now check when and to what degree
the matter produced in collisions like those in Figs.~\ref{fig:wide3D} and \ref{fig:narrow3D} hydrodynamizes.
In Figs.~\ref{fig:wide_hydro} and \ref{fig:narrow_hydro} we plot selected components of $\mathcal{J}^{\mu}$ and $\mathcal{T}^{\mu\nu}$ at constant-$z$ slices as obtained directly from 
our holographic calculation together with the values obtained 
from the hydrodynamic constitutive relations.  Recall that in Section \ref{Sec:Analysis} we
described the procedure to start from $\mathcal{E}$ and $\rho$ as obtained
from the full holographic calculation,
obtain the local fluid velocity, and from that determine the local baryon
number density in the local rest frame of the fluid, which we denote by $\rho_{\rm loc}$,
and then use hydrodynamic constitutive relations
to obtain the transverse and longitudinal pressure, denoted 
$\mathcal{P}_T$ and $\mathcal{P}_L$, and the baryon number current $J^z$. 
We then compare these three quantities obtained in this fashion 
via the laws of hydrodynamics to their values as obtained directly from the 
full holographic calculation.  We show this comparison for 
thick (thin) sheets with $\beta=0.05$ ($\beta=0.025$). 
We see from Fig.~\ref{fig:wide_hydro} 
that the transverse and longitudinal pressures, $\mathcal{P}_T = T^{x_{\perp}}_{\ x_{\perp}}$ and $\mathcal{P}_L = T^{z}_{\ z}$, are well described by first-order hydrodynamics for nearly all times during and after a collision between thick sheets, even before the peak energy density has passed.
And, hydrodynamics with a conserved current describes $J^z$ well 
starting around the time at which the energy and baryon number densities
peak. From Fig.~\ref{fig:narrow_hydro}, we see that the transverse
and longitudinal pressure are well described by hydrodynamics
starting about $0.6/m$ after the time $2.0/m$ at which the energy density
produced in a collision of thin sheets peaks, while $J^z$ is described
well starting a little bit later, about $0.8/m$ after the time when the
energy density peaks.  Note that the temperature of the plasma at this
time is $T_{\rm hyd} \approx 0.32\, m$, where we have used the relationship \eqref{eq:edensexp} between
$T$ and $\mathcal{E}$, meaning that
the hydrodynamization process after the collision of our thin
sheets is completed about $0.26/T_{\rm hyd}$ after the 
energy density peaks, whereas the collision of our thick sheets is close to
hydrodynamic throughout.

\subsection{Subleading $\mu/T$ effects for thick sheets}

\begin{figure}
\centering
\includegraphics[width=0.7\textwidth]{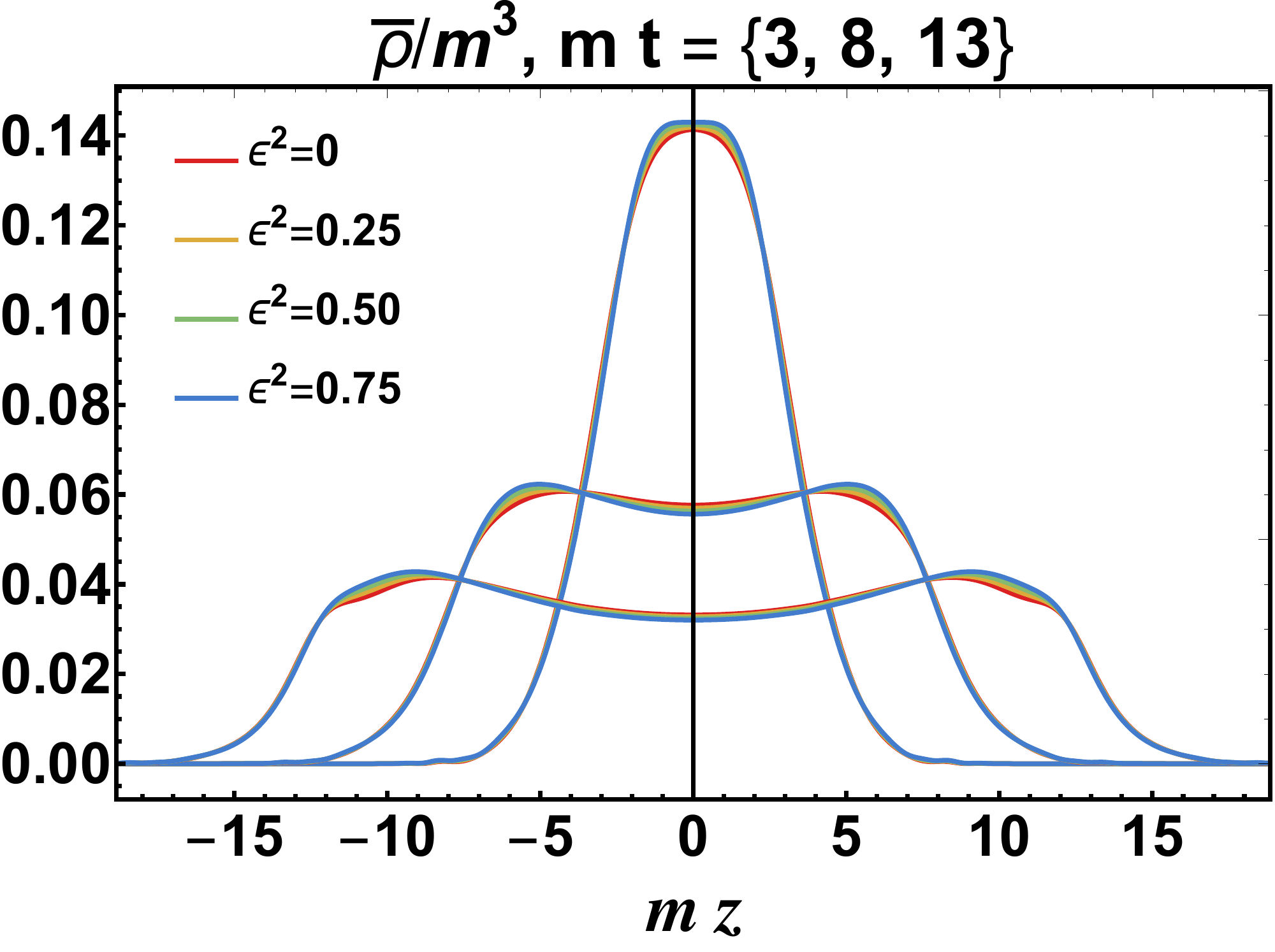}
\caption{The effect of $\epsilon^2$ on the $\bar{\rho}$ distribution at three times
after a collision of thick sheets, for $\lgb=-0.2$, $\beta=0.05$.}
\label{fig:echargeeffect}
\end{figure}

We now turn to the case where $\epsilon \neq 0$, where the lowest order ${\cal O}\left(\frac{ \mu^2 }{ T^2 } \right)$ corrections to $\rho$ and $\mathcal{E}$ are included in our calculation, meaning
in particular that we include the backreaction of $\rho$ on $\mathcal{E}$.  We are at the same time
 including
what can be thought of as the backreaction of $\rho$ on $\rho$.
In bulk terms, $\epsilon \neq 0$ means that we are including the backreaction of 
the Maxwell field on the bulk geometry, to linear order. 
In a collision of 
thick sheets we now find that $\frac{ \bar{\mu} }{ T }\approx 1.5 + {\cal O}(\lgb)$ in the region where hydrodynamics is valid, so that $\left(\frac{ \mu }{ T }\right)^2 = \epsilon^2 \left(\frac{ \bar{\mu} }{ T }\right)^2 \approx 2\epsilon^2 + {\cal O}(\lgb^2)$.
In Fig.~\ref{fig:echargeeffect}, we show the effect of introducing a nonzero $\epsilon^2$, 
for values up to $\epsilon^2 =0.75$, on the baryon number distribution
after a collision of thick sheets. We see that the effects of choosing $\epsilon^2$ as large
as 0.75 on the baryon number  distribution are very small.
If we extend our expression for the fraction of the total baryon number found
near one lightcone or the other,
$R_{\rm LC}(t)$ as written in \eqref{eq:dcoefdef}, to include 
a term $R_{\epsilon^2}(t)\epsilon^2$, we find that $R_{\epsilon^2}(t=13/m)=0.009$. 
From these findings, and also from expectations based upon 
the results of Ref.~\cite{Casalderrey-Solana:2016xfq}, we 
expect the effect of $\epsilon^2$ to be small for thin sheets also.

\subsection{Asymmetric collisions of thin sheets}

When colliding two incident sheets that both carry baryon number,
it is impossible to determine from which sheet the baryon number deposited
in the plasma originally came from. In a strongly coupled theory one could, for instance, imagine a complete `bounce' where all left-moving baryon number on the lightcone after the collision came from the initially right-moving sheet and vice versa. In order to disentangle 
the contribution to the baryon number in the plasma from reflected and non-reflected baryon
number, it is therefore interesting to look at a collision where only one of the incident
sheets carries baryon number~\cite{Casalderrey-Solana:2016xfq}. 
Indeed, we need to do this check in order to confirm our interpretation of our results in
terms of baryon transparency, which is to say in order to confirm that when we see
increasing baryon number on the left-moving lightcone after the collision as we increase $\beta$,
this baryon number does in fact come from the left-moving incident sheet.

\begin{figure}
\vspace{-0.1in}
\centering
\includegraphics[width=0.48\textwidth]{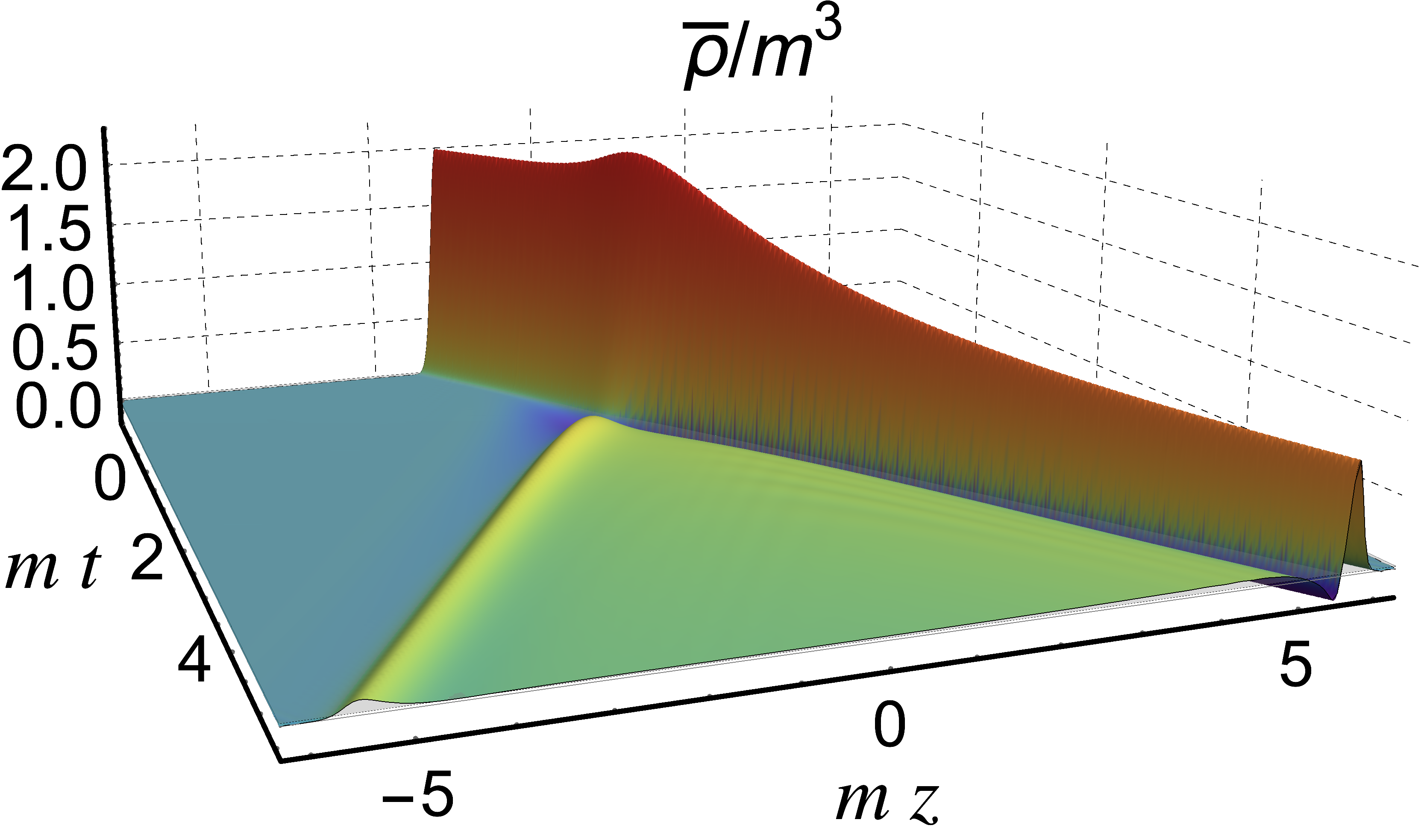}
\includegraphics[width=0.48\textwidth]{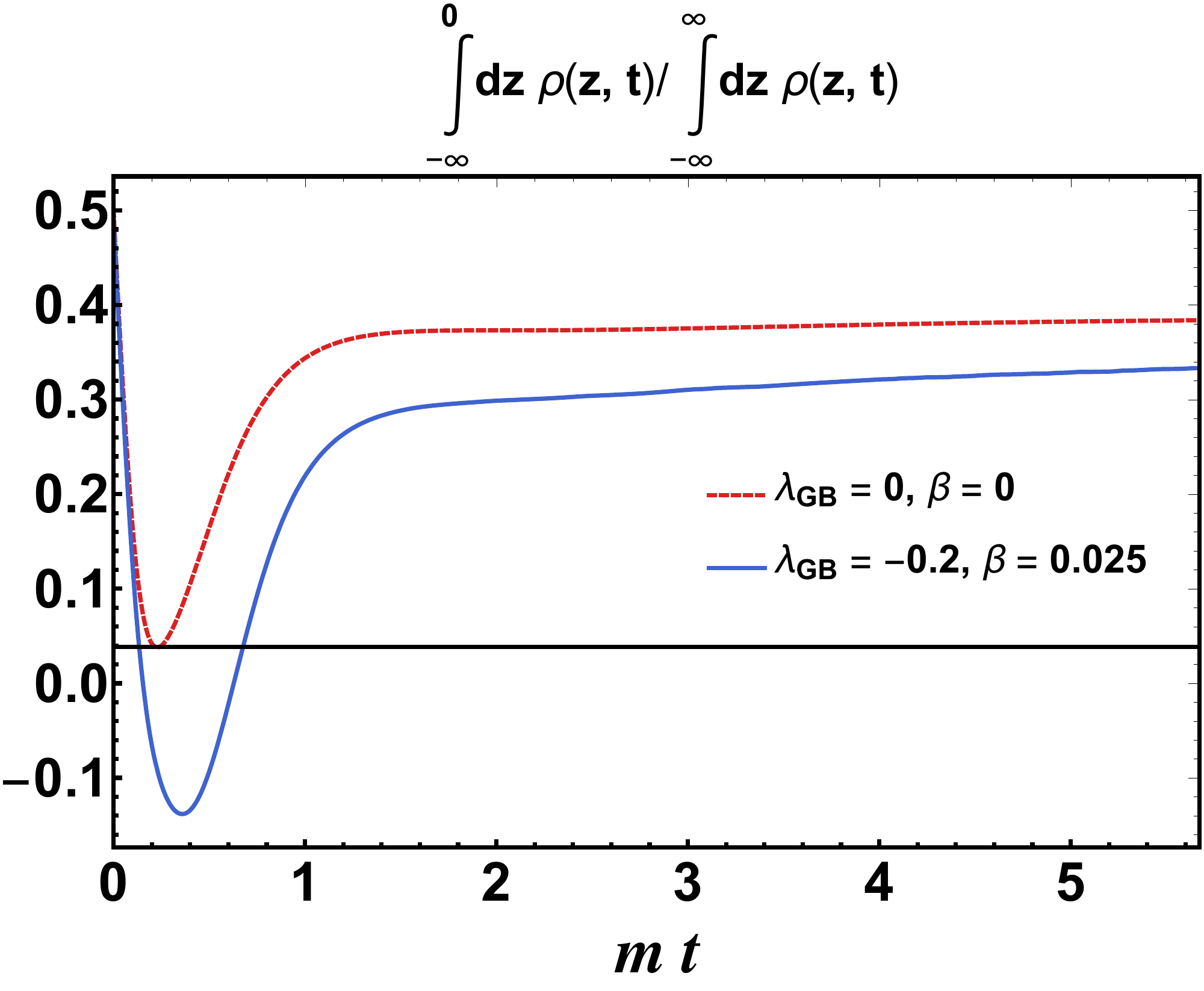}
\newline
\newline
\includegraphics[width=0.75\textwidth]{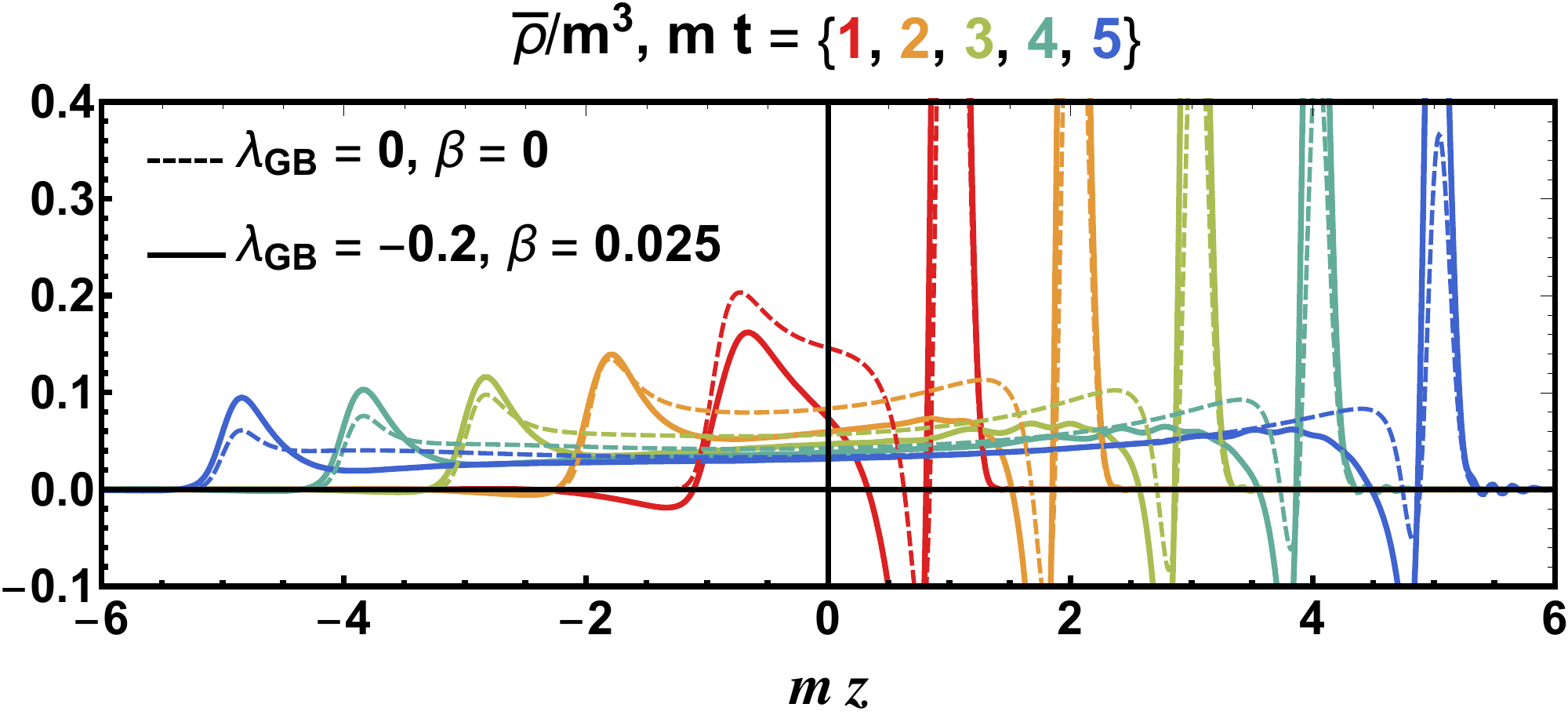}
\caption{The upper-left panel shows the baryon number distribution in the 
collision of two thin sheets, only one of which (the one incident from $z<0$) 
carries baryon number. 
The calculation is done at $\lgb=-0.2$, $\beta=0.025$. 
Relative to what we saw in Fig.~\ref{fig:narrow3D} when both incident sheets 
carried baryon number, the peak in the baryon number density near $z=t=0$
is much smaller here. The absence of the large peak near $z=t=0$ is not surprising, since when both sheets
carry baryon number this comes directly from  summing the two original sheets, but the presence of the small peak here as well as the charge density on the negative-moving lightcone indicates that the sheet of energy density that comes in at the speed of light carrying no baryon number 
immediately obtains some baryon number from the other sheet. Initially, at $m t=1$ this effect is stronger at infinite coupling (see lower panel), but at later times there is more charge on the lightcone for the collision done with nonzero $\lgb$ and $\beta$, 
due to the weaker coupling.  
The amount of baryon number that ends up going through
the collision zone and staying near the positive-going lightcone at late times is
however almost the same as in the case of two colliding charged sheets.
In the lower panel, we look at the time evolution of the baryon number 
density distribution at five time slices.
In the upper-right panel, we compute the fraction of the baryon number
that has been reflected (which is to say that is at $z<0$ after $t=0$) 
as a function of time 
for $\lgb=-0.2$,  $\beta=0.025$ compared to the infinite coupling case, $\lgb=\beta=0$.}
\label{fig:asymmplots}
\end{figure}

In Ref.~\cite{Casalderrey-Solana:2016xfq}, it was found that in a collision of thin
sheets --- at  infinite coupling --- at the time $t=7/m$, 
roughly $40\%$ of the baryon number that arrived on an incident sheet coming from $z<0$
ended up at $z<0$, meaning that it was reflected.
At finite coupling we expect the amount of reflected baryon number
to be reduced, and indeed that is what we find, as shown in Fig.~\ref{fig:asymmplots}. 
At the last plotted time of $t=5.7/m$ we find that in the infinite coupling
case $38\%$ of the baryon number
that came in from $z<0$ has been reflected back, ending up at $z<0$, 
while $33\%$ of the baryon number is reflected for $\lgb=-0.2$ and $\beta=0.025$.
What is perhaps even more interesting is that if we compare the upper-left
panel of Fig.~\ref{fig:asymmplots} to the lower-right panel of 
Fig.~\ref{fig:narrow3D}, we see that the baryon number that ends up
on the positive-going lightcone at late times  is almost the same in
both cases, meaning that it comes from baryon number going through the collision zone not
from reflection.  This confirms the interpretation of our results in terms of 
baryon transparency, and increasing baryon transparency with increasing $\beta$.

\section{Discussion and Outlook}\label{Sec:Discussion}

We have achieved our goal of doing an initial analysis of the 
dynamics of baryon number, as well as energy-momentum, in a holographic
model of heavy ion collisions at a finite value of the gauge coupling, 
to lowest order in the inverse coupling.
In the dual gravitational theory this means that we have worked
to linear order in the two new bulk couplings 
that appear in the most general four-derivative theory with a dynamical metric and gauge field 
in a 4+1-dimensional asymptotically anti-de Sitter spacetime. 
We have also assumed a small baryon number chemical potential compared to the 
equilibrium temperature, $\mu/ T \ll 1$, consistent with heavy ion collisions at RHIC and at the LHC.

The two higher-derivative gravity couplings ($\lgb$ and $\beta$) are both thought of as being proportional to the  inverse 't Hooft coupling in our model gauge theory, with proportionality
coefficients which could only be uniquely fixed had we derived our bulk theory from string theory. 
Rather than doing so, we have used a bottom-up approach to developing the model studied here,
working with 
the most general gravitational action to quartic order in
derivatives of the bulk metric and gauge fields, given 
our choice of discrete symmetries and scalings. 
We have then motivated a qualitative sense for what the 
reasonable range of values for the bulk four-derivative couplings $\lgb$ and $\beta$
may be.  This cannot be done quantitatively as in an effective field theory because we do
not know which deformation of ${\cal N}=4$ SYM theory our gravitational action with
its inverse-coupling corrections is dual to, but at a qualitative level 
we have done so by computing the shear viscosity to entropy
density ratio $\eta/s$ and the baryon susceptibility normalized by its value
in free field theory $\chi/\chi_0$ in our model gauge theory and comparing
the former to what we know from heavy ion collision phenomenology and
the latter to what we know from lattice QCD.
 Based upon many previous studies~\cite{Grozdanov:2014kva,Grozdanov:2016fkt,Grozdanov:2016vgg,Grozdanov:2016zjj,Solana:2018pbk} we expect 
 qualitatively similar behavior to arise from higher-derivative gravitational actions whether
 they have been derived top-down from string theory or bottom-up, as here.

 Our central qualitative result is that as we reduce the gauge coupling from infinity (by increasing $\lgb$ and $\beta$ from zero) the holographic collisions whose dynamics we
 have calculated become increasingly transparent to baryon number.
 The baryon transparency turns out to be much more sensitive to $\beta$.
 As $\beta$ is increased from zero through its reasonable range of 
 values, significantly more of the incident baryon number ends up near the
 lightcones after the collision --- having passed through the collision zone --- 
 and less of it ends up stopped in the strongly coupled plasma produced
 at mid-rapidity.   This is in stark contrast with previous
 results at infinite coupling~\cite{Casalderrey-Solana:2016xfq}. 
 Our present findings show that introducing finite gauge coupling
 corrections to lowest order changes the dynamics of baryon number
 transport in these collisions in a direction that makes the final state
 look more similar to what is seen in heavy ion collisions in QCD.
 This strongly
 supports the hypothesis from Section~\ref{Sec:Intro} that this previously observed discrepancy between the dynamics of baryon number in heavy ion collisions 
 and holographic models for collisions
 can be attributed to the strength of the coupling in the early moments of the collision.
 Furthermore, we find that for a given nonzero value of $\beta$ the increase in baryon
 transparency relative to $\beta=0$ is much more significant for holographic
 collisions of thin sheets
 than for collisions of thick sheets.  This is plausibly consistent with the observation that
 in QCD as the beam energy is increased and the incident nuclei become more
 Lorentz contracted the distribution of the baryon number density after the collision
 becomes more and more different from that in holographic collisions with infinite
 coupling (with $\mu$ at mid-rapidity dropping and the rapidity at which the baryon
 number ends up increasing).  This observation indicates that finite coupling corrections
 to the holographic collisions of thin sheets have ``farther to go'' if they are to take the
 infinite coupling results and make them more realistic, and indeed we find that 
 the finite coupling corrections are larger for collisions of thin sheets.

 We note that we have checked that holographic collisions of incident sheets of
 energy and baryon number hydrodynamize rapidly, even with the introduction of
 finite coupling corrections.  And, we have checked that for values of $\mu/T$ as large as around 1
 the backreaction of the Maxwell field in the bulk action on the gravitational geometry is small,
 indicating that the back reaction of the baryon number dynamics on the dynamics of the
energy density is small. Finally, we have further confirmed the interpretation of our
results in terms of increasing baryon transparency with increasing $\beta$
by analyzing the collision of two incident sheets only one of which carries
baryon number.

Significant open questions remain. 
The most obvious next step would be to include corrections that are higher order in $\lgb$ and $\beta$
in the gravitational action in the analysis of holographic collisions of incident sheets carrying baryon number as well as energy density, as well as other higher-than-four-derivative corrections
whose contribution to the equilibrium thermal spectrum were first addressed in 
Ref.~\cite{Grozdanov:2018gfx}.
For collisions of incident sheets carrying only energy density, without any baryon number, 
i.e.~in the Einstein-Gauss-Bonnet theory with $\beta = 0$, 
the first question was addressed in Ref.~\cite{Grozdanov:2016zjj} where it was shown that, 
at $\lgb = -0.2$ (the value we have also used in this work), corrections at ${\cal O}(\lgb^2)$ 
are indeed small. 
The magnitude of sub-leading $\beta$-dependent terms remains to be determined
in future work, but because the leading $\beta$-dependent terms that we have computed
are so significant in their effects (in particular on the dynamics of baryon number transport in the 
collisions of thin sheets) we expect to see effects of the 
sub-leading $\beta$-dependent terms.  Working to linear order in $\beta$ as we have done,
the increase in the fraction of the total baryon number that ends up near the lightcones
is linear in $\beta$.  In collisions of  thin sheets, with $\beta=0.05$ and $\lgb=-0.2$ that fraction
is already more than 50\%, and with $\beta=0.1$ and $\lgb=-0.2$ it would be more than 80\% if we work to linear order in $\beta$.  We therefore expect that in this range of values of $\beta$ the effect of contributions that are higher-order in $\beta$ should become significant.

A more challenging, and more important, future goal is to find a holographic model setting in
which one can analyze collisions in which the coupling is weak (or at least finite and intermediate in value, not infinite) at very early times while the coupling at later times, in particular 
during the hydrodynamic expansion, is strong.
Our interpretation of why the baryon number distribution in heavy ion collisions in QCD is
so different from that in holographic collisions at infinite coupling is that in QCD the coupling
is weak during the first moments of the collision, and if the nuclei are sufficiently 
highly Lorentz contracted then
by the time QGP starts to form the incident baryon number has already passed through
the collision zone.  Later, the QGP that forms in QCD is indeed a strongly coupled liquid.
So, to improve the holographic modelling of heavy ion collisions in QCD, instead
of reducing the gauge coupling (by introducing inverse-coupling corrections) 
at all times as we have done it would be better to explore methods of doing so only at early times.
This challenge of course reflects the asymptotic freedom of QCD, but addressing
this challenge framed in these terms may be easier than implementing asymptotic
freedom in a holographic gauge theory in full.  It remains a challenge for future work, however.

\acknowledgments

We are grateful to Peter Arnold, Jorge Casalderrey-Solana, Christian Ecker, Guy Moore, Andrei Starinets and Urs Wiedemann for helpful conversations. KR and WS gratefully acknowledge the hospitality of the CERN Theory group. We also thank Iain Stewart for making computational resources available. This work was supported by the U.S.~Department of Energy, Office of Science, Office of Nuclear Physics through grant
DE-SC0011090. WS is supported by the Netherlands Organisation for Scientific Research (NWO) under the grant 680-47-458 (Veni). The work of \r{A}F was supported by an Aker Scholarship.

\begin{appendix}
\section{General four-derivative action and field redefinitions}\label{Sec:ActFD}

In this Appendix, we review the construction of the most general four-derivative action involving a metric tensor $g_{\mu\nu}$ and a vector gauge field $A_\mu$ in a five-dimensional asymptotically AdS spacetime, 
and show how to use field redefinitions and a few simplifying choices to cast this action in the form \eqref{ActFin} that we use throughout this paper.
The presentation largely follows Ref.~\cite{Myers:2009ij}. The most general action can be written as
\begin{align}
S = \frac{1}{2 \kappa_5^2} \int d^5 x \sqrt{-g} \CL ,
\end{align}
where we split $\CL$ into two-derivative and four-derivative terms $\CL_2$ and $\CL_4$, respectively. They are
\begin{align}
    \CL_2 = R + \frac{12}{\tilde L^2} - \frac{\tilde{L}^2}{4} F_{\mu\nu} F^{\mu\nu} + \frac{\kappa_1 \tilde{L}^3}{3} \varepsilon^{\mu\nu\rho\sigma\lambda} A_\mu F_{\nu\rho} F_{\sigma\lambda},
    \label{twoderiv}
\end{align}  
where $\tilde L$ is a length scale that we shall later relate to the AdS radius and
\begin{align}
\label{fourderiv}
\CL_4 / \tilde L^2&= \, \alpha_1  R^2  + \alpha_2   R_{\mu\nu} R^{\mu\nu} + \alpha_3  R_{\mu\nu\rho\sigma} R^{\mu\nu\rho\sigma} \nn
& +  \beta_1 \tilde L^2 R F_{\mu\nu} F^{\mu\nu}  + \beta_2 \tilde L^2  R^{\mu\nu} F_{\mu\rho} F_\nu^{~\rho} + \beta_3 \tilde L^2  R^{\mu\nu\rho\sigma} F_{\mu\nu} F_{\rho\sigma}  \nn
& + \delta_1 \tilde L^4 \left(F_{\mu\nu} F^{\mu\nu} \right)^2 +  \delta_2 \tilde L^4 F^{\mu\nu} F_{\nu\rho} F^{\rho\sigma} F_{\sigma\mu}  +  \delta_3  \tilde L^2 \nabla_\mu F^{\mu\nu} \nabla^{\rho} F_{\rho\nu}  \nn
&+ \gamma_1 \tilde L^3 \varepsilon^{\mu\nu\rho\sigma\lambda} F_{\mu\nu} F_{\rho\sigma} \nabla^\chi F_{\chi\lambda}  + \kappa_2 \tilde L \varepsilon^{\mu\nu\rho\sigma\lambda} A_\mu R_{\nu\rho\chi\zeta} R_{\sigma\lambda}^{~~\,\chi\zeta},
\end{align} 
where we treat all dimensionless coupling constants in $\CL_4$ as perturbatively small.
Other possible terms in the action can be removed by using standard metric identities and integration by parts. Because doing so leaves the dual field theory unchanged, we can now perform first-order field redefinitions $g_{\mu\nu} \to g_{\mu\nu} + \delta g_{\mu\nu}$ and $A_\mu \to A_{\mu} + \delta A_\mu$, where
\begin{align}
\delta g_{\mu\nu} & \equiv \mu_1 \tilde L^2 R_{\mu\nu} + \mu_2 \tilde L^4 F_{\mu\lambda} F_{\nu}^{~\lambda} + \mu_3 \tilde L^2 R g_{\mu\nu} + \mu_4 \tilde L^4 F_{\rho\sigma} F^{\rho\sigma} g_{\mu\nu} + \mu_5 g_{\mu\nu} , \label{FRd-1} \\
\delta A_\mu &\equiv \lambda_1 A_\mu + \lambda_2 \tilde L^2 \nabla^\nu F_{\nu\mu} + \lambda_3 \tilde L^3 \varepsilon_{\mu\nu\rho\sigma\lambda} F^{\nu\rho} F^{\sigma\lambda} . \label{FRd-2}
\end{align}
The $\mu_i$'s and the $\lambda_i$'s are constants that we may choose in order to specify a particular field redefinition.
After this redefinition, the gravitational action transforms as $S \to S + \delta S$, where we write the transformations of the two- and four-derivative parts as  
\begin{align}
\delta S_2 = \frac{1}{2 \kappa_5^2} \int d^5 x \sqrt{-g} \,\delta \CL_2  , \qquad \delta S_4 = \frac{1}{2 \kappa_5^2} \int d^5 x \sqrt{-g}\, \delta \CL_4 . 
\end{align}
The two-derivative part of the variation is
\begin{align}
\CL_2 + \delta \CL_2 &= \left(1 + 6\mu_1 + 30 \mu_3 + \frac{3 \mu_5}{2} \right) R+  \frac{12}{{\tilde L}^2} \left( 1+ \frac{5\mu_5}{2}\right) +   \nn
&+ \left(- \frac{1}{4} + 6\mu_2 + 30 \mu_4 - \frac{\mu_5}{8} - \frac{\lambda_1}{2} \right) \tilde L^2 F_{\mu\nu} F^{\mu\nu} + \frac{\kappa_1 \tilde L^3}{3}  \left( 1 + 3\lambda_1 \right) \varepsilon^{\mu\nu\rho\sigma\lambda} A_\mu F_{\nu\rho} F_{\sigma\lambda} ,
\end{align}
while at the four-derivative order, 
\begin{align}
\left( \CL_4 + \delta \CL_4 \right) / \tilde L^2 & = \CC_1 R^2 + \CC_2 R_{\mu\nu} R^{\mu\nu} + \CC_3 R_{\mu\nu\rho\sigma} R^{\mu\nu\rho\sigma} \nn
& +  \CC_4 \tilde L^2 R F_{\mu\nu} F^{\mu\nu}  + \CC_5 \tilde L^2 R^{\mu\nu} F_{\mu\rho} F_\nu^{~\rho} + \CC_6  \tilde L^2R^{\mu\nu\rho\sigma} F_{\mu\nu} F_{\rho\sigma}  \nn
& + \CC_7\tilde L^4 \left(F_{\mu\nu} F^{\mu\nu} \right)^2 +  \CC_8 \tilde L^4 F^{\mu\nu} F_{\nu\rho} F^{\rho\sigma} F_{\sigma\mu}  +  \CC_9  \tilde L^2 \nabla_\mu F^{\mu\nu} \nabla^{\rho} F_{\rho\nu}  \nn
&+ \CC_{10}\tilde L^3  \varepsilon^{\mu\nu\rho\sigma\lambda} F_{\mu\nu} F_{\rho\sigma} \nabla^\chi F_{\chi\lambda}  + \CC_{11}  \tilde L \varepsilon^{\mu\nu\rho\sigma\lambda} A_\mu R_{\nu\rho\chi\zeta} R_{\sigma\lambda}^{~~\,\chi\zeta},
\end{align}
with
\begin{align}
\CC_1 &= \alpha_1 + \frac{\mu_1}{2} + \frac{3\mu_3}{2} , && &\CC_2 &= \alpha_2 - \mu_1 , \\
\CC_3 &= \alpha_3, && &\CC_4 &= \beta_1 - \frac{\mu_1}{8} + \frac{\mu_2}{2} - \frac{\mu_3}{8} + \frac{3\mu_4}{2} ,\\
\CC_5 &= \beta_2 + \frac{\mu_1}{2} - \mu_2 , && & \CC_6 &= \beta_3, \\
\CC_7 &= \delta_1 - \frac{\mu_2}{8} - \frac{\mu_4}{8} + 8 \kappa_1 \lambda_3, && & \CC_8 &= \delta_2 + \frac{\mu_2}{2} - 16 \kappa_1 \lambda_3  , \\
\CC_9&= \delta_3 + \lambda_2 , && & \CC_{10} &= \gamma_1 + \kappa_1\lambda_2 + \lambda_3 , \\
\CC_{11} & = \kappa_2 .&&&&
\end{align}

The coupling constants $\CC_n$ in the action obtained after the field redefinition depend on the
coupling constants in the original action and on the constants  ($\mu_i$'s and $\lambda_i$'s) that
specify the field redefinition.  We can now choose these constants, specifying the field redefinition.
First, though, since we are only interested in theories without any Chern-Simons-type terms (terms with the Levi-Civita symbol that are odd in $A_\mu$), we set 
\begin{align}
\kappa_1 = \kappa_2 = \gamma_1 = \lambda_3 = 0. \label{KillCS}
\end{align} 
Next, in order to keep the two-derivative part of the action in `canonical' form after the field redefinition, we choose
\begin{align}
6\mu_1 + 30 \mu_3 + \frac{3 \mu_5}{2} &= 0, \label{FRD-2nd-1} \\
 \frac{12}{{\tilde L}^2} \left( 1+ \frac{5\mu_5}{2}\right) & = \frac{12}{L^2}, \label{FRD-2nd-2}   \\
\left(- \frac{1}{4} + 6\mu_2 + 30 \mu_4 - \frac{\mu_5}{8} - \frac{\lambda_1}{2} \right) \tilde L^2   & = -\frac{L^2}{4} .\label{FRD-2nd-3} 
\end{align}
We then choose to fix $\mu_5$ and $\lambda_1$ as
\begin{align}
\mu_5 &= -4 \mu_1 - 20 \mu_3 ,\\
\lambda_1 &= 6\left( \mu_1 + 2 \mu_2 + 5 \mu_3 + 10 \mu_4\right) ,
\end{align}
and define the AdS radius $L$ in terms of $\tilde L$ by 
\begin{align}
    L^2 = \tilde L^2  \left(1 + 10\mu_1 + 50 \mu_3 \right).\label{eq:Ltransform}
\end{align}
It is important to notice that neither $\CC_3=\alpha_3$ nor $\CC_6=\beta_3$ can be changed by using the field redefinitions \eqref{FRd-1}--\eqref{FRd-2}. These constants in the action after field redefinition are (related to) the Gauss-Bonnet coupling $\lgb$ and the new coupling $\beta$:
\begin{align}
\CC_3 &= \alpha_3 \equiv \frac{\lgb}{2}, \\
\CC_6 &= \beta_3 \equiv \beta.
\end{align}
In the absence of any terms with the Levi-Civita symbol in the theory, i.e.~with \eqref{KillCS}, we have seven remaining unfixed $\CC_n$ with $n = \{1,2,4,5,7,8,9\}$ and the remaining 
freedom to choose five coefficients, $\mu_1$, $\mu_2$, $\mu_3$, $\mu_4$ and $\lambda_2$.  First, we choose 
$\mu_1$, $\mu_2$, $\mu_3$ and $\mu_4$ so as to satisfy
the following set of linear equations, because doing so brings the action into the form that gives only two-derivative equations of motion:
\begin{align}
\CC_1 &= \alpha_1 + \frac{\mu_1}{2} + \frac{3\mu_3}{2}  = \frac{\lgb}{2} ,\\
\CC_2 &= \alpha_2 - \mu_1 = - 2\lgb ,\\
\CC_4 &= \beta_1 - \frac{\mu_1}{8} + \frac{\mu_2}{2} - \frac{\mu_3}{8} + \frac{3\mu_4}{2} = \beta, \\
\CC_5 &= \beta_2 + \frac{\mu_1}{2} - \mu_2 = - 4 \beta.
\end{align}
These equations imply that 
\begin{align}
\mu_1 &= \alpha_2 + 2 \lgb ,\\
\mu_2 &= \frac{\alpha_2}{2} + \beta_2 + \lgb + 4 \beta , \\
\mu_3 &= - \frac{1}{3} \left( 2 \alpha_1 + \alpha_2 + \lgb  \right) ,\\
\mu_4 &= - \frac{1}{36} \left( 2 \alpha_1 + 4 \alpha_2 + 24 \beta_1 + 12 \beta_2 + 7 \lgb + 24 \beta \right).
\end{align}
We are left with a single free coefficient, $\lambda_2$, and three $\CC_n$'s. We specify $\lambda_2$ so as to make $\CC_9$ to vanish:
\begin{align}
\CC_9 = \delta_3 + \lambda_2  = 0 , 
\end{align}
which gives
\begin{align}
\lambda_2 = - \delta_3.
\end{align}
There is then no remaining freedom, and the two remaining $\CC_n$'s are given by
\begin{align}
\CC_7 &= \delta_1 - \frac{\mu_2}{8} - \frac{\mu_4}{8} = \frac{1}{288} \left( 2 \alpha_1 - 14 \alpha_2 + 24 \beta_1 - 24 \beta_2 + 288 \delta_1 - 29 \lgb - 120 \beta  \right) \equiv \zeta_1 , \\
\CC_8 &= \delta_2 + \frac{\mu_2}{2} =\frac{1}{4} \left( \alpha_2 + 2 \beta_2 + 4 \delta_2 + 2\lgb + 8 \beta \right) \equiv \zeta_2,
\end{align}
where we have defined two new couplings $\zeta_1$ and $\zeta_2$. 
To first order in four-derivative couplings, this gives the action $S' = S + \delta S$:
\begin{align}
S' &=  \frac{1}{2 \kappa_5^2} \int d^5 x \sqrt{-g}  \biggr[  R + \frac{12}{L^2} - \frac{L^2}{4} F_{\mu\nu} F^{\mu\nu} +  \frac{\lgb L^2}{2} \left(R^2  - 4 R_{\mu\nu} R^{\mu\nu} +  R_{\mu\nu\rho\sigma} R^{\mu\nu\rho\sigma}  \right) \nn
& +  \beta L^4  \left( R F_{\mu\nu} F^{\mu\nu}   - 4 R^{\mu\nu} F_{\mu\rho} F_\nu^{~\rho} + R^{\mu\nu\rho\sigma} F_{\mu\nu} F_{\rho\sigma} \right) + \zeta_1 L^6 \left(F_{\mu\nu} F^{\mu\nu} \right)^2 +  \zeta_2 L^6 F^{\mu\nu} F_{\nu\rho} F^{\rho\sigma} F_{\sigma\mu}  \biggr] .
\end{align}
Finally, the scaling of four-derivative couplings and the size of the gauge field that we employ in this work (see Section \ref{Sec:HolographicSetup}) are $\lgb \sim \beta \sim \zeta_1 \sim \zeta_2  \sim \epsilon^2$ and $|A_\mu| \sim  \epsilon $, with small dimensionless $\epsilon \ll 1$. It is convenient to explicitly rescale $A_\mu \to \epsilon A_\mu$, so that to order $\epsilon^4 \sim \lgb^2 \sim \ldots $,
\begin{align}
S' &=  \frac{1}{2 \kappa_5^2} \int d^5 x \sqrt{-g}  \biggr[  R + \frac{12}{L^2} - \frac{\epsilon^2 L^2}{4} F_{\mu\nu} F^{\mu\nu} +  \frac{\lgb L^2}{2} \left(R^2  - 4 R_{\mu\nu} R^{\mu\nu} +  R_{\mu\nu\rho\sigma} R^{\mu\nu\rho\sigma}  \right) \nn
& +  \beta \epsilon^2 L^4  \left( R F_{\mu\nu} F^{\mu\nu}   - 4 R^{\mu\nu} F_{\mu\rho} F_\nu^{~\rho} + R^{\mu\nu\rho\sigma} F_{\mu\nu} F_{\rho\sigma} \right)  \biggr] .
\end{align}
Both terms with four powers of $F_{\mu\nu}$ are proportional to $\zeta_n \epsilon^4 \sim \epsilon^6$ and can therefore be neglected. Thus, we have arrived at the final form of the action \eqref{ActFin} that we use in this paper, which results in purely second-order equations of motion.

Using the above field redefinition, we can connect the thermodynamic relations used in this work with the results discussed in Ref.~\cite{Myers:2009ij}. The Lagrangian of Ref.~\cite{Myers:2009ij}, up to a rescaling of $A_{\mu}\rightarrow A_{\mu} \tilde{L} $, is obtained from \eqref{twoderiv} and \eqref{fourderiv} by setting 
\begin{align}
    \alpha_1 &= \alpha_2 = \beta_1 = \beta_2 = \gamma_1 = \delta_3 = 0, \nonumber \\ 
    \alpha_3 &= c_1=\frac{ \lgb }{ 2 },\quad \beta_3 = c_2 =\beta ,\quad \delta_1 = c_3, \quad \delta_2 = c_4,\quad \kappa_2 = c_5\quad \kappa_1 = \kappa.\label{eq:myersCoefs}
\end{align}
Since $c_5$ and $\kappa$ do not affect thermodynamics, we can set them to zero. Working with the assumption of a small chemical potential, i.e. that $A_{\mu}\sim O(c_i)$, the results 
of Ref.~\cite{Myers:2009ij} for the baryon number and conductivity to the order we are working at are 
\begin{equation}
    \tilde{\rho} = \frac{ \pi^2  \tilde{L}^3 }{\kappa_5^2}\tilde{\mu} \tilde{T}^2\left(1 + \frac{ 11 }{ 3 }c_1 + \frac{ \tilde{\mu}^2 }{ 3\pi^2 \tilde{T}^2 } \right),     \qquad \tilde{\sigma} = \frac{ \pi \tilde{L}^3 \tilde{T}}{ 2\kappa_5^{2}  }\left( 1 + \frac{ 5 }{ 3 }c_1 + 16 c_2 - \frac{5 \tilde{\mu}^2 }{6\pi^2 \tilde{T}^2 }\right),\label{eq:MyersSols}
\end{equation}
where we have adjusted for a relative factor of $\pi$ in our definition of the chemical potential.
We use tildes to indicate the temperature and the chemical potential for the theory in Ref.~\cite{Myers:2009ij}.
Now, it turns out that at the boundary, the field redefinitions \eqref{FRd-1} and \eqref{FRd-2} with the choice of the coefficients in \eqref{eq:myersCoefs} simply reduce to global rescalings 
\begin{equation}
    g_{\mu\nu} \rightarrow \left(1 - \frac{ 8 }{ 3 }\lgb\right) g_{\mu\nu} \equiv e^{-2\omega_1}g_{\mu\nu},\qquad
    A_{\mu} \rightarrow \left(1 + 8\beta + \frac{ 7 }{ 3 }\lgb\right) A_{\mu} \equiv e^{-2\omega_2}A_{\mu},
\end{equation}
with 
\begin{equation}
\omega_1 = \frac{ 4 }{ 3 }\lgb, \qquad \omega_2 = - 4 \beta - \frac{ 7 }{ 6 }\lgb.
\end{equation}
In terms of the boundary metric, this is a constant Weyl transformation, under which the transformation properties of thermodynamic and hydrodynamic quantities are well known (see Refs.~\cite{Baier:2007ix,Loganayagam:2008is,Grozdanov:2014kva}). They follow from the condition that the generating functional of the conformal field theory must remain invariant under such transformations, giving
\begin{equation}
    T \rightarrow e^{\omega_1}T, \quad \mu \rightarrow e^{\omega_1}\mu,\quad \rho \rightarrow e^{3\omega_1}\rho,\quad \mathcal{E} \rightarrow e^{4\omega_1}\mathcal{E}, \quad \sigma \rightarrow e^{\omega_1}\sigma, \quad \cdots\, .
\end{equation}
An alternative way to state this, from the point of view of gauge-gravity duality, is that the generating functional (or the partition function) must remain invariant under field redefinitions. Following analogous logic, we impose that the generating functional must also remain invariant under the rescaling transformation of the gauge field, which gives
\begin{equation}
\mu \rightarrow e^{-2\omega_2} \mu, \quad \rho  \rightarrow e^{2\omega_2}\rho, \quad \sigma \rightarrow e^{4 \omega_2}\sigma, \quad \cdots\,.
\end{equation}
The transformation of $\mu$ follows directly from the identification of the chemical potential with the gauge field on the boundary. 
The transformation of the local baryon number density can be found from the fact that $\mu \rho$ should transform the same way as $\mathcal{E}$, which does not change under the rescaling of the gauge field. The conductivity transformation follows from the Kubo formula, which schematically relates $\sigma \sim [\rho, \rho]$. Thus, we find the following relation between thermodynamic quantities and transport coefficients in our theory and the theory of Ref.~\cite{Myers:2009ij}:
\begin{equation}
    \tilde{T} = e^{\omega_1}T, \quad  \tilde{\mu} =  e^{\omega_1-2\omega_2} \mu, \quad \tilde{\rho} = e^{3\omega_1+2\omega_2}\rho,
    \quad \tilde{\sigma} = e^{\omega_1+4\omega_2}\sigma , \quad \cdots\,.
\end{equation}
Furthermore, we solve \eqref{eq:Ltransform} to give 
\begin{equation}
    \tilde{L} = L\left( 1- \frac{ 5 }{ 3 }\lgb\right). 
\end{equation}
With the above relations we can now use \eqref{eq:MyersSols} to solve for $\sigma$ and $\rho$ in terms of $L$, $ T$, and $\mu$, finding 
\begin{equation}
    \rho = \frac{ \pi^2 L^3}{ \kappa_5^2 } \mu T^2 \left(1+16\beta + \frac{ 3 }{ 2 }\lgb + \frac{ \mu^2 }{ 3\pi^2 T^2 }\right), \qquad 
    \sigma = \frac{\pi  L^3 T }{ 2\kappa_5^2 }\left(1+32\beta + \frac{ 1 }{ 2 }\lgb - \frac{5  \mu^2 }{ 6\pi^2 T^2 }\right).
\end{equation}
Plugging in $L=1 +  \lgb/2$, $\mu = \epsilon \bar{\mu}$ and setting $\kappa_5^2 = 2$ to conform with the conventions used in this work (cf. Eq. \eqref{rescaledCurrents}), we arrive at equations \eqref{eq:rhomuexp} and \eqref{eq:conductivitydef}. Continuing further, we can also show agreement between the entropy, free energy and energy density from Ref.~\cite{Myers:2009ij} after the field redefinitions that we have specified and the same quantities in our theory. We note that the agreement between our entropy, calculated using the formula $A/ 4G_N $, and the entropy obtained from the result of Ref.~\cite{Myers:2009ij} after field redefinitions, which is calculated from the (higher-derivative) Wald formula~\cite{Wald:1993nt,Iyer:1994ys}, confirms that our theory (in the case of a black brane) 
has the property that the higher derivative corrections to $A/4 G_N$ appearing in 
the Wald formula vanish, meaning that the entropy is given by the Bekenstein-Hawking 
result~\cite{Wald:1993nt,Iyer:1994ys}.

\end{appendix}

\bibliography{biblio}
\end{document}